\documentclass[journal]{IEEEtran}
\ifCLASSINFOpdf
\else
   \usepackage{graphicx}
   \graphicspath{{../eps/}}
   \DeclareGraphicsExtensions{.eps}
\fi

\usepackage{amsmath}
\usepackage{amsfonts,amssymb}
\usepackage{amssymb}
\usepackage{wrapfig}
\usepackage{psfrag}
\usepackage{epstopdf}
\usepackage{cite}
\usepackage{graphicx}
\usepackage{subfigure}
\usepackage{threeparttable}
\usepackage{cases}
\usepackage{subeqnarray}
\usepackage{color}
\usepackage{underscore}
\usepackage{verbatim}
\usepackage{bm}
\usepackage{stfloats}
\usepackage{longtable}
\usepackage{rotating}
\usepackage{diagbox}
\usepackage{pifont}
\usepackage{xcolor}
\usepackage{multicol}       % table
\usepackage{multirow}       % table
\usepackage{array}          % table
\usepackage{colortbl}
\usepackage{makecell}
\definecolor{crimson}{RGB}{192,0,0}         % color crimson
\definecolor{navy}{RGB}{47,85,151}         % color crimson
\usepackage{bbding}
\usepackage{booktabs}
\usepackage{xpatch}
\usepackage{xcolor}

\usepackage{algorithm}
\usepackage{algorithmic}

\begin{document}
% paper title
\title{A Tutorial on Near-Field XL-MIMO Communications Towards 6G}

%% author names and IEEE memberships
%% note positions of commas and nonbreaking spaces ( ~ ) LaTeX will not break
%% a structure at a ~ so this keeps an author's name from being broken across
%% two lines.
%% use \thanks{} to gain access to the first footnote area
%% a separate \thanks must be used for each paragraph as LaTeX2e's \thanks
%% was not built to handle multiple paragraphs
%%
%
\author{
        Haiquan~Lu,
        Yong~Zeng,~\IEEEmembership{Senior Member,~IEEE,}
        Changsheng~You,~\IEEEmembership{Member,~IEEE,}
        Yu~Han,~\IEEEmembership{Member,~IEEE,}
        Jiayi~Zhang,~\IEEEmembership{Senior Member,~IEEE,}
        Zhe~Wang,
        Zhenjun~Dong,
        Shi~Jin,~\IEEEmembership{Fellow,~IEEE,}
        Cheng-Xiang~Wang,~\IEEEmembership{Fellow,~IEEE,}
        Tao~Jiang,~\IEEEmembership{Fellow,~IEEE,}
        Xiaohu~You,~\IEEEmembership{Fellow,~IEEE,}

        and Rui~Zhang,~\IEEEmembership{Fellow,~IEEE}
        %{\it (Invited Paper)}
%        % <-this % stops a space
\thanks{This work was supported by the National Key R\&D Program of China with Grant number 2019YFB1803400. (\emph{Corresponding author: Yong Zeng.}) }
\thanks{Haiquan Lu, Yong Zeng, Yu Han, Zhenjun Dong, Shi Jin, Cheng-Xiang Wang, and Xiaohu You are with the National Mobile Communications Research Laboratory, Southeast University, Nanjing 210096, China. Haiquan Lu, Yong Zeng, Cheng-Xiang Wang, and Xiaohu You are also with the Purple Mountain Laboratories, Nanjing 211111, China (e-mail: \{haiquanlu, yong_zeng, hanyu, zhenjun_dong, jinshi, chxwang, xhyu\}@seu.edu.cn). }
\thanks{Changsheng You is with the Department of Electronic and Electrical
Engineering, Southern University of Science and Technology (SUSTech),
Shenzhen 518055, China (e-mail: youcs@sustech.edu.cn).}
\thanks{Jiayi Zhang and Zhe Wang are with the School of Electronic and Information Engineering, Beijing Jiaotong University, Beijing 100044, China (e-mail: \{jiayizhang, zhewang_77\}@bjtu.edu.cn).}
\thanks{Tao Jiang is with the Research Center of 6G Mobile Communications, School of Cyber Science and Engineering, and the School of Electronic Information and Communications, Huazhong University of Science and Technology, Wuhan 430074 (e-mail: Tao.Jiang@ieee.org).}
\thanks{Rui Zhang is with School of Science and Engineering, Shenzhen Research Institute of Big Data, The Chinese University of Hong Kong, Shenzhen, Guangdong 518172, China (e-mail: rzhang@cuhk.edu.cn). He is also with the Department of Electrical and Computer Engineering, National University of Singapore, Singapore 117583 (e-mail: elezhang@nus.edu.sg).}
%
% %<-this % stops a space
}

% make the title area
\maketitle

% As a general rule, do not put math, special symbols or citations
% in the abstract or keywords.
\begin{abstract}
 Extremely large-scale multiple-input multiple-output (XL-MIMO) is a promising technology for the sixth-generation (6G) mobile communication networks. By significantly boosting the antenna number or size to at least an order of magnitude beyond current massive MIMO systems, XL-MIMO is expected to unprecedentedly enhance the spectral efficiency and spatial resolution for wireless communication. The evolution from massive MIMO to XL-MIMO is not simply an increase in the array size, but faces new design challenges, in terms of near-field channel modelling, performance analysis, channel estimation, and practical implementation. In this article, we give a comprehensive tutorial overview on near-field XL-MIMO communications, aiming to provide useful guidance for tackling the above challenges. First, the basic near-field modelling for XL-MIMO is established, by considering the new characteristics of non-uniform spherical wave (NUSW) and spatial non-stationarity. Next, based on the near-field modelling, the performance analysis of XL-MIMO is presented, including the near-field signal-to-noise ratio (SNR) scaling laws, beam focusing pattern, achievable rate, and degrees-of-freedom (DoF). Furthermore, various XL-MIMO design issues such as near-field beam codebook, beam training, channel estimation, and delay alignment modulation (DAM) transmission are elaborated. Finally, we point out promising directions to inspire future research on near-field XL-MIMO communications.
\end{abstract}

% Note that keywords are not normally used for peerreview papers.
\begin{IEEEkeywords}
 Extremely large-scale MIMO, near-field modelling, non-uniform spherical wave, spatial non-stationarity, near-field SNR scaling law, beam focusing pattern, near-field codebook, near-field beam training, near-field inter-user interference.
\end{IEEEkeywords}

\IEEEpeerreviewmaketitle
% >>>>>>>>>>>>>SECTIONS I -  here >>>>>>>>>>>>
\section{Introduction}
\subsection{Background}
 While the fifth-generation (5G) mobile communication networks are being deployed worldwide, both academia and industry have envisioned the roadmap to the future sixth-generation (6G) wireless systems to accommodate diverse foreseeable applications such as immersive reality, metaverse, and fully autonomous vehicles  \cite{tataria20216g,tariq2020speculative,letaief2019roadmap,you2021towards,wang2023road}. In June 2023, the International Telecommunication Union (ITU) has released its visions for 6G, together with the timeline, future technology trends, recommended frameworks, which marks the official kick-off of the journey towards 6G standardization \cite{ITU}. In particular, six major usage scenarios are defined, including immersive communication (eMBB+), massive communication (mMTC+), hyper reliable and low-latency communication (URLLC+), which are extensions of the usage scenarios defined in 5G, as well as three new items that will flourish in the new era of 6G, namely, integrated sensing and communication (ISAC), integrated artificial intelligence (AI) and communication, and ubiquitous connectivity. Moreover, customized key performance indicators (KPIs) for IMT-2030 (6G) are presented, which consist of nine enhanced capabilities and six new capabilities. Compared to 5G, 6G is expected to achieve a $100$-fold increase in peak data rate (from Gbps to Tbps), a $10$-fold latency reduction with a hyper-reliability requirement of $99.99999\%$, a $10$-fold improvement in connection density \cite{SamsungWhitePaper6G,wang2023road}. These stringent requirements, however, may not be achieved by existing 5G technologies, hence calling for new and disruptive technologies for 6G.

 In this context, several promising 6G candidate technologies have been proposed, such as extremely large-scale multiple-input multiple-output (XL-MIMO) \cite{bjornson2019massive,lu2021how,lu2022communicating,cui2022channel}, ISAC \cite{liu2022integrated}, and Terahertz (THz) communications \cite{akyildiz2014teranets,ITU1}. In particular, as a natural evolution of the contemporary massive MIMO technology, XL-MIMO further boosts the number of antennas by at least an order of magnitude, e.g., several hundreds or even thousands of antennas \cite{bjornson2019massive,tong20216g,wang20206g,xiao2023error}, thus unprecedentedly improving the spectral efficiency and spatial resolution for wireless communication and sensing. As such, XL-MIMO is perceived to be a key enabling technology for 6G to fulfill several stringent KPIs, such as peak data rate, spectral efficiency, reliability, positioning and sensing accuracy \cite{tong20216g,xiao2022overview}. Note that the recent evolutions of mobile communication networks are accompanied by the advances in MIMO technology. Initially, the Third Generation Partnership Project (3GPP) standardized the first MIMO specification in Release 7 at the tail end of the third generation (3G) era \cite{3GPPTR25.913}. Subsequently, MIMO technology flourished in the fourth-generation (4G) mobile networks and was recognized as the essential transmission technology, where the 4G long-term evolution (LTE) advanced network supported up to $8 \times 8$ MIMO. Hitherto, MIMO technology had evolved to massive MIMO in 5G, whose typical configuration at the base station (BS) is 64 antenna elements \cite{bjornson2019massive,zhang2020prospective}. Looking forward to the forthcoming beyond 5G (B5G) and 6G era, massive MIMO is expected to evolve towards XL-MIMO, so as to propel the aforementioned usage scenarios into a reality. For example, the drastically improved beamforming gain and spectral efficiency of XL-MIMO are believed to be essential for eMBB+ applications, such as the augmented/virtual/mixed reality and holographic display \cite{tong20216g,you2021towards,kang2022personalized,du2023attention}. Besides, deploying an extremely large number of antennas results in large array aperture, thus enhancing the array spatial resolution unprecedentedly, which is beneficial to mMTC+, as well as the high-accuracy localization and sensing \cite{Latvaaho2019Keydrivers,tong20216g,you2021towards}. Besides XL-MIMO, other similar terminologies used in the literature include extremely large aperture array (ELAA) \cite{bjornson2019massive}, ultra-massive MIMO (UM-MIMO) \cite{wang2023road,akyildiz2016realizing}, and extremely large aperture massive MIMO (xMaMIMO) \cite{amiri2018extremely}.

\subsection{XL-MIMO: New Channel Characteristics}\label{subSectionNewCharacteristics}
 However, the evolution from massive MIMO to XL-MIMO is not a simple increase in antenna number or size, but fundamentally changes the channel characteristics, e.g., shifting from the conventional far-field uniform plane wave (UPW) to the new non-uniform spherical wave (NUSW) propagation \cite{lu2021how,lu2022communicating}, and from the conventional spatial stationarity to spatial non-stationarity \cite{decarvalho2020nonstationarities,dong2022near,Han2023Towards,zheng2023novel}, as discussed below.

 \subsubsection{NUSW}
\begin{figure}[!t]
  \centering
  \subfigure[Rayleigh distance versus the physical dimension $D$.]{
    \includegraphics[width=3.0in,height=2.25in]{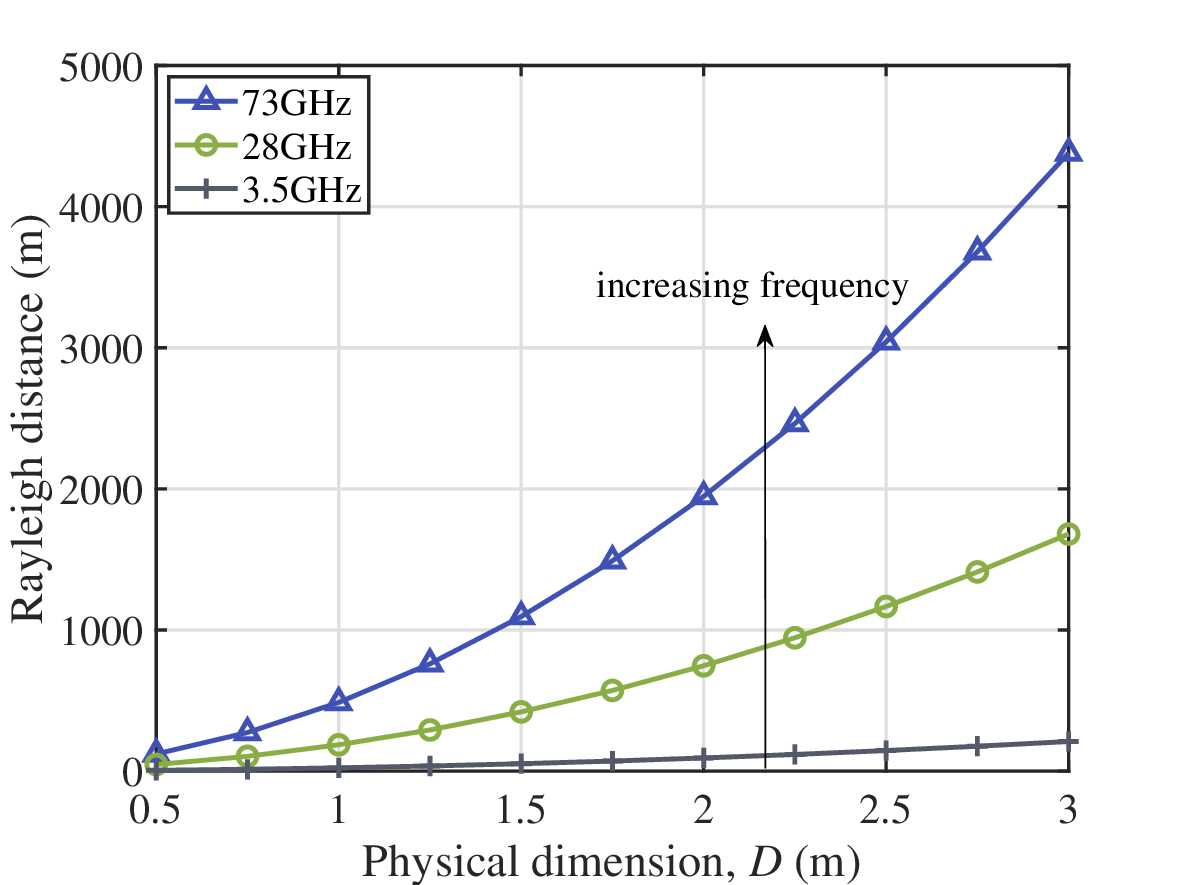}
  }\hspace{1cm}
  \subfigure[Rayleigh distance versus the antenna number $M$.]{
    \includegraphics[width=3.0in,height=2.25in]{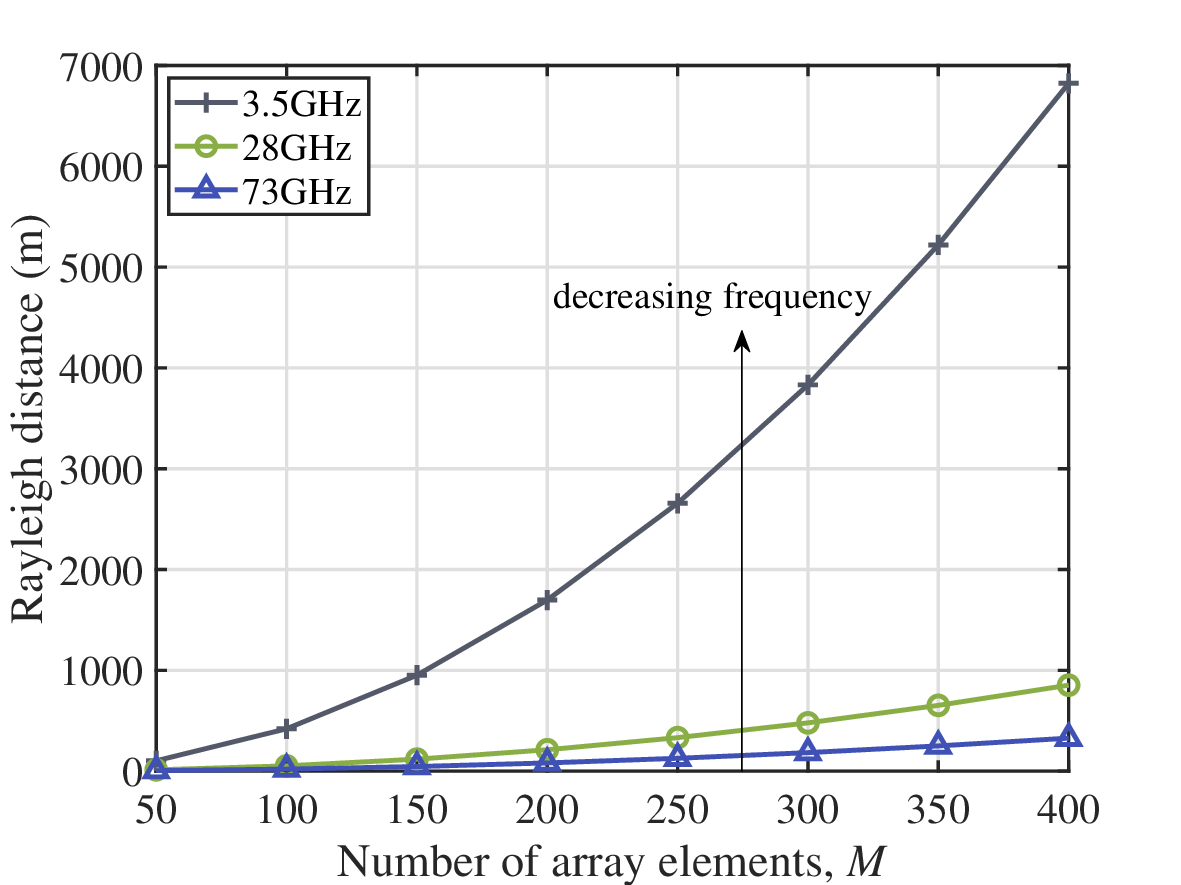}
  }
  \caption{An example of the Rayleigh distance. For fixed $D$, Rayleigh distance increases with the increase of frequency, whereas for fixed $M$, it increases as the frequency gets lower.}
  \label{fig:rayleighDistance}
 \end{figure}

 The deployment of XL-MIMO at the BS, along with the progressively shrinking cell size, renders the users/ scatterers more likely to be located in the near-field region, where the conventional UPW assumption is no longer valid. Note that the classic criterion for distinguishing the near- and far-field regions is the Rayleigh/Fraunhofer distance, given by ${r_{{\rm{Rayl}}}} \triangleq 2{D^2}/\lambda  = 2{D^2}f/c$ \cite{stutzman2012antenna,balanis2016antenna,sherman1962properties,selvan2017fraunhofer,chang2023novel}, where $D$ and $\lambda$ denote the array physical dimension and the signal wavelength, $f$ and $c$ denote the carrier frequency and speed of light, respectively. Consider a uniform linear array (ULA) of $M$ array elements, where adjacent elements are separated by $d = I\frac{\lambda }{2}$, with $I$ denoting the antenna separation parameter. In general, $I \ge 1$ is considered to avoid mutual coupling among elements. Note that for standard arrays, the elements are separated by half wavelength, and we have $I = 1$. The physical dimension is $D = \left( {M - 1} \right)d = \left( {M - 1} \right)I\lambda /2$. As a result, the Rayleigh distance can also be expressed as ${r_{{\rm{Rayl}}}} = {\left( {M - 1} \right)^2}{I^2}c/\left(2f\right)$. The above two alternative expressions for $r_{\rm{Rayl}}$ imply that for any given frequency, the Rayleigh distance increases quadratically with the array physical dimension $D$, antenna number $M$, and antenna separation parameter $I$. On the other hand, the relation of $r_{\rm{Rayl}}$ on the frequency $f$ depends on whether the physical dimension $D$ or the number of antennas $M$ is a specified parameter. For the former, Rayleigh distance increases linearly with the frequency $f$ when the physical dimension $D$ is fixed, whereas for the latter, larger Rayleigh distance is resulted at lower frequency instead when the antenna number $M$ is fixed. Fig.~\ref{fig:rayleighDistance} illustrates the Rayleigh distances of a ULA versus $D$ and $M$, respectively, by considering three carrier frequencies $f =3.5$, $28$, and $73$ GHz, with $I=1$. Two important observations are made from the figure. Firstly, in the XL-MIMO regime, the Rayleigh distance can be up to hundreds or even thousands of meters, which is comparable or even larger than typical cell size. This implies that the near-field region that has been previously ignored should be considered for XL-MIMO systems. Secondly, different from some existing misconceptions that near-field effect only exists in high-frequency or low-frequency systems, Fig.~\ref{fig:rayleighDistance} shows that it may exist for all frequency bands, depending on whether an array with large physical dimension or large number of array elements is deployed.

 \begin{figure}[!t]
 \centering
 \centerline{\includegraphics[width=3.15in,height=2.0in]{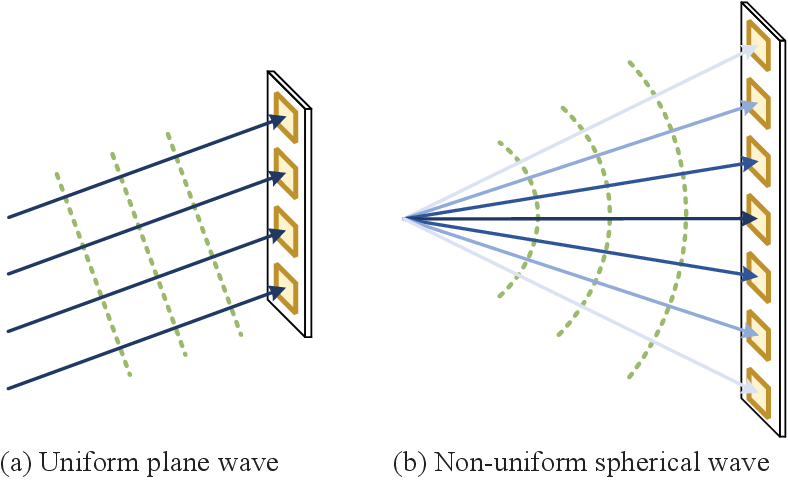}}
 \caption{Illustration of far-field UPW versus near-field NUSW.}
 \label{fig:UPWVersusSW}
 \end{figure}
 As a result, when moving towards the XL-MIMO regime, the more general NUSW is required to accurately characterize both the phase and amplitude variations across array elements. Fig.~\ref{fig:UPWVersusSW} illustrates the differences between the far-field UPW and near-field NUSW. For the far-field UPW model, for each signal source, all the array elements are assumed to share identical angle of arrival/departure (AoA/AoD), and the phases vary linearly across array elements, with the phase gradient depending on the AoA/AoD. Besides, all array elements are assumed to have equal signal amplitude for the same signal path. By contrast, for the more general near-field NUSW model, the phases vary nonlinearly across array elements and the assumption of equal AoA/AoD become invalid in general. Moreover, for the same signal path, the amplitudes of different array elements may no longer be equal in general, due to the non-uniform waves, as illustrated by the faded lines from the center to the edge in Fig.~\ref{fig:UPWVersusSW}(b).

 \subsubsection{Spatial Non-Stationarity}
 On the other hand, spatial non-stationarity means that different from existing spatially stationary MIMO or massive MIMO systems, different portions of the XL-MIMO array may undergo distinct propagation environment, such as visible cluster sets and/or obstacles. Besides, even when all array elements share the same visibility regions (VRs), XL-MIMO also exhibits spatial non-stationarity since the channel correlation across each pair of array elements depends on their actual locations \cite{dong2022near}, instead of their relative locations only as in conventional MIMO systems.

 It is also worth mentioning that there exist differences between near- and far-field regions in mutual coupling and polarization. Specifically, the mutual coupling effect refers to the interaction or coupling between the antenna elements within the array. When an extremely large-scale number of array elements are packed in a dense area with quite small antenna spacing, the distorted radiation pattern introduced by mutual coupling should be considered, which may result in a low radiation efficiency \cite{yuan2023effects}. On the other hand, polarization is the orientation of the electric field vector in an electromagnetic (EM) wave, and the mismatch between the polarization of receive antenna and that of the incident wave is another important factor for near-field communications  \cite{dardari2020communicating,bjornson2020power,bjornson2021primer,ramezani2022near}. In contrast to the far-field UPW model where all array elements have the identical polarization mismatch, the array elements will experience different mismatch due to the distinct AoAs in the near-field region \cite{bjornson2020power}.

%\begin{table}[t]
%	\caption{{Comparison of mmWave MIMO and XL-MIMO}}
%	\label{table:mmWaveXLMIMOComparison}
%	\centering
%	\begin{tabular}{|c|c|c|}
%		\hline
%		&{\bf{mmWave MIMO}} & {\bf{XL-MIMO}}\\
%		\hline
%		\textbf{Antenna number}&64/128& $\gg 64$ \\
%		\hline
%		\multirow{2}{*}{\textbf{Channel characteristics}}&UPW & NUSW \\
%		\cline{2-3}
%			&Spatial stationarity & Spatial non-stationarity \\
%		\hline
%
%	\end{tabular}
%\end{table}

\subsection{Different Categories of XL-MIMO}
 XL-MIMO can be classified according to different criteria, as elaborated below.
 \begin{figure}[!t]
  \centering
  \subfigure[Collocated XL-MIMO]{
    \includegraphics[width=1.4in,height=1.1in]{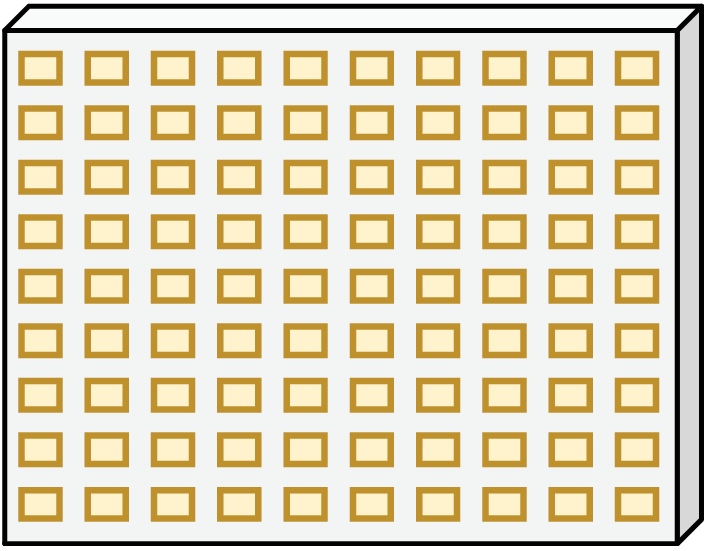}
  }\hspace{0.5cm}
  \subfigure[Sparse XL-MIMO]{
    \includegraphics[width=1.4in,height=1.1in]{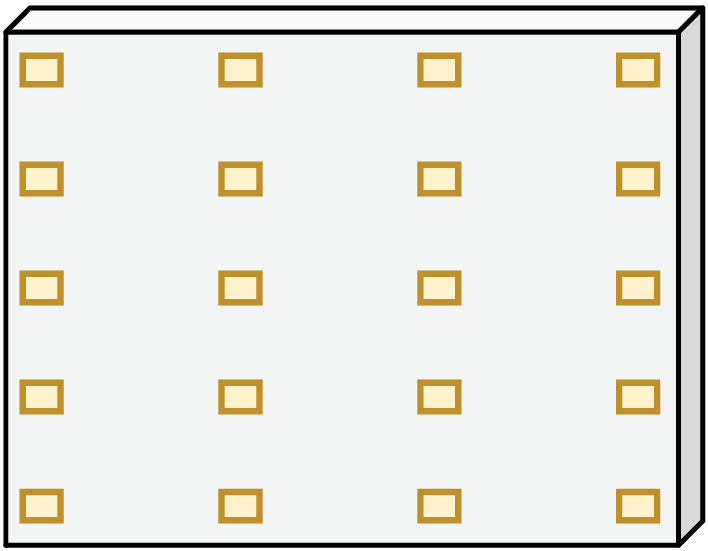}
  }
  \subfigure[Modular XL-MIMO]{
    \includegraphics[width=1.4in,height=1.1in]{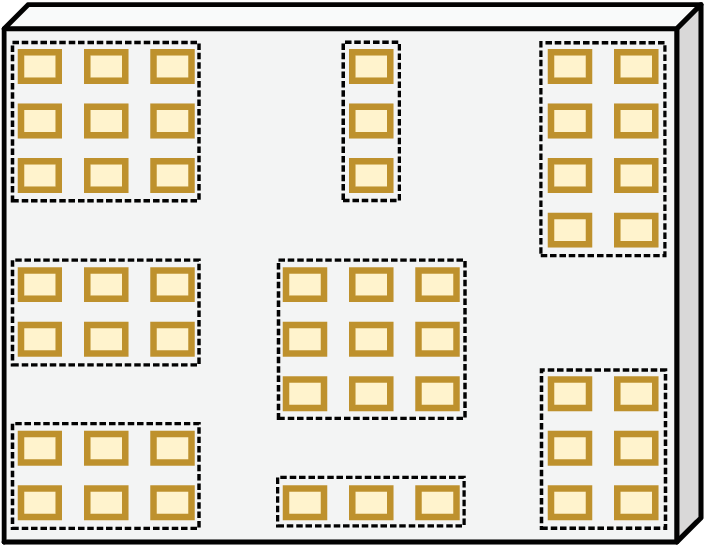}
  }
  \subfigure[Distributed XL-MIMO]{
    \includegraphics[width=1.6in,height=1.35in]{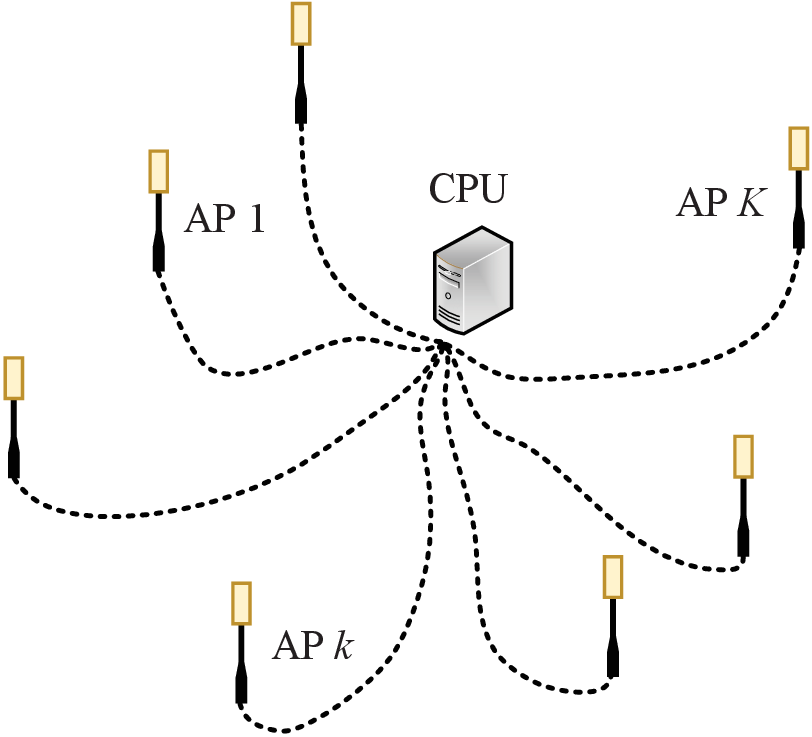}
  }
  \caption{Different architectures of XL-MIMO.}
  \label{fig:XLMIMOArchitecture}
 \end{figure}
\subsubsection{Discrete Versus Continuous-Aperture XL-MIMO}
 Based on the implementation methods, XL-MIMO can be divided into discrete antenna array and continuous-aperture surface. The state-of-the-art MIMO and massive MIMO systems are realized with discrete antenna arrays, where the array elements are connected to radio-frequency (RF) chains and analog-to-digital converters/digital-to-analog converters (ADCs/DACs) \cite{heath2016overview}. Typically, the adjacent array elements are separated by half wavelength, so as to circumvent the impact of mutual coupling among elements and reap the spatial diversity gain. However, thanks to the recent advances in metamaterials and metasurfaces, the element separation may be reduced to sub-wavelength, rendering it possible to pack more array elements in the same physical dimension  \cite{gong2022holographic}. In particular, when an uncountably infinite number of antennas are packed in a compact surface, the continuous-aperture or quasi continuous-aperture antenna array can be realized, also known as Holographic MIMO \cite{pizzo2020spatially,gong2022holographic} or large intelligent surface (LIS) \cite{hu2018beyond,hu2018beyondpositioning}. Different from the conventional discrete antenna array, the whole continuous-aperture surface is capable of transmitting/receiving signals, thus enabling a signal processing paradigm shift from the conventional hybrid digital-analog domain to the EM domain.
 On the other hand, the sub-wavelength architecture renders the effect of mutual coupling and spatial correlation among elements non-negligible for practical modelling and communications \cite{gong2022holographic,an2023tutorial}. It is also worth mentioning that metasurface based XL-MIMO differs from the extensively studied intelligent reflecting surfaces (IRSs) or reconfigurable intelligent surfaces (RISs) \cite{wu2019intelligent,wu2021intelligent,huang2019reconfigurable,du2022reconfigurable,li2023reconfigurable}, where the active metasurface based XL-MIMO, such as dynamic metasurface antenna (DMA) \cite{shlezinger2021dynamic} and reconfigurable holographic surface (RHS) \cite{deng2021reconfigurable,deng2022hdma}, possesses the capabilities of transmitting/receiving signals, while the semi-passive IRS/RIS without requiring RF chains is usually used for signal reflection. Unless otherwise stated, this article focuses on XL-MIMO based on discrete array architecture.
 \begin{figure*}[!t]
 \centering
 \centerline{\includegraphics[width=7.0in,height=2.0in]{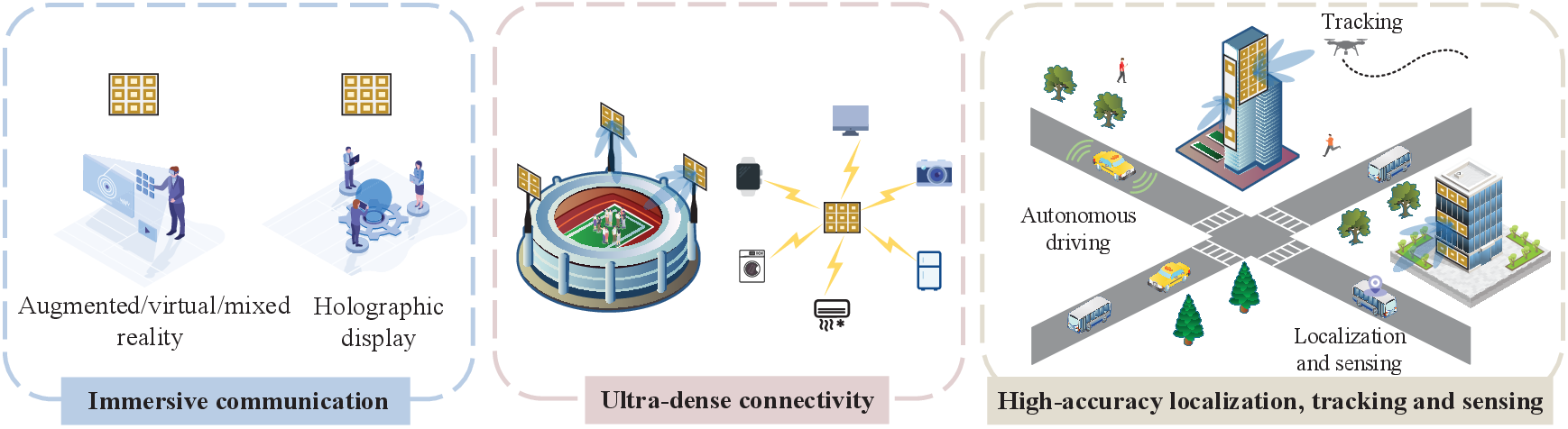}}
 \caption{Illustration of XL-MIMO application scenarios in future wireless networks.}
 \label{fig:XLMIMOApplicationScenarios}
 \end{figure*}
\subsubsection{Collocated, Sparse, Modular, and Distributed XL-MIMO}

 Depending on the array element spacing, XL-MIMO may be implemented in collocated, sparse, modular, and distributed architectures, as illustrated in Fig.~\ref{fig:XLMIMOArchitecture}. The collocated XL-MIMO is the standard array architecture where all antenna elements are placed on a common continuous platform, with adjacent elements typically separated by half wavelength, as illustrated in Fig.~\ref{fig:XLMIMOArchitecture}(a). As the antenna number drastically increases, the collocated XL-MIMO may face practical deployment difficulty since a large continuous platform may not always be available. A similar issue is also faced by uniform sparse XL-MIMO, whose inter-element spacing is larger than half wavelength \cite{vaidyanathan2010sparse,wang2023can}, as illustrated in Fig.~\ref{fig:XLMIMOArchitecture}(b). Compared to the collocated counterpart, sparse XL-MIMO is able to increase the total array aperture without increasing the number of antennas. However, sparse XL-MIMO will lead to undesired grating lobes, which refer to the additional major lobes that have an equal or comparable intensity to the main lobe  \cite{wang2023can,stutzman2012antenna,balanis2016antenna}. Moreover, for ease of practical deployment, a novel modular XL-MIMO architecture was proposed, for which antenna elements are arranged in a modular manner \cite{jeon2021mimo,li2022near}, like Lego-type building blocks, as illustrated in Fig.~\ref{fig:XLMIMOArchitecture}(c). The array elements within each module are regularly arranged like the collocated array, while the inter-module spacing depends on the actual deployment environment that can be much larger than the wavelength scale. For example, when the modular XL-MIMO is mounted to the facades of buildings, the neighbouring modules can be separated by windows, thus achieving the conformity with actual deployment structure. Thanks to the flexible structure among modules, the modular XL-MIMO relaxes the requirement of large continuous deployment platform as in the collocated and sparse XL-MIMO. On the other hand, it also gives rise to undesired grating lobes, similar to the sparse XL-MIMO.

 Instead of accommodating all array elements on a common continuous/discontinuous platform, distributed XL-MIMO architecture consists of multiple distributed sites over a large geographical region, as illustrated in Fig.~\ref{fig:XLMIMOArchitecture}(d). Some typical distributed antenna systems include coordinated multipoint (CoMP) \cite{irmer2011coordinated}, cloud radio access network (C-RAN) \cite{peng2016recent}, network MIMO \cite{hosseini2014large}, and cell-free massive MIMO \cite{ngo2017cell,lei2023uplink,wang2023uplink}. Distributed XL-MIMO systems usually require stringent synchronization and frequent information exchange among different sites. Therefore, this article will mainly focus on collocated, sparse or modular XL-MIMO architectures.

\subsection{Application Scenarios for XL-MIMO}

 In Fig.~\ref{fig:XLMIMOApplicationScenarios}, we envision several promising application scenarios of XL-MIMO in future wireless networks, where different architectures of XL-MIMO can be mounted on the facades of buildings, advertising boards, and indoor walls. The leap of data throughput brought by XL-MIMO facilitates the eMBB+ scenarios, which brings users unprecedentedly immersive experiences and multi-sensory interactions in the augmented/virtual/mixed reality and holographic display applications, and supports services like real-time ultra-high definition videos. Benefiting from the significant increase in network throughput, XL-MIMO can achieve coverage enhancement in hotspot scenarios, e.g., big sports events, concerts, and railway stations. Thanks to the enhancement of spatial resolution, XL-MIMO empowers ultra-dense connectivity in Internet of Things (IoT) applications with extreme reliability, such as smart home, wearables, and agricultures. The super spatial resolution is also a prerequisite for achieving high-accuracy wireless sensing and tracking. For instance, in outdoor environments, XL-MIMO mounted on the facades of buildings is capable of sensing the surrounding environment, e.g., pedestrians, vehicles and buildings, or tracking the position and velocity of objects, and thus a higher level of autonomous driving is expected. For indoor smart factories, XL-MIMO can provide the services of high-accuracy localization and navigation for robots, and support the prompt information exchange among the same production or cross-production lines, which is beneficial to the realization of automated manufacturing. Moreover, the spherical wavefront characteristic in the near-field communication endows XL-MIMO with the capability of beam focusing, which reduces the power/informtaion leakage to the neighboring regions, thus providing new opportunities for the applications of wireless power transfer (WPT) and physical layer security.

\subsection{Motivation and Organization}
 Despite the appealing advantages, XL-MIMO faces many new challenges. For example, the new channel characteristics of the NUSW and spatial non-stationarity render the conventional far-field UPW based channel modelling and performance analysis no longer valid, thus calling for the near-field modelling and performance analysis. Furthermore, accurate channel state information (CSI) is pivotal to achieving super beamforming gain brought by XL-MIMO. Thus, developing efficient near-field channel estimation methods or beam training algorithms are needed. Besides, XL-MIMO with fully digital beamforming entails more RF chains than massive MIMO, rendering the issues of high hardware cost and power consumption more severe \cite{zheng2023flexible}. The large-dimensional channel exacerbates the signal processing complexity in both the digital and analog domains, which hinders efficient implementation of signal precoding and combining.

 \begin{table*}[!t]
	\centering
	\caption{List of Representative Overview/Survey/Tutorial Papers on XL-MIMO}\label{table:XLMIMOpapers}
	\vspace{-0.3cm}
	\resizebox{\textwidth}{!}{
		\begin{tabular}{|m{1.5cm}<{\centering}|m{5cm}<{\centering}|m{18cm}|}
			\hline
			{\bf Reference }                                  & {\bf Topics/Theme}                                                                    & \qquad\qquad\qquad\qquad\qquad\qquad\qquad\qquad{\bf Major Contributions}                                                                                                                                                                                                               \\ \hline\hline
			\cite{bjornson2019massive}                           & Research directions \& open problems                             & Envision five promising research directions for antenna arrays and discuss the open problems.
\\ \hline
            \cite{Han2023Towards}                     & Channel properties
\& low-cost designs & Discuss the new channel properties and low-cost designs of XL-MIMO systems, with an emphasis on the hardware cost, signal processing, computation complexity, and overhead.                                                                               \\ \hline
            \cite{gong2022holographic}                &  Physical aspects
\& theoretical foundations \& enabling technologies & Provide a systematic overview of the continuous-aperture holographic MIMO communications, including physical aspects, theoretical foundations, and enabling technologies, and discuss technical challenges and open research directions.
\\ \hline
            \cite{deng2021reconfigurable}             & Basic concept \&
holographic beamforming               & Introduce the basic concept of RHS and present a hybrid beamforming scheme for RHS-aided communications, and discuss the key challenges.
\\ \hline
			\cite{cui2022near}                             & Near-field spherical wave \& technical challenges                                & Discuss the principle, recent progress, and future directions of near-field communications.                                                                \\ \hline
			\cite{zhang20236g}                           & Physical characteristics \& applications                                        & Introduce the near-field beam focusing of XL-MIMO and discuss the appealing applications of multi-user communications, accurate localization and sensing, and WPT.
\\ \hline
			\cite{zhang2022near}                           & System structure \& applications                                        & Provide an overview of near-field WPT in future Internet of Everything networks and discuss the potential research directions.
\\ \hline
			\cite{wang2022extremely}                       & Hardware design \& channel modelling \& effective degrees-of-freedom                                        & Briefly summarize four general XL-MIMO hardware designs and characteristics of XL-MIMO, including channel modelling, performance analysis, and signal processing.
\\ \hline
			\cite{you2023near}                  & Near-field beam management                                                              & Provide an overview of near-field beam management including near-field beam training, beam tacking and beam scheduling, and discuss promising research directions.
   \\ \hline
			\cite{cong2023near}                 & Near-field ISAC designs                                                              & Provide an overview of near-field ISAC designs including joint near-field communication and sensing, sensing-assisted near-field communication, and communication- assisted near-field sensing.
\\ \hline
			\cite{liu2023near}                           & Channel modelling \& antenna architectures \& performance analysis & Provide a tutorial overview of near-field communications, with an emphasis on spherical wave based channel modelling, hybrid beamforming architectures, and power scaling laws.                                                                       \\ \hline
			\cite{wang2023tutorial}                  & Channel modelling \& signal processing \& applications                                                              & Provide a survey overview of XL-MIMO communications, including hardware architectures, channel modelling, low-complexity signal processing schemes, and main application scenarios.                                                                                      \\ \hline
			This article                  & Near-field modelling \& performance analysis \& practical design issues                                                              & Provide a tutorial overview of XL-MIMO communications, with an emphasis on the near-field modelling, performance analysis, and practical design issues, and point out promising directions for future work.\\ \hline
		\end{tabular}
	}
\end{table*}

 The appealing benefits of XL-MIMO have spurred active research recently, and several overviews  \cite{deng2021reconfigurable,cui2022near,zhang20236g,zhang2022near,wang2022extremely,you2023near,cong2023near} and survey/tutorial \cite{bjornson2019massive,gong2022holographic,Han2023Towards,liu2023near,wang2023tutorial} papers on XL-MIMO have appeared,
 %Among them, the works \cite{gong2022holographic} focus on the continuous-aperture holographic MIMO communications, and the magazine paper \cite{deng2021reconfigurable} mainly introduces the RHS to achieve holographic beamforming. Moreover, there are several works that have provided the state-of-the-art results on XL-MIMO \cite{bjornson2019massive,cui2022near,zhang20236g,zhang2022near,wang2022extremely,Han2023Towards,liu2023near,wang2023tutorial},
 which are summarized in Table~\ref{table:XLMIMOpapers} for ease of reference.  Compared to existing overview/survey/tutorial papers listed in Table~\ref{table:XLMIMOpapers}, this article aims to provide a comprehensive tutorial overview on near-field XL-MIMO communications, with an emphasis on addressing the practical challenges in near-field modelling, fundamental performance analysis and XL-MIMO designs. Specifically, in addition to providing a state-of-the-art literature review on XL-MIMO, this paper provides technically in-depth results and discussions on near-field modelling that comprehensively considers the NUSW and spatial non-stationarity characteristics. Based on such models, the fundamental performance of near-field communications is discussed, which is compared with the conventional far-field communications in detail. Furthermore, practical XL-MIMO design issues are systematically overviewed, including near-field beam codebook, beam training, channel estimation and a novel transmission technology termed delay alignment modulation (DAM) that exploits the super spatial resolution of XL-MIMO, as well as the issues of hardware cost and signal processing complexity.

  \begin{figure}[!t]
 \centering
 \centerline{\includegraphics[width=3.2in,height=5.8in]{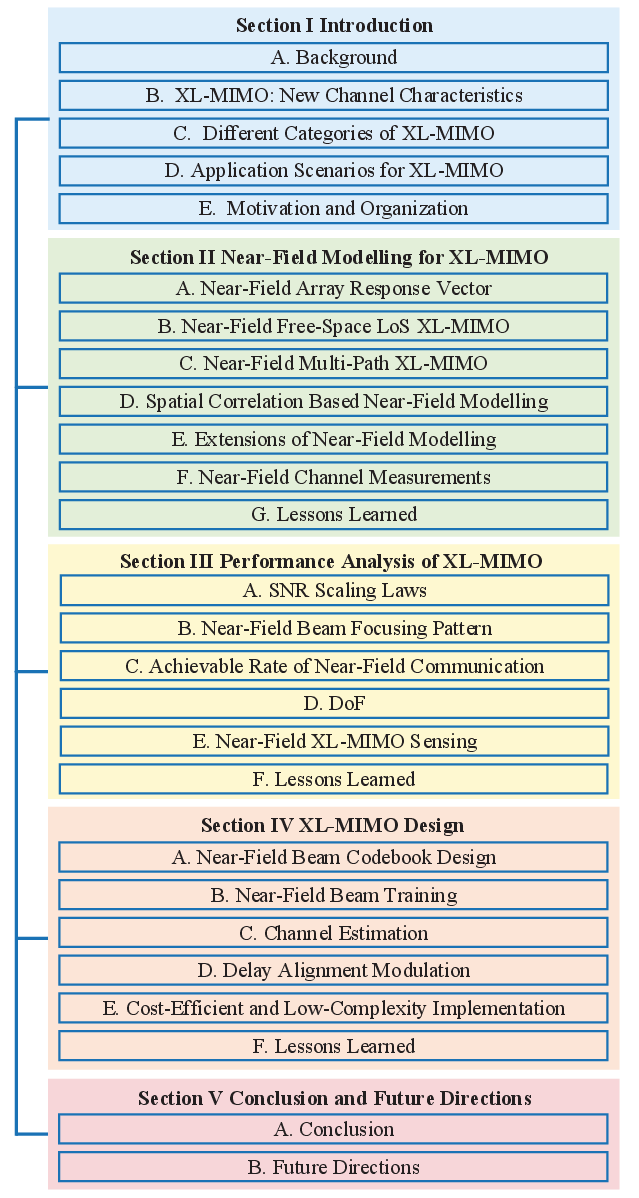}}
 \caption{Organization of this paper.}
 \label{fig:articleOrganization}
 \end{figure}
 As shown in Fig.~\ref{fig:articleOrganization}, the rest of this paper is organized as follows. Section \ref{sectionChannelModellingForXLMIMO} presents the basic near-field modelling for XL-MIMO, including the array response vector, XL-MIMO line-of-sight (LoS) and multi-path modelling. Section \ref{sectionPerformanceAnalysis} presents the fundamental performance analysis of XL-MIMO communications, including signal-to-noise ratio (SNR) scaling laws, near-field beam focusing pattern, achievable rate, and degrees-of-freedom (DoF). In Section \ref{sectionXLMIMODesign}, we discuss the XL-MIMO design issues in near-field beam codebook, beam training, channel estimation, and the novel transmission technology DAM, together with the cost-efficient and low-complexity design. Finally, we conclude this paper in Section \ref{sectionFutureDirections}, and highlight promising research directions for future work. For ease of reference, the definitions of the main acronyms are summarized in Table~\ref{table:acronyms}.
\begin{table*}[!t]
	\centering
	\caption{List of Main Acronyms}\label{table:acronyms}
	\vspace{-0.3cm}
	\resizebox{0.9\textwidth}{!}{
	\begin{tabular}{l|l||l|l}
		\hline
		\textbf{Acronyms} & \qquad\qquad\qquad\textbf{Definition}                         & \textbf{Acronyms} & \qquad\qquad\qquad\textbf{Definition}  \\ \hline\hline
    2D     & Two-dimensional                           & LS      & Least-squares                                          \\ \hline
    3D     & Three-dimensional                         & LTE     & Long-term   evolution                                   \\ \hline
    3G     & Third-generation                          & MIMO    & Multiple-input   multiple-output                        \\ \hline
    3GPP   & The Third Generation Partnership Project  & MISO    & Multiple-input single-output                            \\ \hline
    4G     & Fourth-generation                         & MMSE    & Minimum   mean-square error                             \\ \hline
    5G     & Fifth-generation                          & mMTC+   & Massive communication                                   \\ \hline
    6G     & Sixth-generation                          & mmWave  & Millimeter wave                                         \\ \hline
    ADC    & Analog-to-digital   converter             & MRC/MRT & Maximal-ratio combining/transmission                       \\ \hline
    AI     & Artificial   intelligence                 & MRDN    & Multiple   residual dense network                       \\ \hline
    AoA    & Angle of   arrival                        & NLoS    & Non-line-of-sight                                       \\ \hline
    AoD    & Angle of departure                        & NMSE    & Normalized   mean-square error                          \\ \hline
    AS     & Angle   spread                            & NOMA    & Non-orthogonal multiple access                          \\ \hline
    AWGN   & Additive white Gaussian noise             & NUPW    & Non-uniform   plane wave                                \\ \hline
    B5G    & Beyond 5G                                 & NUSW    & Non-uniform spherical   wave                            \\ \hline
    BS     & Base station                              & OFDM    & Orthogonal   frequency-division multiplexing            \\ \hline
    CFO    & Carrier   frequency offset                & OOB     & Out-of-band                                             \\ \hline
    CKM    & Channel knowledge map                     & OTFS    & Orthogonal time   frequency space                       \\ \hline
    CoMP   & Coordinated   multipoint                  & PADP    & Power angle delay profile                               \\ \hline
    CP     & Cyclic prefix                             & PAPR    & Peak-to-average-power   ratio                           \\ \hline
    CPU    & Central   processing unit                 & PAS     & Power angular spectrum                                  \\ \hline
    C-RAN  & Cloud radio access network                & PBW     & Parabolic wave                                          \\ \hline
    CRB    & Cram\'{e}r-Rao bound                      & PDP     & Power delay profile                                     \\ \hline
    CS     & Compressed sensing                        & PLS     & Power location   spectrum                               \\ \hline
    CSI    & Channel state information                 & RA      & Random access                                           \\ \hline
    DAC    & Digital-to-analog converter               & RCS     & Radar cross   section                                   \\ \hline
    DAM    & Delay alignment modulation                & RF      & Radio-frequency                                         \\ \hline
    DDAM   & Delay-Doppler alignment modulation        & RHS     & Reconfigurable   holographic surface                    \\ \hline
    DDRayl & Direction-dependent Rayleigh distance     & RIS     & Reconfigurable intelligent surface                      \\ \hline
    DFT    & Discrete Fourier transform                & rKA     & Randomized   Kaczmarz algorithm                         \\ \hline
    DMA    & Dynamic metasurface antenna               & RMS     & Root mean square                                        \\ \hline
    DoF    & Degrees-of-freedom                        & SF      & Shadow fading                                        \\ \hline
    DS     & Delay spread                              & SIMO    & Single-input multiple-output                            \\ \hline
    EDoF   & Effective degrees-of-freedom              & SINR    & Signal-to-interference-plus-noise   ratio               \\ \hline
    EH     & Energy-harvesting                         & SNR     & Signal-to-noise   ratio                                 \\ \hline
    EIT    & Electromagnetic information theory        & SWIPT   & Simultaneous wireless information and power transfer    \\ \hline
    ELAA   & Extremely large aperture array            & SWSS    & Spatial wide-sense stationarity                         \\ \hline
    EM     & Electromagnetic                           & THz     & Terahertz                                               \\ \hline
    eMBB+  & Immersive communication                   & TTD     & True-time-delay                                         \\ \hline
    FAS    & Flexible antenna selection                & UCA     & Uniform cylindrical array                               \\ \hline
    FBMC   & Filter bank multi-carrier                 & UE      & User equipment                                          \\ \hline
    FSS    & Fixed  subarray selection                 & ULA     & Uniform linear array                                    \\ \hline
    GA     & Genetic algorithm                         & UM-MIMO & Ultra-massive MIMO                                      \\ \hline
    GBSM   & Geometry-based  stochastic model          & UPA     & Uniform planar array                                    \\ \hline
    HRNP   & Highest received normalized   power       & UPD     & Uniform-power   distance                                \\ \hline
    IoT    & Internet of Things                        & UPW     & Uniform plane wave                                      \\ \hline
    IRS    & Intelligent reflecting surface            & URLLC+  & Hyper reliable and low-latency communication            \\ \hline
    ISAC   & Integrated sensing and communication      & USW     & Uniform spherical wave                                  \\ \hline
    ISI    & Inter-symbol interference                 & VMP     & Variational   message passing                           \\ \hline
    ITU    & International   Telecommunication Union   & VR      & Visibility region                                       \\ \hline
    IUI    & Inter-user   interference                 & WPT     & Wireless power   transfer                               \\ \hline
    KPI    & Key performance   indicator               & XL-MIMO & Extremely large-scale   multiple-input multiple-output  \\ \hline
    LIS    & Large   intelligent surface               & xMaMIMO & Extremely large aperture massive MIMO                   \\ \hline
    LoS    & Line-of-sight                             & ZF      & Zero-forcing \\ \hline
    LPU    & Local   processing unit                   \\ \hline
\end{tabular}
}
\end{table*}

\emph{Notations:} Scalars are denoted by italic letters. Vectors and matrices are denoted by bold-face lower- and upper-case letters, respectively. ${{\mathbb{C}}^{M \times N}}$ and ${{\mathbb{R}}^{M \times N}}$ represent the space of $M \times N$ complex-valued and real-valued matrices, respectively. For a vector ${\bf{x}}$, $\left\| {\bf{x}} \right\|$ denotes its Euclidean norm. For a matrix ${\bf A}$, its complex conjugate, transpose, and Hermitian transpose are denoted by ${\bf A}^*$, ${\bf A}^T$, ${\bf A}^H$, respectively, and ${\left\| {\bf{A}} \right\|_F}$ denotes the Frobenius norm. The distribution of a circularly symmetric complex Gaussian random vector with mean $\bf{x}$ and covariance matrix $\bf{\Sigma}$ is denoted by ${\cal CN}\left( {\bf{x},\bf{{\Sigma}}} \right)$; and $\sim$ stands for ``distributed as". The symbol $j$ denotes the imaginary unit of complex numbers, with ${j^2} =  - 1$. For real numbers $x$ and $y$, $\left\lceil x \right\rceil $ denotes the ceiling operation, and ${\rm mod} \left( {x,y} \right)$ returns the remainder after division of $x$ by $y$. The notations $\odot$ and $ \circledast$ represent the Hadamard product and linear convolution operations, respectively. ${\mathbb E}\left[{\cdot}\right]$ denotes the statistical expectation. For a set $\cal S$, $\left| {\cal S} \right|$ denotes its cardinality. ${\cal O}\left({\cdot}\right)$ denotes the standard big-O notation.

% >>>>>>>>>>>>>SECTIONS II -  here >>>>>>>>>>>>
\section{Near-Field Modelling For XL-MIMO}\label{sectionChannelModellingForXLMIMO}

 In this section, we present the basic near-field modelling for XL-MIMO. To this end, the modelling for near-field array response vector is first discussed, followed by the XL-MIMO LoS modelling and multi-path modelling. Furthermore, the spatial correlation based near-field modelling and some extensions are presented. Finally, we provide a review on recent near-field channel measurement campaigns.

\subsection{Near-Field Array Response Vector}\label{subsectionNearFieldArrayResponseVector}

  \begin{figure}[!t]
  \centering
  \centerline{\includegraphics[width=3.0in,height=2.25in]{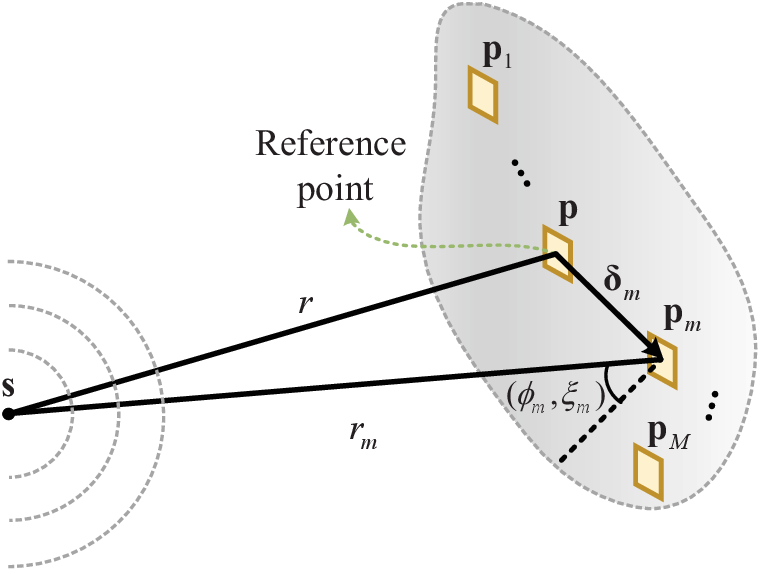}}
  \caption{Wireless link between a signal source ${\bf s}$ and an antenna array of arbitrary architecture, where the source ${\bf s}$ can be either an active transmitter or a passive scatterer.}
  \label{fig:transmitReceiverArray}
  \end{figure}

 The characteristics of the EM field vary with the distance from the antenna, which can be partitioned into {\it reactive field} and {\it radiative field} \cite{stutzman2012antenna,balanis2016antenna}. The reactive field region is the region close to the antenna, where the electric and magnetic fields are out of phase by ${90^ \circ }$ to each other, with the energy being stored in capacitive and inductive reactance. By contrast, in the radiative field, the electric and magnetic fields begin to become radiative. Typically, the boundary for separating the reactive and radiative field is $r = 0.62\sqrt {{D^3}/\lambda }$ \cite{stutzman2012antenna,balanis2016antenna,sherman1962properties,selvan2017fraunhofer}. In this article, we are mainly interested in the radiative field region by assuming $r \ge 0.62\sqrt {{D^3}/\lambda}$ in the rest of the paper.

 \subsubsection{Generic Near-Field Model Based on Exact Distances}
 We first consider a generic wireless link between an isotropic signal source and an antenna array of arbitrary architecture with $M$ elements, as illustrated in Fig.~\ref{fig:transmitReceiverArray}. The signal source can be either an active antenna element or a passive scatterer in the environment. Let $\bf s$ and ${\bf p}$ denote the locations of the signal source and a reference point of the antenna array, respectively. The array architecture can be completely specified by the locations of the array elements ${{\bf{p}}_m} = {{\bf{p}}} + {{\bm{\delta }}_m}$, where ${{\bm{\delta }}_m}$ denotes the relative location of the $m$-th element with respect to the reference location ${\bf p}$. The link distance between the source and antenna $m$ is
 \begin{equation}\label{generalLinkDistanceSIMO}
 \begin{aligned}
 {r_m} & = \left\| {{{\bf{p}}_m} - {\bf{s}}} \right\| = \left\| {{\bf{p}} - {\bf{s}} + {{\bm{\delta }}_m}} \right\|\\
 & = \sqrt {{r^2} + 2{{\left( {{\bf{p}} - {\bf{s}}} \right)}^T}{{\bm{\delta }}_m} + {{\left\| {{{\bm{\delta }}_m}} \right\|}^2}},
 \end{aligned}
 \end{equation}
 where ${r} = \left\| {{{\bf{p}}} - {\bf{s}}} \right\|$ is the distance between $\bf s$ and the reference point ${\bf p}$. The complex-valued channel coefficient from the source to antenna $m$ can be expressed as
 \begin{equation}\label{complexChannelCoefficient}
 {h_m} = {\alpha _m}{e^{j{\varphi _m}}},\ m = 1, \cdots, M,
 \end{equation}
 where ${\alpha _m}$ and ${\varphi_m}$ denote the channel amplitude and phase, respectively. For free-space LoS propagation between ${\bf s}$ and ${{\bf{p}}_m}$, both ${\alpha _m}$ and ${\varphi_m}$ depend on the link distance $r_m$, given by
 \begin{equation}\label{freeSpaceAmplitudePhase}
 \left\{ \begin{split}
 &{\alpha _m} = \frac{{\sqrt {{U_m}} }}{{{r_m}}},\\
 &{\varphi _m} = -\frac{{2\pi }}{\lambda }{r_m},
 \end{split} \right.
 \end{equation}
 where ${U_m}$ accounts for the characteristic of the individual array element, such as directional gain pattern \cite{stutzman2012antenna,balanis2016antenna}. Specifically, let ${G}\left( {{\phi},{\xi}} \right)$ denote the directional gain pattern of each array element, where $\left( {{\phi} ,{\xi} } \right)$ is the local elevation-azimuth signal direction viewed from the element's boresight. Then ${U_m}$ in \eqref{freeSpaceAmplitudePhase} can be modelled as \cite{balanis2016antenna}
 \begin{equation}\label{ParameterGm}
 {U_m} = {G}\left( {{{\phi}_m} ,{{\xi}_m} } \right){\left( {\frac{\lambda }{{4\pi }}} \right)^2},
 \end{equation}
 where the subscript $m$ is needed for $\left( {{{\phi}_m} ,{{\xi}_m} } \right)$ since when the array aperture is large and/or the distance $r$ is small, different array elements may observe distinct signal directions. The commonly used directional gain patterns are given as follows.

 \begin{itemize}[\IEEEsetlabelwidth{12)}]
 \item \textbf{Isotropic model:} When the array elements are modelled as isotropic, we have ${G}\left( {{{\phi}} ,{{\xi}} } \right) = 1$, and ${U_m} = {\left( {\lambda /4\pi } \right)^2}$, $\forall m$ \cite{balanis2016antenna,friis1946note}.
 \item \textbf{Cosine pattern model:} The directional gain pattern of this model is \cite{balanis2016antenna,ellingson2021path,zhang2022beam,feng2023near}
     \begin{equation}\label{antennaDirectionalGainPattern}
     {G}\left( {{\phi},{\xi}} \right)  = \left\{ \begin{split}
     &2\left(2q+1\right){\cos ^{2q}}\left(\phi  \right),\\
     &\ \ \ \ \ \ \ \ \ \ \phi  \in \left[ {0,\frac{\pi }{2}} \right),{\xi}  \in \left[ {0,2\pi } \right],\\
     &0,\ \ \ \ \ \ \ {\rm{otherwise}},
     \end{split} \right.
     \end{equation}
     where $q$ is a parameter determining the directivity of the array element \cite{stutzman2012antenna,ellingson2021path}. In practice, the value of $q$ depends on the specific technology adopted, and a larger $q$ corresponds to a higher directivity \cite{feng2023near}.
  \item \textbf{3GPP element model:} Another widely used directional gain pattern for each array element is the 3GPP model, given by \cite{3GPPTR38.901,rebato2019stochastic}
      \begin{equation}\label{antennaDirectionalGainPattern3GPP}
      \begin{aligned}
      &G\left( {\phi ,\xi } \right) = \\
      &\ {G_{\max }} - \min \left\{ { - \left( {{G_{e,V}}\left( \phi  \right) + {G_{e,H}}\left( \xi  \right)} \right),{A_{\max }}} \right\},
      \end{aligned}
      \end{equation}
      where ${G_{\max}}$ denotes the maximum directional gain of each array element, ${A_{\max}} = 30$ dB is the front-to-back attenuation \cite{3GPPTR38.901,rebato2019stochastic}, and ${G_{e,V}}\left( \phi  \right)$ and ${G_{e,H}}\left( \xi  \right)$ denote the vertical and horizontal cuts of the radiation pattern, respectively.
 \end{itemize}

 By substituting \eqref{freeSpaceAmplitudePhase} into \eqref{complexChannelCoefficient} and after some simple manipulations, we have
 \begin{equation}\label{generalComplexChannelCoefficient}
 {h_m} = \underbrace {\frac{{\sqrt U }}{r}{e^{ - j\frac{{2\pi r}}{\lambda }}}}_\alpha \underbrace {\sqrt {\frac{{{U_m}}}{U}} \frac{r}{{{r_m}}}{e^{ - j\frac{{2\pi }}{\lambda }\left( {{r_m} - r} \right)}}}_{{a_m}\left( {\bf{s}} \right)},\ m = 1, \cdots, M,
 \end{equation}
 where $U$ is a parameter corresponding to the reference location ${\bf p}$. Thus, the channel vector ${\bf{h}} \in {\mathbb C}^{M \times 1}$ between the source ${\bf s}$ and the antenna array can be expressed as ${\bf h} = \alpha {\bf{a}}\left( {\bf{s}} \right)$, where $\alpha \triangleq \frac{{\sqrt U }}{r}{e^{ - j\frac{{2\pi r}}{\lambda }}}$ is a common coefficient for all array elements denoting the complex-valued channel gain at the reference point ${\bf p}$, and ${\bf{a}}\left( {\bf{s}} \right) \in {\mathbb C}^{M\times 1}$ denotes the general near-field array response vector that depends on the exact source location $\bf s$, given by
 \begin{equation}\label{generalChannelSIMO}
 {\bf{a}}\left( {\bf{s}} \right) {\rm =} {\left[ {\sqrt {\frac{{{U_1}}}{U}} \frac{r}{{{r_1}}}{e^{ - j\frac{{2\pi }}{\lambda }\left( {{r_1} - r} \right)}}, {\rm \cdots} ,\sqrt {\frac{{{U_M}}}{U}} \frac{r}{{{r_M}}}{e^{ - j\frac{{2\pi }}{\lambda }\left( {{r_M} - r} \right)}}} \right]^T}.
 \end{equation}

 Note that in the conventional far-field region where UPW model is used, the following three approximations are made for simplifying the array response vector in \eqref{generalChannelSIMO}:
 \begin{itemize}[\IEEEsetlabelwidth{12)}]
 \item ${r_m} \approx r$, $\forall m$, for modelling the free-space path loss across array elements.
 \item ${U_m} \approx U$, $\forall m$, i.e., all array elements have the same gain coefficients. When all array elements are placed with the same orientation, this implies that the signals from the source ${\bf s}$ impinge all array elements with approximately the same direction, i.e., ${\phi _m} \approx \phi $ and ${\xi _m} \approx \xi $, $\forall m$.
 \item The first-order Taylor approximation of the link distance $r_m$ in \eqref{generalLinkDistanceSIMO} is utilized for phase modelling, i.e., $r_m \approx r_m^{{\rm{first}}} \triangleq r + {\left( {{\bf{p}} - {\bf{s}}} \right)^T}{{\bm{\delta }}_m}/r$.
 \end{itemize}

 Under the above three approximations, the array response vector in \eqref{generalChannelSIMO} reduces to the UPW model
 \begin{equation}\label{generalChannelSIMOUPW}
 {{\bf{a}}^{\rm{UPW}}}\left( {\bf{s}} \right) = {\left[ {{e^{ - j\frac{{2\pi }}{\lambda }\frac{{{{\left( {{\bf{p}} - {\bf{s}}} \right)}^T}{{\bm{\delta }}_1}}}{r}}}, \cdots ,{e^{ - j\frac{{2\pi }}{\lambda }\frac{{{{\left( {{\bf{p}} - {\bf{s}}} \right)}^T}{{\bm{\delta }}_M}}}{r}}}} \right]^T}.
 \end{equation}

 \begin{figure}[!t]
 \centering
 \centerline{\includegraphics[width=3.25in,height=2.0in]{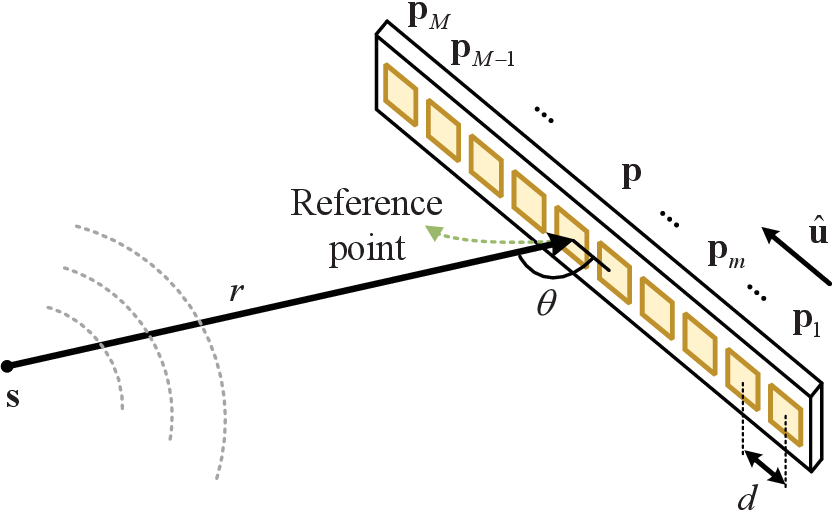}}
 \caption{Wireless link between an isotropic signal source and an $M$-element ULA.}
 \label{fig:transmitReceiverArrayULA}
 \end{figure}
 For the special case of ULA shown in Fig.~\ref{fig:transmitReceiverArrayULA}, let the array center be the reference point ${\bf p}$, we have ${{\bm{\delta }}_m} = {\delta _m}d{\bf{\hat u}}$, where ${\delta _m} = \left( {2m - M - 1} \right)/2$, $m = 1, \dots, M$, $d$ and ${{\bf{\hat u}}}$ denote the antenna spacing and the direction vector of the ULA, respectively, with $\left\| {{\bf{\hat u}}} \right\| = 1$. The physical dimension of the array is $D = \left(M-1\right)d$. By substituting ${{\bm{\delta }}_m}$ into \eqref{generalLinkDistanceSIMO}, the distance between $\bf s$ and antenna $m$ is
 \begin{equation}\label{antennaDistanceSIMO}
 {r_m} = \sqrt {{r^2} + 2{\delta _m}dr\cos \theta  + \delta _m^2{d^2}},
 \end{equation}
 where $\theta$ denotes the AoA at the reference point, i.e., the angle between vectors ${\bf{p}} - {\bf{s}}$ and $\bf{\hat u}$, with $\cos \theta  = \frac{{{{\left( {{\bf{p}} - {\bf{s}}} \right)}^T}{\bf{\hat u}}}}{{\left\| {{\bf{p}} - {\bf{s}}} \right\|\left\| {{\bf{\hat u}}} \right\|}} = \frac{{{{\left( {{\bf{p}} - {\bf{s}}} \right)}^T}{\bf{\hat u}}}}{r}$. The first-order Taylor approximation of ${r_m}$ is
 \begin{equation}\label{firstOrderTaylorApproximation}
 r_m^{{\rm{first}}} = r + {\delta _m}d\cos \theta.
 \end{equation}
 Thus, for ULA, the far-field array response vector in \eqref{generalChannelSIMOUPW} reduces to
 \begin{equation}\label{arrayResponseVectorUPWModel}
 {{\bf{a}}^{{\rm{UPW}}}}\left( \theta  \right) = {\beta}{\left[ {1, \cdots ,{e^{ - j\frac{{2\pi }}{\lambda }\left( {M - 1} \right)d\cos \theta }}} \right]^T},
 \end{equation}
 where ${\beta} = {e^{j\frac{{\pi \left( {M - 1} \right)}}{\lambda }d\cos \theta }}$. It is observed that for the far-field UPW model, all array elements experience the identical signal amplitude and AoA. Besides, the array response vector only depends on $\theta$, and the variation of phase across array elements exhibits a linear relationship with respect to antenna index $m$.

 However, the deployment of XL-MIMO and the shrinking cell size render users/scatterers less likely to be located in the far-field region, and the conventional far-field UPW model is no longer valid. As a classic criterion for separating the near- and far-field regions, the Rayleigh distance corresponds to the minimum link distance so that if the array is used for reception, the maximum phase difference of the received signals across array elements is no greater than $\pi /8$ by assuming {\it normal incidence}  \cite{stutzman2012antenna,balanis2016antenna}. Besides, the Rayleigh distance only concerns about the phase variations across array elements, while ignoring the amplitude variations. As a result, the classic Rayleigh distance is insufficient for separating the near- and far-field regions. In the following, we try to answer the following questions: {\it i) When channel phases can be modelled linearly across array elements? ii) When channel amplitudes can be modelled uniformly across array elements?}

 \subsubsection{Near-Field Phase Modelling}
 In the far-field region $r \ge r_{\rm Rayl}$, the channel phases are modelled linearly across array elements under the approximation of UPW model. On the other hand, in the near-field region $r < r_{\rm Rayl}$, the phase across array elements are no longer linear. Instead, the more accurate spherical wave is required for phase modelling, i.e., the exact link distance $r_m$. Besides, the second-order Taylor approximation of link distance is another common method for near-field phase modelling. For the case of ULA, it follows from \eqref{antennaDistanceSIMO} that the second-order Taylor approximation of $r_m$ is
 \begin{equation}\label{secondOrderTaylorApproximation}
 r_m^{{\rm{second}}} = r + {\delta _m}d\cos \theta  + \frac{{\delta _m^2}{d^2}{{\sin }^2}\theta }{2r}+ O\left( {\frac{1}{{{r^2}}}} \right),
 \end{equation}
 where $O\left( {1/{r^2}} \right)$ denotes the higher-order terms that can be ignored, and such an approximation is also referred to as {\it Fresnel approximation} \cite{stutzman2012antenna,friedlander2019localization},

 Furthermore, to reflect the impact of signal direction on the phase variations across array elements, a new {\it direction-dependent Rayleigh distance} (DDRayl) was introduced in \cite{lu2022communicating}. Let $\Delta \left( {r,\theta } \right)$ denote the maximum phase error across array elements, i.e., the phase difference between the exact value and that based on the first-order Taylor distance approximation, given by
 \begin{equation}\label{phaseErrorFirstOrderApproximation}
 \Delta \left( {r,\theta } \right) \triangleq \mathop {\max }\limits_m \frac{{2\pi }}{\lambda }\left( {{r_m} - r_m^{{\rm{first}}}} \right).
 \end{equation}
 In order to be compatible with the classic Rayleigh distance, the direction-dependent Rayleigh distance, denoted as ${r_{{\rm{DDRayl}}}}\left( \theta  \right)$, is then defined as the minimum distance $r$ satisfying $\Delta \left( {r,\theta } \right) \le \pi /8$ \cite{lu2022communicating}, i.e.,
 \begin{equation}\label{directionDependentRayleighDistanceDefinition}
 {r_{{\rm{DDRayl}}}}\left( \theta  \right) \triangleq \arg \mathop {\min }\limits_r \Delta \left( {r,\theta } \right) \le \frac{\pi }{8}.
 \end{equation}
 It is difficult to directly obtain the closed-form solution to \eqref{directionDependentRayleighDistanceDefinition}, but its value can be found numerically. To gain useful insights, by replacing the exact distance ${r_m}$ with its second-order Taylor approximation in \eqref{phaseErrorFirstOrderApproximation}, the direction-dependent Rayleigh distance can be obtained in closed-form as
 \begin{equation}\label{directionDependentRayleighDistanceApproxiamtion}
 {r_{{\rm{DDRayl}}}}\left( \theta  \right) \approx \frac{{2{D^2}{{\sin }^2}\theta }}{\lambda }.
 \end{equation}
 In particular, for the normal incidence with $\theta  = \pi /2$, the direction-dependent Rayleigh distance reduces to the classic Rayleigh distance, i.e., ${r_{{\rm{DDRayl}}}}\left( {\pi /2} \right) = 2{D^2}/\lambda$ \cite{lu2022communicating}. It is observed that the direction-dependent Rayleigh distance is affected by the signal direction via ${{{\sin }^2}\theta }$, and the classic Rayleigh distance exaggerates the boundary of the near- and far-field regions, since ${\sin ^2}\theta  \le 1$.

 The comparison of the classic Rayleigh distance and the direction-dependent Rayleigh distance is shown in Fig.~\ref{fig:directionDependentRayleighDistance}. The carrier frequency is $f = 2.4$ GHz. A ULA with $M=128$ array elements is placed along the $y$-axis, with its center located at the origin, and the adjacent elements are separated by half wavelength. It is observed that the boundary corresponding to the classic Rayleigh distance is a quarter circle, which is expected since its definition ignores the actual direction of the signal source. It is also observed that the approximate direction-dependent Rayleigh distance in \eqref{directionDependentRayleighDistanceDefinition} matches well with the exact value, whose boundaries exhibit a semi-ellipse shape. Note that when the signal source is at inclined directions, the classic Rayleigh distance exaggerates the near-field region, i.e., it is rather conservative from the perspective of phase modelling.

 \subsubsection{Near-Field Amplitude Modelling}\label{subsubSecNearFieldAmplitudeModelling}
 In the conventional UPW modelling, the channel amplitude is modelled uniformly across array elements with the approximations of ${U_m} \approx U$ and ${r_m} \approx r$, $\forall m$. When the array physical dimension increases and/or the link distance decreases, the impacts of direction gain pattern and link distance variations across array elements on the channel amplitude may become non-negligible, rendering the assumption of ``uniform'' wave invalid.

\begin{figure}[!t]
 \centering
 \centerline{\includegraphics[width=2.85in,height=2.6in]{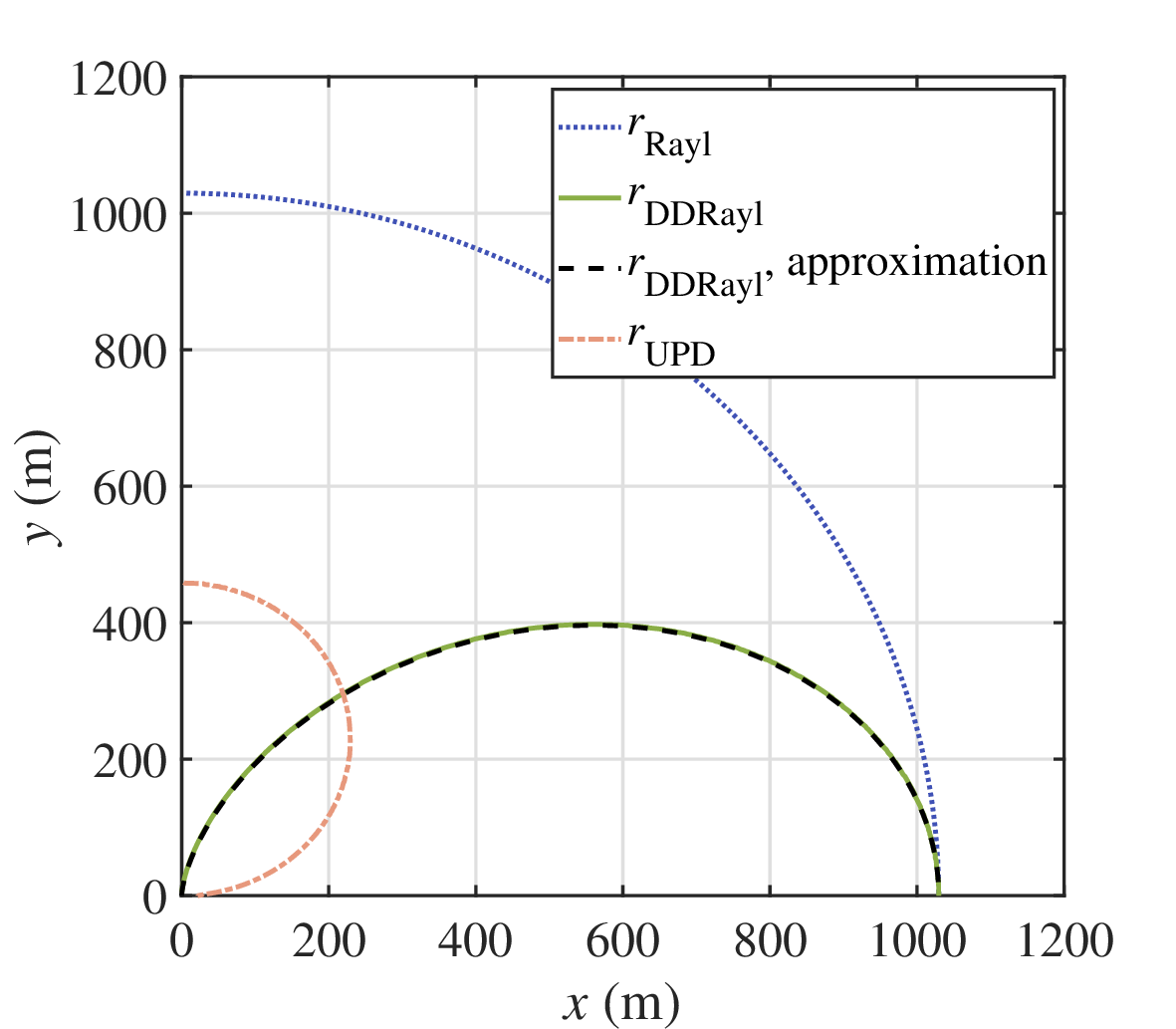}}
 \caption{Comparison of the classic Rayleigh distance, the direction-dependent Rayleigh distance, and the UPD by considering the cosine pattern model with $q=2$. The ULA is placed along the $y$-axis and centered at the origin. Due to symmetry, only the first quadrature is shown.}
 \label{fig:directionDependentRayleighDistance}
 \end{figure}

 To determine whether the EM waves are uniform or not, a new distance criterion termed {\it uniform-power distance} (UPD) was introduced in \cite{lu2022communicating}, which focuses on the variation of amplitude across array elements. Specifically, let $\Upsilon  \left( {r,\theta }\right)$ denote the ratio of the weakest and the strongest power across array elements, given by
 \begin{equation}\label{powerRatioUPD}
 \Upsilon  \left( {r,\theta } \right) \triangleq \frac{{\mathop {\min }\limits_m\ {{\left| {{\alpha _m}} \right|}^2}}}{{\mathop {\max }\limits_m \ {{\left| {{\alpha _m}} \right|}^2}}} = \frac{{\mathop {\min }\limits_m\  {U_m}/r_m^2}}{{\mathop {\max }\limits_m\ {U_m}/r_m^2}}.
 \end{equation}
 The UPD is then defined as
 \begin{equation}\label{definitionOfUPD}
 {r_{\rm{UPD}}}\left( \theta \right) \triangleq \arg \mathop {\min }\limits_r \Upsilon  \left( {r,\theta } \right) \ge {\Upsilon  _{{\rm{th}}}},
 \end{equation}
 where ${\Upsilon  _{{\rm{th}}}}$ is a certain threshold, and the value of UPD can be obtained numerically. As such, for any given signal direction $\theta$, when the link distance $r \ge {r_{\rm{UPD}}}\left( \theta \right)$, the variation of signal amplitude across array elements is negligible, i.e., the assumption of ``uniform" waves holds. When $r < {r_{\rm{UPD}}}\left( \theta \right)$, the significant power difference renders the assumption of ``uniform" waves no longer valid. The result of the UPD is also shown in Fig.~\ref{fig:directionDependentRayleighDistance}, by adopting the cosine pattern model with $q=2$. The power ratio threshold is ${\Upsilon  _{{\rm{th}}}} = 0.9$. It is observed that compared to the classic and the direction-dependent Rayleigh distances, the UPD yields quite different curve, which is expected since different criteria are used. For example, the direction-dependent Rayleigh distance achieves the maximum value for the normal incidence, under which the value of UPD is the minimum. In other words, compared to the case of normal incidence, a larger link distance is required to neglect the amplitude variations for the general case with inclined incidence.

\begin{figure*}
\subfigure[Direction-dependent near- and far-field separation]{
\begin{minipage}[t]{0.48\textwidth}
\centering
\centerline{\includegraphics[width=3.2in,height=3.6in]{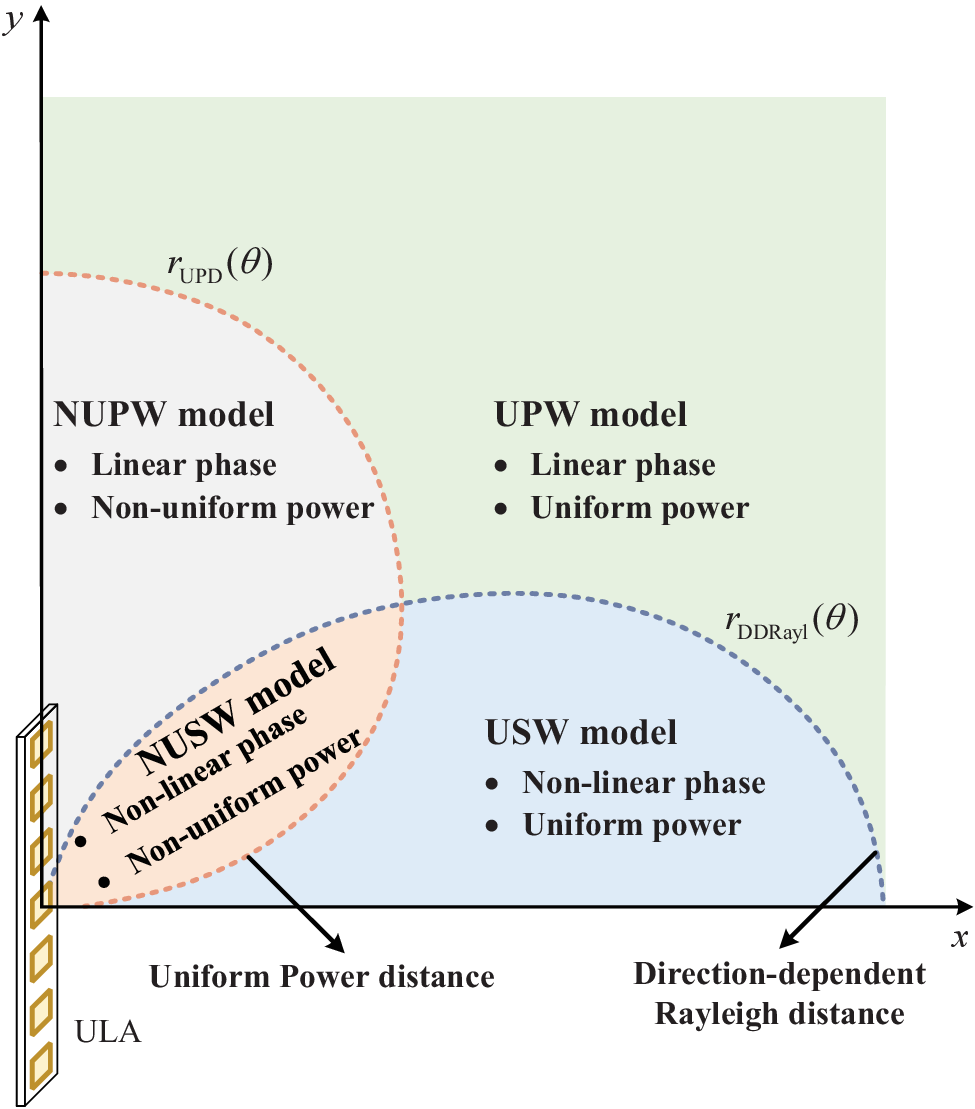}}
\end{minipage}
}
\subfigure[A simplified and conservative near- and far-field separation without direction dependency]{
\begin{minipage}[t]{0.48\textwidth}
\centerline{\includegraphics[width=3.2in,height=3.6in]{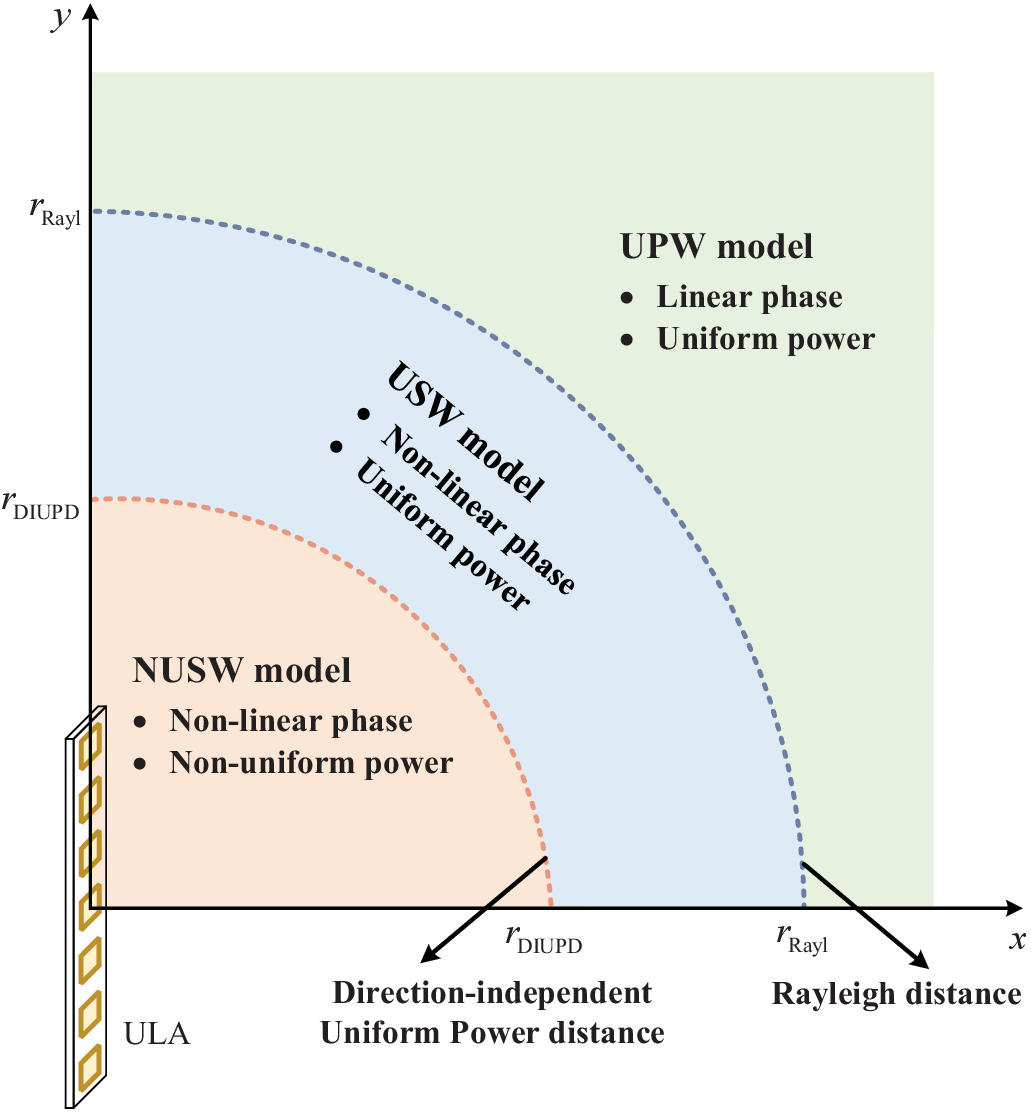}}
\end{minipage}
}
\caption{An illustration of the refined near- and far-field separation, where the ULA is placed along the $y$-axis, with its center located at the origin. Due to symmetry, only the first quadrant is shown.}
\label{fig:NearFarFieldSeparation}
\end{figure*}

 Based on the above observations, a refined direction-dependent near- and far-field separation criterion is illustrated in Fig.~\ref{fig:NearFarFieldSeparation}(a). Due to symmetry, only the first quadrant is shown. It is observed that the space is partitioned into four parts, each corresponding to one model, i.e., UPW, non-uniform plane wave (NUPW), uniform spherical wave (USW), and NUSW models, as elaborated below.
 \begin{itemize}
  \item \textbf{UPW model:} For the given signal direction $\theta$, when $r \ge \max \left\{ {{r_{{\rm{UPD}}}}\left( \theta  \right),{r_{{\rm{DDRayl}}}}\left( \theta  \right)} \right\}$, the EM waves can be regarded as UPW, and the array response vector is given by \eqref{arrayResponseVectorUPWModel}.
  \item \textbf{NUPW model:} When ${r_{{\rm{DDRayl}}}}\left( \theta  \right) \le r < {r_{{\rm{UPD}}}}\left( \theta  \right)$, though the plane wave approximation is valid, the non-negligible power variation across array elements renders the EM waves no longer uniform. Such a surprising result is mainly due to the fact that the conventional Rayleigh distance that ignores the signal direction is a conservative criterion for separating the plane and spherical waves, as can be seen in Fig.~\ref{fig:directionDependentRayleighDistance}. For ULA, the array response vector for NUPW model is
      \begin{equation}\label{generalNUPWModel}
      \begin{aligned}
      {{\bf{a}}^{{\rm{NUPW}}}}\left( {r,\theta } \right) &= \beta \left[ {\sqrt {\frac{{{U_1}}}{U}} \frac{r}{{{r_1}}}, \cdots ,} \right.\\
      &\ \ \ \ {\left. {\sqrt {\frac{{{U_M}}}{U}} \frac{r}{{{r_M}}}{e^{ - j\frac{{2\pi }}{\lambda }\left( {M - 1} \right)d\cos \theta }}} \right]^T},
      \end{aligned}
      \end{equation}
      which depends on both the distance $r$ and the angle $\theta$.
  \item \textbf{USW model:} When ${r_{{\rm{UPD}}}}\left( \theta  \right) \le r < {r_{{\rm{DDRayl}}}}\left( \theta  \right)$, the amplitude variation across array elements is negligible, while the more accurate spherical wave is required for phase modelling. The array response vector for USW model is \cite{cui2022channel,liu2023deep,wei2022channel}
      \begin{equation}\label{arrayResponseVectorUSWModel}
      {{\bf{a}}^{{\rm{USW}}}}\left( {r,\theta } \right) =  {\left[ {{e^{ - j\frac{{2\pi }}{\lambda }\left( {{r_1} - r} \right)}}, \cdots ,{e^{ - j\frac{{2\pi }}{\lambda }\left( {{r_M} - r} \right)}}} \right]^T}.
      \end{equation}

      For moderately large array physical dimension and/or moderately short link distance, $r_m^{{\rm{second}}}$ in \eqref{secondOrderTaylorApproximation} is a valid approximation for phase modelling, which is known as {\it parabolic wave} (PBW) model \cite{lopez2016novel,lopez2018novel,le2019massive}. By replacing ${r_m}$ with $r_m^{{\rm{second}}}$ in \eqref{arrayResponseVectorUSWModel}, the array response vector for PBW model is
      \begin{equation}\label{arrayResponseVectorPBWModel}
      \begin{aligned}
      {{\bf{a}}^{{\rm{PBW}}}}\left( {r,\theta } \right) &= \Big[ {{e^{ - j\frac{{2\pi }}{\lambda }\left( {{\delta _1}{d}\cos \theta  + \frac{{\delta _1^2d^2{{\sin }^2}\theta }}{{2r}}} \right)}}, \cdots ,} \Big.\\
      &\ \ \ \ \ \ \ \ \ \ \ \Big. {{e^{ - j\frac{{2\pi }}{\lambda }\left( {{\delta _M}{d}\cos \theta  + \frac{{\delta _M^2d^2{{\sin }^2}\theta }}{{2r}}} \right)}}} \Big]^T.
      \end{aligned}
      \end{equation}
  \item \textbf{NUSW model:} When $r < \min \left\{ {{r_{{\rm{UPD}}}}\left( \theta  \right),{r_{{\rm{DDRayl}}}}\left( \theta  \right)} \right\}$, neither the linear phase approximation nor the uniform amplitude approximation is valid. Instead, the accurate model is required for both the phase and amplitude modelling. By replacing the location ${\bf s}$ with $\left( {r,\theta } \right)$ in \eqref{generalChannelSIMO}, the array response vector of ULA for NUSW model is
      \begin{equation}\label{arrayResponseVectorNUSWModel}
      \begin{aligned}
      {{\bf{a}}^{{\rm{NUSW}}}}\left( {r,\theta } \right)&= \left[ {\sqrt {\frac{{{U_1}}}{U}} \frac{r}{{{r_1}}}{e^{ - j\frac{{2\pi }}{\lambda }\left( {{r_1} - r} \right)}}, \cdots ,} \right.\\
      &\ \ \ \ \ \ \ \ \ \ {\left. {\sqrt {\frac{{{U_M}}}{U}} \frac{r}{{{r_M}}}{e^{ - j\frac{{2\pi }}{\lambda }\left( {{r_M} - r} \right)}}} \right]^T}.
      \end{aligned}
      \end{equation}
  \end{itemize}

 It is worth mentioning that NUSW is the most general and accurate model, which includes the USW, NUPW, and UPW models as special cases, i.e., NUSW $\supset$ USW $\supset$ UPW, and NUSW $\supset$ NUPW $\supset$ UPW. Motivated by this, we propose a conservative and simplified near- and far-field separation that does not depend on the direction, as illustrated in Fig.~\ref{fig:NearFarFieldSeparation}(b), where the direction-independent UPD is given by ${r_{{\rm{DIUPD}}}} \triangleq \mathop {\max }\limits_\theta  {r_{{\rm{UPD}}}}\left( \theta  \right)$. Such a conservative separation approach stems from the fact that the direction-independent UPD exaggerates the non-uniform wave region from the perspective of amplitude modelling, and the Rayleigh distance exaggerates the spherical wave region from the perspective of phase modelling. Furthermore, we typically have ${r_{{\rm{DIUPD}}}} < {r_{{\rm{Rayl}}}}$. As such, while still preserving the modelling accuracy, we reduce the number of space partitions from four to three, corresponding to three models, i.e., UPW, USW, and NUSW models.
  \begin{itemize}
  \item When $r \ge {r_{{\rm{Rayl}}}}$, the NUSW, USW, and UPW models are in fact all valid, but UPW gives the simplest model without notably compromising the accuracy.
  \item When ${r_{{\rm{DIUPD}}}} \le r < {r_{{\rm{Rayl}}}}$, both NUSW and USW models are valid, and the simpler USW model can be used.
  \item When $0.62\sqrt {{D^3}/\lambda }  \le r < {r_{{\rm{DIUPD}}}}$, NUSW model is needed for the accurate phase and amplitude modelling.
  \end{itemize}

 To draw some insights, we consider the commonly used isotropic array elements, with ${U_m} = U =  {\left( {\lambda /4\pi } \right)^2}$, $\forall m$. In the following, the special signal directions with ${\theta} = {\pi}/2$, and ${\theta} = 0, {\pi}$ are respectively discussed.
 \begin{itemize}[\IEEEsetlabelwidth{12)}]
 \item ${\theta} = {\pi}/2$, i.e., the normal incidence. The UPD for a given ${\Upsilon  _{\rm{th}}}$ is ${r_{{\rm{UPD}}}}\left( {\pi /2} \right) = \frac{D}{2}\sqrt {{\Upsilon  _{{\rm{th}}}}/\left( {1 - {\Upsilon  _{{\rm{th}}}}} \right)}$. By choosing ${\Upsilon  _{\rm{th}}} = 0.9$, we have
     \begin{equation}\label{normalIncidenceUPD}
     {r_{{\rm{UPD}}}}\left( {\frac{\pi }{2}} \right) = 1.5D.
     \end{equation}
     On the other hand, if we choose ${\Upsilon  _{{\rm{th}}}} = {\cos ^2}\left( {\pi /8} \right)$, the UPD is ${r_{{\rm{UPD}}}}\left( {\pi /2} \right) \approx 1.2D$, which is the distance criterion corresponding to the negligible amplitude difference derived in \cite{kay1960near,sherman1962properties}.
 \item ${\theta} = 0, {\pi}$, i.e., the source is located along the line spanned by the ULA. In fact, for any given $r$, $\Upsilon  \left({r,\theta} \right)$ in \eqref{powerRatioUPD} achieves the minimum value when ${\theta} = 0$ or ${\pi}$ \cite{lu2021how}. For ease of exposition, we show the result of $\theta = 0$, for which the UPD is ${r_{{\rm{UPD}}}}\left( 0 \right) = D\left( {1 + {\Upsilon  _{{\rm{th}}}} + 2\sqrt {{\Upsilon  _{{\rm{th}}}}} } \right)/\left(2\left( {1 - {\Upsilon  _{{\rm{th}}}}} \right)\right)$. Similarly, by choosing ${\Upsilon  _{\rm{th}}} = 0.9$, we have
     \begin{equation}\label{SameDirectionUPD}
     {r_{{\rm{UPD}}}}\left( 0 \right) \approx 19D,
     \end{equation}
     which is much larger than that for the normal incidence.
 \end{itemize}

 It is observed that the commonly used distance criterion $1.2D$ \cite{kay1960near,sherman1962properties,bjornson2021primer} for neglecting the variation of amplitude across array elements cannot be safely applied to the case with inclined incidence. Note that ${r_{{\rm{UPD}}}}\left( 0 \right)$ is smaller than the Rayleigh distance when $D > 9.5{\lambda}$, which is easily satisfied for XL-array. Besides, different from the Rayleigh distance that depends on the {\it electrical size}, the UPD is related to the {\it physical size} of the array \cite{lu2021how}. Furthermore, the existing array models for isotropic antenna elements include UPW, USW, PBW, and NUSW models, which can be obtained based on \eqref{arrayResponseVectorUPWModel}, \eqref{arrayResponseVectorUSWModel}, \eqref{arrayResponseVectorPBWModel}, and \eqref{arrayResponseVectorNUSWModel}, by letting ${U_m}= {\left( {\lambda /4\pi } \right)^2}$, $\forall m$.

\begin{figure}[!t]
 \centering
 \centerline{\includegraphics[width=3.5in,height=2.5in]{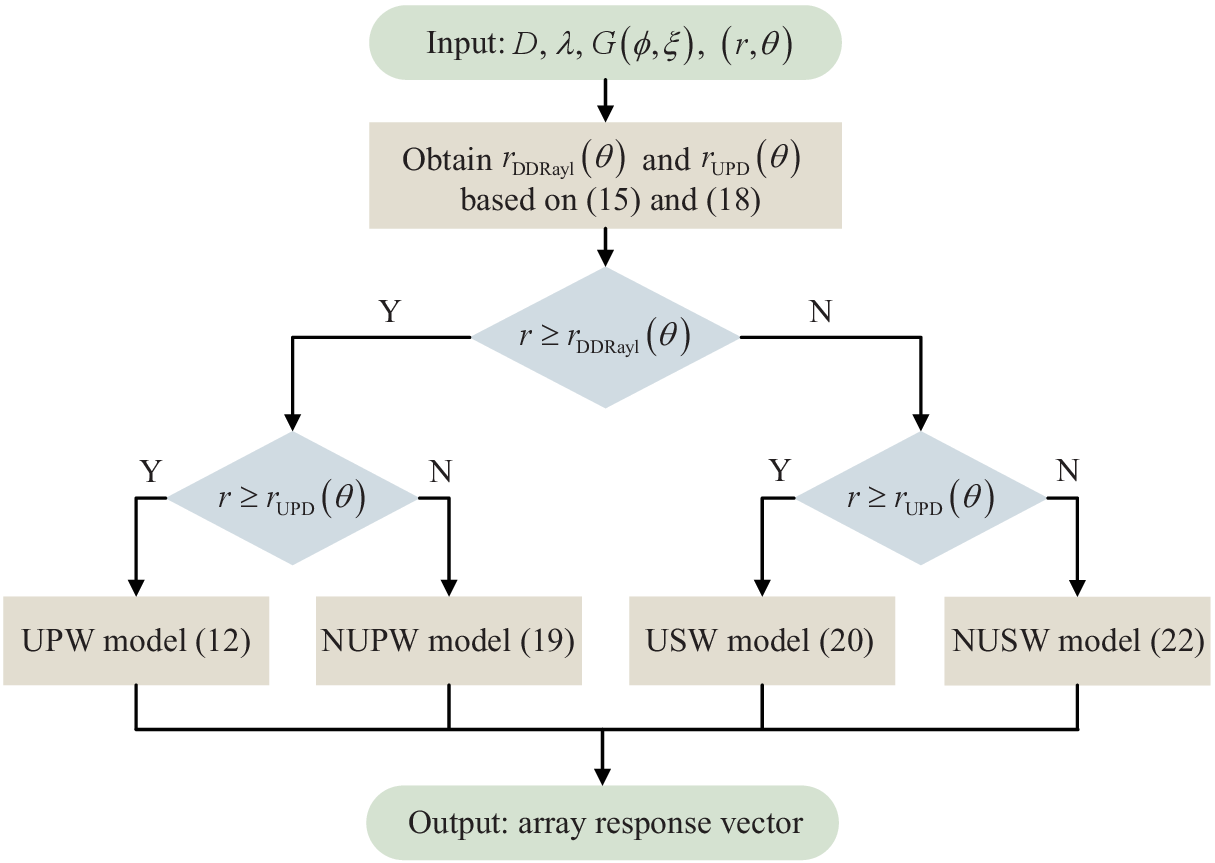}}
 \caption{Summary of near-field array response vector modelling.}
 \label{fig:arrayResponseVectorModelsFlowChart}
 \end{figure}
 In summary, the near-field array response vector modelling involves the near-field phase and amplitude modelling. The direction-dependent Rayleigh distance is a general  criterion to determine whether the channel phases can be modelled linearly across array elements, and the UPD aims to determine whether the channel amplitude can be modelled uniformly across array elements. Such two distance criteria constitute a refined near- and far-field separation approach. The main procedure of generating the appropriate array response vector is summarized in  Fig.~\ref{fig:arrayResponseVectorModelsFlowChart}. In the following, based on the established near-field array response vectors, the near-field free-space LoS and multi-path XL-MIMO modelling are respectively discussed.

\subsection{Near-Field Free-Space LoS XL-MIMO}
\begin{figure}[!t]
 \centering
 \centerline{\includegraphics[width=3.5in,height=1.85in]{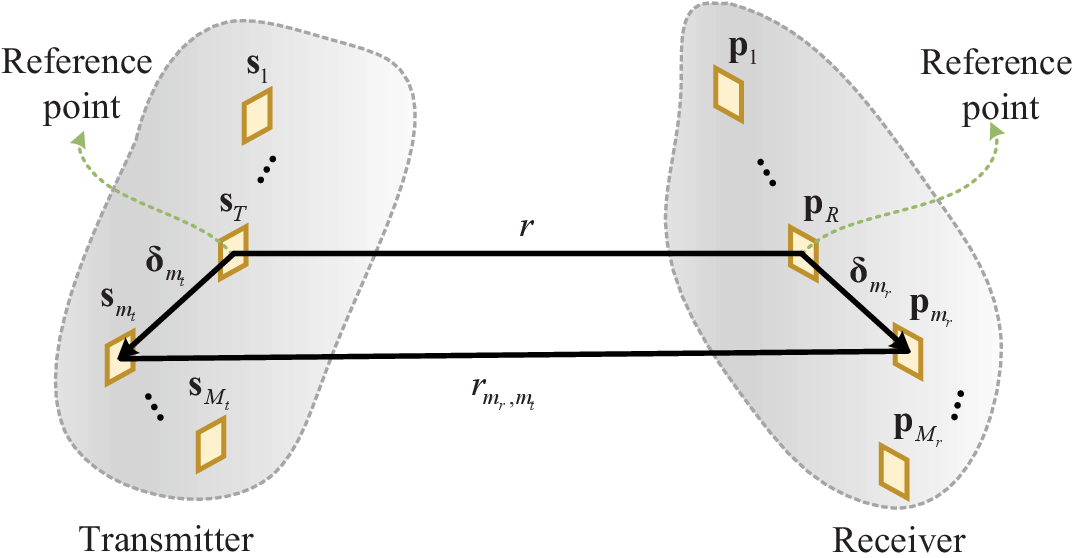}}
 \caption{Illustration of free-space LoS XL-MIMO between a transmitter and a receiver of general architectures.}
 \label{fig:XLMIMOLoSCommunication}
 \end{figure}

 In this subsection, we consider the modelling for near-field free-space LoS XL-MIMO, where the transmitter and receiver are equipped with ${M_t}$ and ${M_r}$ array elements, respectively, as illustrated in Fig.~\ref{fig:XLMIMOLoSCommunication}. Note that for the special cases of LoS XL-MISO or XL-SIMO communications, the near-field channel models can be directly obtained based on \eqref{generalComplexChannelCoefficient} and the various simplifications presented in Section  \ref{subsectionNearFieldArrayResponseVector}. For a general ${M_r} \times {M_t}$ XL-MIMO system with any array architecture, let ${\bf s}_{m_t}$, $1 \le {m_t} \le {M_t}$, denote the location of transmit antenna ${m_t}$, and ${\bf s}_T$ denote the reference point of the transmit array. Further denote by ${{\bm{\delta }}_{{m_t}}}$ the vector from ${\bf s}_T$ to ${\bf s}_{m_t}$, so that ${{\bf{s}}_{{m_t}}} = {{\bf{s}}_T} + {{\bm{\delta }}_{{m_t}}}$, $\forall {m_t}$. Similarly, the locations of the receive array elements are specified by ${{\bf{p}}_{{m_r}}} = {{\bf{p}}_R} + {{\bm{\delta }}_{{m_r}}}$, $1 \le {m_r} \le {M_R}$, where ${\bf p}_R$ denotes the location of the reference point of the receive array, and ${{\bm{\delta }}_{{m_r}}}$ denotes the vector from ${\bf p}_R$ to ${\bf p}_{m_r}$. Let $U_{{m_r},{m_t}}^R$ denote the antenna gain parameter of the receive antenna $m_r$ with respect to the transmit antenna $m_t$, which depends on both $m_r$ and $m_t$ in general since the receive antenna $m_r$ may observe distinct signal directions from different transmit antennas, and vice versa. Besides, let ${U_R}$ denote the antenna gain parameter at ${\bf p}_R$ with respect to ${\bf s}_T$. The notations $U_{{m_r},{m_t}}^T$ and ${U_T}$ at the transmitter side follow similar definitions. Moreover, denote by ${r_{{m_r},{m_t}}}$ the distance between receive-transmit antenna pair $\left({m_r},{m_t}\right)$, and $r$ the link distance between ${\bf p}_R$ and ${\bf s}_T$. In general, there are two methods for near-field LoS XL-MIMO modelling, as elaborated below.

 \textbf{Method 1:}
 The direct method is to model the individual complex-valued channel coefficient  between each transmitter-receiver antenna pair $\left({{m_r},{m_t}} \right)$, so that the LoS channel matrix ${\bf H}^{\rm{LoS}} \in {\mathbb C}^{{M_r} \times {M_t}}$ can be obtained based on all the ${M_r}{M_t}$ channel coefficients, given by
 \begin{equation}\label{generalChannelMIMO}
 {{\bf{H}}^{{\rm{LoS}}}} = \tilde \alpha \left\{ {\sqrt {\frac{{U_{{m_r},{m_t}}^RU_{{m_r},{m_t}}^T}}{{{U_R}{U_T}}}} \frac{r}{{{r_{{m_r},{m_t}}}}}{e^{ - j\frac{{2\pi }}{\lambda }\left( {{r_{{m_r},{m_t}}} - r} \right)}}} \right\},
 \end{equation}
 where $\tilde \alpha  \triangleq \frac{{4\pi \sqrt {{U_R}{U_T}} }}{{\lambda r}}{e^{ - j\frac{{2\pi r}}{\lambda }}}$ is a common coefficient for all channel elements. For the special case of isotropic array elements, the USW-based modelling for LoS XL-MIMO was considered in \cite{bohagen2005construction,bohagen2007design,wu2022distance}. By further modelling the variation of signal amplitude across antenna pairs, the NUSW-based LoS XL-MIMO modelling can be found in \cite{jiang2005spherical,li2023applicable}.

 \textbf{Method 2:}
 Inspired by the far-field UPW-based LoS MIMO modelling, another possible method is to express the near-field LoS XL-MIMO channel matrix as the outer product of the near-field transmit and receive response vectors developed in Section \ref{subsectionNearFieldArrayResponseVector}. Specifically, by treating the transmit array as a point, the near-field receive array response vector can be obtained with respect to the reference location ${\bf s}_T$ of the transmit array, denoted as ${{\bf{a}}_R}\left( {{{\bf{s}}_T}} \right) \in {\mathbb C}^{{M_r} \times 1}$ as in \eqref{generalChannelSIMO}. Similarly, the transmit array response vector with respect to ${\bf p}_R$ is ${{\bf{a}}_T}\left( {{{\bf{p}}_R}} \right) \in {\mathbb C}^{{M_t} \times 1}$, which can be obtained via \eqref{generalChannelSIMO}. Thus, the free-space LoS XL-MIMO channel matrix can be expressed as
 \begin{equation}\label{generalChannelMIMOProduct}
 {{\bf{H}}^{{\rm{LoS}}}} = \tilde \alpha {{\bf{a}}_R}\left( {{{\bf{s}}_T}} \right){\bf{a}}_T^H\left( {{{\bf{p}}_R}} \right).
 \end{equation}
 It is observed that the major difference of the aforementioned two methods lies in that model \eqref{generalChannelMIMOProduct} is always rank-one, while the rank of \eqref{generalChannelMIMO} could be greater than one.

 Note that the classic Rayleigh distance is defined by assuming that either the transmitter or the receiver is equipped with an antenna array, i.e., MISO or SIMO systems. In \cite{lu2023near}, by considering the isotropic elements, the authors derived the {\it MIMO Rayleigh distance} (MIMO-RD) based on the largest phase difference arising from the near-field spherical wavefront and the far-field planar wavefront, given by
 \begin{equation}\label{MIMORayleighDistance}
 {r_{{\rm{MIMO-RD}}}} = \frac{{2{{\left( {{D_T} + {D_R}} \right)}^2}}}{\lambda },
 \end{equation}
 where $D_T$ and $D_R$ denote the array physical dimensions of the transmitter and the receiver, respectively. Similarly, in the far-field region $r \ge {r_{\rm{MIMO-RD}}}$, the three assumptions are made:
 \begin{itemize}[\IEEEsetlabelwidth{12)}]
 \item ${r_{{m_r},{m_t}}} \approx r$, $\forall {m_r},{m_t}$,  for modelling the free-space path loss across array elements.
 \item $U_{{m_r},{m_t}}^R \approx {U_R}$ and $U_{{m_r},{m_t}}^T \approx {U_T}$, $\forall {m_r},{m_t}$, are used for amplitude modelling.
 \item The first-order Taylor approximation of the link distance ${r_{{m_r},{m_t}}}$ is utilized for phase modelling.
 \end{itemize}

 Thus, the free-space LoS XL-MIMO channel matrix for UPW model is
 \begin{equation}\label{generalChannelMIMOUPW}
 \begin{aligned}
 {{\bf{H}}^{{\rm{LoS}}}} &= \tilde \alpha \left\{ {{e^{ - j\frac{{2\pi }}{\lambda }\left( {\frac{{{{\left( {{{\bf{p}}_R} - {{\bf{s}}_T}} \right)}^T}\left( {{{\bm{\delta }}_{{m_r}}} - {{\bm{\delta }}_{{m_t}}}} \right)}}{r}} \right)}}} \right\}\\
 & = \tilde \alpha {\bf{a}}_R^{{\rm{UPW}}}\left( {{{\bf{s}}_T}} \right){\left[ {{\bf{a}}_T^{{\rm{UPW}}}\left( {{{\bf{p}}_R}} \right)} \right]^H},
 \end{aligned}
 \end{equation}
 which is a rank-one matrix. For the case of ULAs equipped at the transmitter and receiver, the locations of antenna $m_r$ and $m_t$ are ${{\bf{p}}_{{m_r}}} = {{\bf{p}}_R} + {\delta _{{m_r}}}d{{\bf{\hat u}}_R}$ and ${{\bf{s}}_{{m_t}}} = {{\bf{s}}_T} + {\delta _{{m_t}}}d{{\bf{\hat u}}_T}$, respectively, where ${\delta _{{m_r}}} = \left( {2{m_r} - {M_r} - 1} \right)/2$ and ${\delta _{{m_t}}} = \left( {2{m_t} - {M_t} - 1} \right)/2$, respectively, and ${{{\bf{\hat u}}}_R}$ and ${{{\bf{\hat u}}}_T}$ denote the direction vectors of the receiver and transmitter, respectively. Besides, denote by ${\theta _T}$ the angle between vectors ${{\bf{p}}_R}-{{\bf{s}}_T} $ and ${{{\bf{\hat u}}}_T}$, and ${\theta _R}$ the angle between vectors ${{\bf{p}}_R}-{{\bf{s}}_T} $ and ${{{\bf{\hat u}}}_R}$. Based on the geometric relationship between the transmitter and receiver, the channel matrix in \eqref{generalChannelMIMOUPW} reduces to
 \begin{equation}\label{MIMOChannelUPW}
 {{\bf{H}}^{{\rm{LoS}}}}\left( {{\theta _T},{\theta _R} } \right) = \tilde \alpha {\bf{a}}_R^{{\rm{UPW}}}\left( {{\theta _R}} \right){\left[ {{\bf{a}}_T^{{\rm{UPW}}}\left( {{\theta _T}} \right)} \right]^H},
 \end{equation}
 where ${{\bf{a}}_T^{{\rm{UPW}}}\left( {{\theta _T}} \right)}$ and ${\bf{a}}_R^{{\rm{UPW}}}\left( {{\theta _R}  } \right)$ denote the far-field transmit and receive array response vectors, respectively, as defined in \eqref{arrayResponseVectorUPWModel}.

 In addition, the UPD in Section \ref{subsubSecNearFieldAmplitudeModelling} can be extended to the MIMO case. Specifically, for the channel vector from transmit antenna $m_t$ to all the receive array elements, the corresponding UPD, denoted as ${r_{\rm{UPD}}}\left({\theta _{{m_t}}}\right)$, can be obtained based on \eqref{definitionOfUPD}, where ${\theta _{{m_t}}}$ follows the similar definition as $\theta$ in \eqref{antennaDistanceSIMO}. By considering all the $M_t$ channel vectors, the MIMO UPD is given by
 \begin{equation}\label{MIMOUPD}
 {r_{{\rm{MIMO - UPD}}}}\left( {\left\{ {{\theta _{{m_t}}}} \right\}} \right) {\rm=} \mathop {\max }\limits_{{m_t}} \left\{ {{r_{{\rm{UPD}}}}\left( {{\theta _{{m_1}}}} \right), \cdots ,{r_{{\rm{UPD}}}}\left( {{\theta _{{M_t}}}} \right)} \right\},
 \end{equation}
 i.e., the distance ensuring all the $M_t$ channel vectors satisfying the uniform wave approximation.

 \begin{figure}[!t]
 \centering
 \centerline{\includegraphics[width=3.5in,height=2.625in]{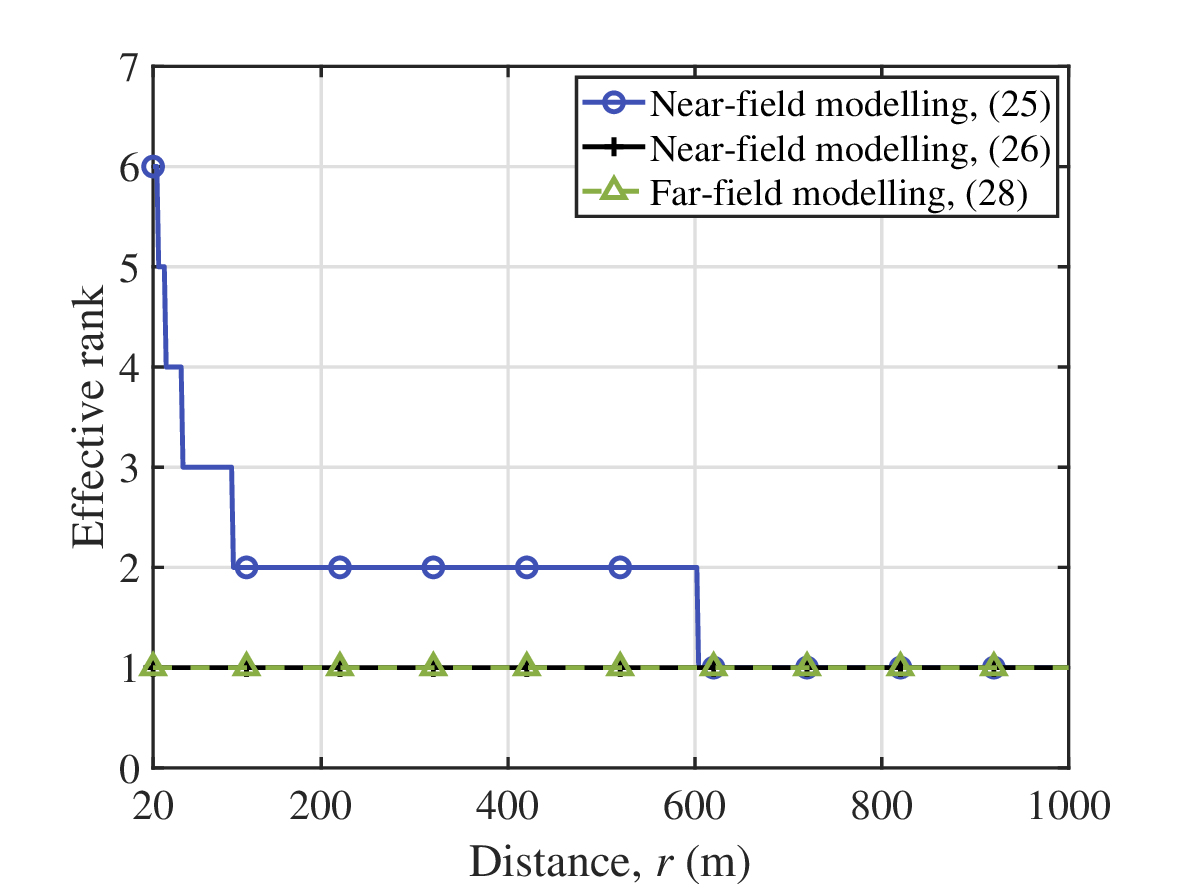}}
 \caption{Effective rank of the LoS XL-MIMO channel matrix versus the link distance $r$ for the near- and far-field modelling.}
 \label{fig:rankComparisonForDifferentModellingApproach}
 \end{figure}

 Fig.~\ref{fig:rankComparisonForDifferentModellingApproach} compares the effective rank of the LoS XL-MIMO channel matrix for the near-field and far-field modelling, where the effective rank is defined as the number of significant singular values that are no smaller than 10$\%$ of the sum of all singular values. The transmit and receive array elements are $M_t = 128$ and $M_r = 32$, respectively, and the cosine element pattern model with $q=2$ is adopted. The transmitter is placed along the $y$-axis, with its center located at the origin, and the receiver is placed perpendicular to the $x$-axis, with its center being $\left(r,0\right)$. It is observed that when the distance $r$ is relatively small, the near-field LoS XL-MIMO modelling \eqref{generalChannelMIMO} yields a rank much greater than one, thus enabling the possibility of spatial multiplexing even in free-space environment. %Besides, the expression of DoF in \eqref{ApproximationDoFULA} gives a valid approximation in this region.
 As the distance increases, the rank of the LoS XL-MIMO channel matrix \eqref{generalChannelMIMO} gradually decreases to one, which is expected since it will reduce to the far-field UPW based channel matrix. On the other hand, for both the near-field LoS XL-MIMO modelling \eqref{generalChannelMIMOProduct} and far-field UPW modelling \eqref{generalChannelMIMOUPW}, we always have a rank-one channel matrix, as expected.

 The impact of the spherical wavefront on properties of LoS MIMO channels was studied in \cite{bohagen2009spherical}, and it was shown that rank greater than one is achieved when applying the USW model for the LoS XL-MIMO channel in the near-field region. By exploiting such a property, the authors in \cite{wu2022distance} proposed a distance-aware precoding architecture, where the number of RF chains is dynamically adjusted according to the distance-related rank.

\subsection{Near-Field Multi-Path XL-MIMO}
 Due to signal reflection, diffraction, and scattering, wireless signals usually undergo the multi-path propagation actually. Thus, proper multi-path modelling is necessary for XL-MIMO communications. A common approach is to model the LoS channel component ${\bf{H}}^{\rm{LoS}}$ and the non-line-of-sight (NLoS) channel component ${\bf{H}}^{\rm{NLoS}}$ separately, and the multi-path XL-MIMO channel matrix can be obtained by superimposing these components \cite{dong2022near,lu2023near}.

 Let $\zeta$ be the indicator variable, with $\zeta = 1$ and $\zeta = 0$ denoting the existence and absence of the LoS component, respectively. Denote by $Q$ the number of scatterers, and ${{\bf{e}}_q}$ the location of scatterer $q$. By regarding scatterers as isotropic points, the NLoS channel component can be expressed as the product of the transmit and receive array response vectors. Thus, the multi-path XL-MIMO channel matrix ${\bf H} \in {\mathbb C}^{{M_r} \times {M_t}}$ can be expressed as
 \begin{equation}\label{generalMultiPathChannelMIMO}
 \begin{aligned}
 {\bf{H}} &= \zeta {{\bf{H}}^{{\rm{LoS}}}} + {{\bf{H}}^{{\rm{NLoS}}}}\\
 &= \zeta {{\bf{H}}^{{\rm{LoS}}}} + \sum\limits_{q = 1}^Q {{\alpha _q}{{\bf{a}}_R}\left( {{\bf{e}}_q} \right){\bf{a}}_T^H\left( {{\bf{e}}_q} \right)},
 \end{aligned}
 \end{equation}
 where ${\alpha}_q$ denotes the complex-valued gain of the NLoS channel path $q$, and ${{{\bf{a}}_T}\left( {{\bf{e}}_q} \right)}$ and ${{{\bf{a}}_R}\left( {{\bf{e}}_q} \right)}$ denote the near-field transmit and receive response vectors with respect to scatterer $q$, respectively, which can be modelled based on Section \ref{subsectionNearFieldArrayResponseVector}. For the special case of XL-SIMO communications, the multi-path channel matrix reduces to a vector
 \begin{equation}\label{generalMultiPathChannelMISOSIMO}
 {\bf{h}} = \zeta {{\bf{h}}^{{\rm{LoS}}}} + {{\bf{h}}^{{\rm{NLoS}}}} = \zeta \alpha {\bf{a}}\left( {\bf{s}} \right) + \sum\limits_{q = 1}^Q {{\alpha _q}{\bf{a}}\left( {{\bf{e}}_q} \right)},
 \end{equation}
 where ${{\bf{a}}\left( {{\bf{e}}_q} \right)}$ denotes the receive array response vector with respect to scatterer $q$. The multi-path channel vector for XL-MISO communications is similar, which is omitted for brevity.

 \begin{figure}[!t]
 \centering
 \centerline{\includegraphics[width=3.3in,height=1.85in]{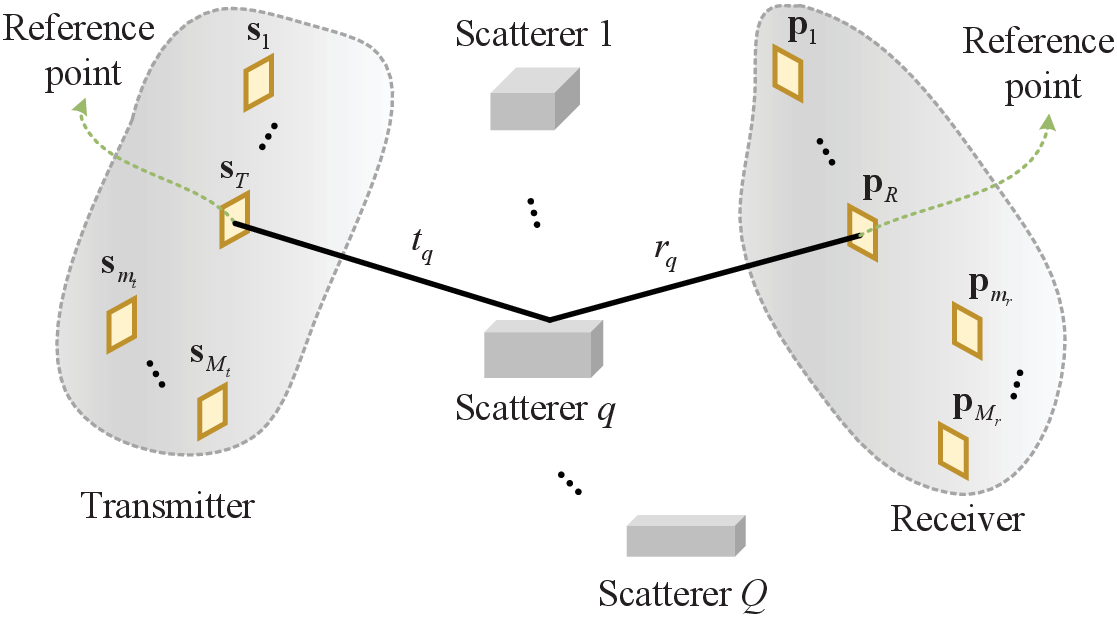}}
 \caption{Illustration of multi-path XL-MIMO between a transmitter and a receiver of general architectures.}
 \label{fig:XLMIMOMultiPathCommunication}
 \end{figure}
 Another effective modelling of the NLoS channel component is using the bistatic radar equation due to scattered rays \cite{skolnik1980introduction,dong2022near,dong2023characterizing,dong2023nearfield}. Specifically, let ${\sigma _q} > 0$ and ${\psi _q}$ denote the radar cross section (RCS) of scatterer $q$ and the additional phase shift arising from scatterer $q$, respectively, ${t_q}$ denote the link distance between the reference point of the transmitter and scatterer $q$, and ${r_q}$ denote the link distance between scatterer $q$ and the reference point of the receiver, as illustrated in Fig.~\ref{fig:XLMIMOMultiPathCommunication}. Then the NLoS channel component can be expressed as \cite{dong2023characterizing}
 \begin{equation}\label{NLoSChannelMatrixBistatic}
 {{\bf{H}}^{{\rm{NLoS}}}} = \sqrt {\frac{{{\beta ^{{\rm{NLoS}}}}}}{Q}} \sum\limits_{q = 1}^Q {{g_q}{e^{ - j\frac{{2\pi }}{\lambda }\left( {{t_q} + {r_q}} \right) + j{\psi _q}}}{{\bf{a}}_R}\left( {{\bf{e}}_q} \right){\bf{a}}_T^T\left( {{\bf{e}}_q} \right)},
 \end{equation}
 where ${\beta ^{{\rm{NLoS}}}} = \sum\nolimits_{q = 1}^Q {\frac{{{\lambda ^2}{\sigma _q}}}{{{{(4\pi )}^3}t_q^2r_q^2}}}$ denotes the total power of the NLoS channel paths between the reference points at the transmitter and receiver, and ${g_q}$ is a random variable accounting for the signal amplitude between the reference point pair that is contributed by scatterer $q$, with ${Q^{ - 1}}\sum\nolimits_{q = 1}^Q {{\mathbb E}\left[ {g_q^2} \right]}  = 1$ \cite{dong2022near,dong2023characterizing}. Besides, for the XL-MISO or XL-SIMO communications, the multi-path channel vector can be similarly modelled \cite{dong2022near,dong2023nearfield}.

 Note that the above models assume that all the array elements are visible to the same set of user equipments (UEs)/scatterers. However, as the array size significantly increases, the spatial non-stationarity appears across the array, i.e., different portions of the array may experience distinct propagation environment, such as cluster sets and/or obstacles \cite{gao2013massive,amiri2018extremely,decarvalho2020nonstationarities,feng2022classification,bian2023novel}. Therefore, VR can be utilized to characterize such spatial non-stationarity. Initially, VR was introduced at the UE side to achieve smooth time evolution \cite{Liu2012COST2100}. Specifically, when the UE moves inside the UE-side VR, the associated scatterers will be active and are visible to the UE. Furthermore, if the UE moves inside the area where multiple VRs overlap, multiple associated scatterers will be visible to the UE simultaneously. It is worth mentioning that one scatterer is at least associated with one VR, while one VR determines the visibility of only one scatterer. For XL-MIMO, the concept of VR is further extended to the BS side \cite{bjornson2019massive,decarvalho2020nonstationarities}. In the following, we give a general overview on the basic concept and modelling methods.

\begin{figure}[!t]
 \centering
 \centerline{\includegraphics[width=3.5in,height=2.1in]{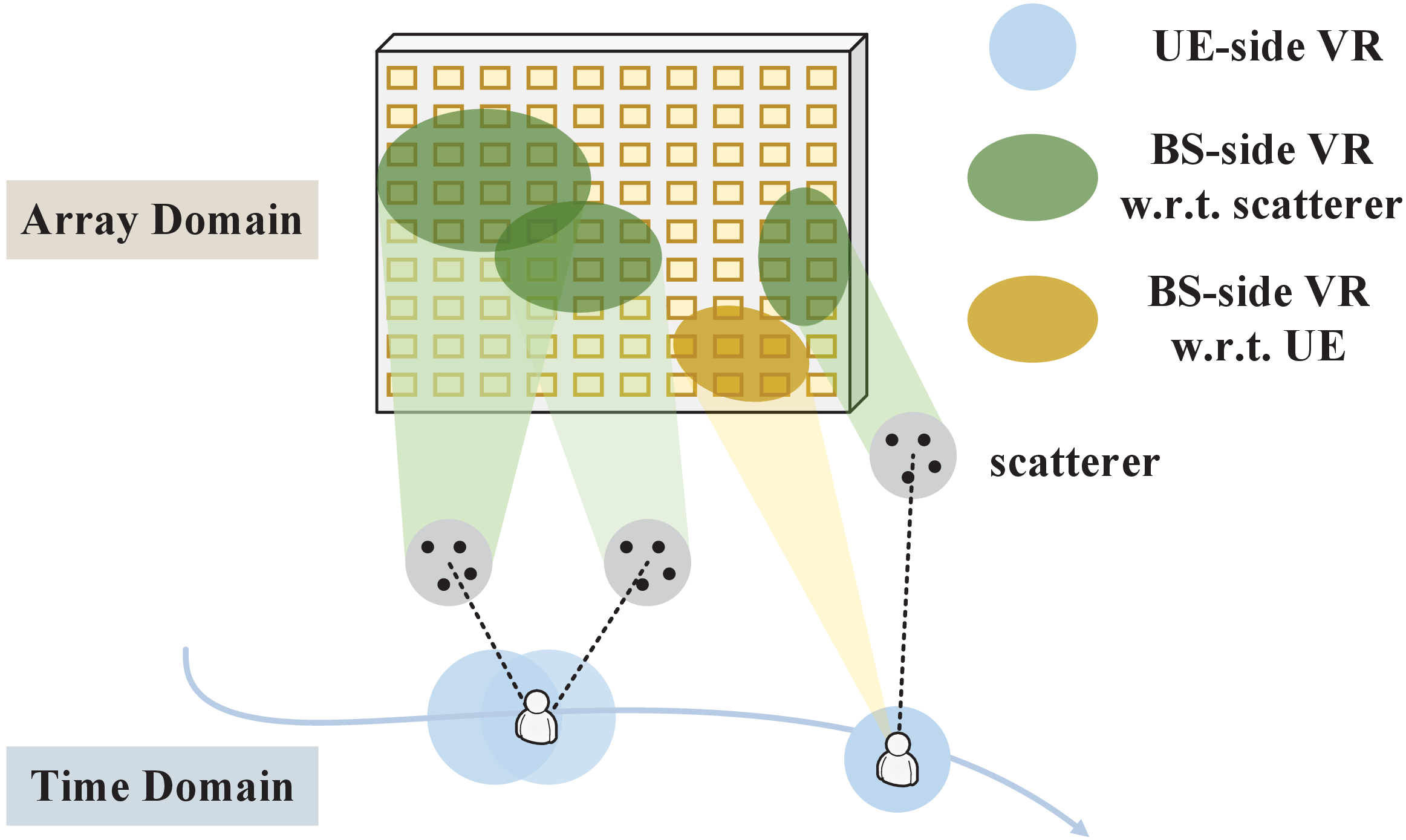}}
 \caption{Illustration of different kinds of VR.}
 \label{fig:VR}
 \end{figure}

 In \cite{decarvalho2020nonstationarities}, the authors summarized the evolution of VR from massive MIMO systems to XL-MIMO systems, and grouped the VR into two categories, i.e., UE-side VR and BS-side VR. The former refers to a geographical area at the UE side, corresponding to the scatterers that can be seen from the UE. The latter stands for the portion of BS side array that are visible to the scatterers or UE, as illustrated in Fig.~\ref{fig:VR}.

 Regarding the UE-side VR, it is derived from geometry-based stochastic model (GBSM) such as COST 2100 \cite{Liu2012COST2100}, which are defined as geographical areas shown in Fig.~\ref{fig:VR}. The UE-side VR is modelled in the time domain to achieve smooth time evolution. For the time-varying scenarios, the visibility of the UE may change due to its movement, thus resulting in the switching of VR mapping relationships \cite{Liu2012COST2100,flordelis2020massive}. Let $\Psi$ denote the set of scatterers visible to the UE, which is determined by the UE location.
 On the other hand, BS-side VR for XL-MIMO systems can be divided into BS-side VR with respect to (w.r.t.) UE and BS-side VR w.r.t. scatterer. They are defined as the portion of array visible to the UE and scatterers, corresponding to the LoS link and NLoS links, respectively, as illustrated in Fig. \ref{fig:VR}. Specifically, BS-side VR w.r.t. UE, denoted as $\Phi^{\rm{LoS}}$, is the portion of array that is visible to the UE. Besides, BS-side VR w.r.t. scatterer, denoted as $\Phi^{\rm{NLoS}}$, is the portion of array that is visible to the scatterer. The occurrence of BS-side VR can be viewed as two major manifestations \cite{decarvalho2020nonstationarities}: 1) unequal pathloss across the array, and 2) signal blockage stemming from obstacles between the UE and the array.

\begin{itemize}
\item \textit{Unequal Pathloss:} When moving towards the XL-MIMO regime, the UE and scatterers are less likely to be located in the far-field region, and the NUSW characteristic may result in significant variations of signal amplitude across array elements, as discussed in Section \ref{subsectionNearFieldArrayResponseVector}. Thus, an unequal pathloss will be observed, and channel measurements have reported the variations of the channel power across the array \cite{martinez2014towards,wang2022characteristics,yuan2022spatial}.
\item \textit{Signal Blockage: }For XL-MIMO systems, the XL-array can be widely spread on the facades of a building to serve densely distributed UEs in hotspot scenarios. Therefore, part of the array elements may be blocked due to the existence of obstacles, such as vehicles, tree, buildings, thus leading to the uneven channel power distribution across the array.
\end{itemize}

 Note that to achieve the smooth time evolution, the UE-side VR can be modelled as a birth-death process at the time axis \cite{flordelis2020massive}. The birth and death rates of scatterers are assumed to be $\lambda_G$ and $\lambda_R$, respectively. At time $t$, an initial set of scatterers, denoted as ${\Psi}\left(t\right)$, are generated, and $N\left( t \right) \triangleq \left| {\Psi \left( t \right)} \right|$ denotes the cardinality of the set ${\Psi}(t)$. At time $t + \Delta t$, scatterer evolves on the time axis, and a random number of new scatterers are generated according to a Poisson distribution, given by
 \begin{equation}
 {\mathbb E}\left[ {N\left( {t + \Delta t} \right)} \right] = \frac{{{\lambda _G}}}{{{\lambda _R}}}\left( {1 - {P_T}\left( {\Delta t} \right)} \right),
 \end{equation}
 where ${{P_T}\left( {\Delta t} \right)}$ denotes the survival probability of a scatterer after $\Delta t$, as defined in \cite{wu2017general}.

 Furthermore, denote by ${\Phi ^{{\rm{NLOS}}}}\left( {{\bf{e}}_q} \right)$ the BS-side VR w.r.t. scatterer $q$, $\forall q\in \Psi$ \cite{Han2023Towards}. By taking into account VR, the multi-path channel vector can be modelled as \cite{han2020channel,han2022localization,han2020deep,tian2023low}
 \begin{equation}\label{multiPathChannelVR}
 \begin{aligned}
 {\bf{h}} &= \zeta \alpha {\bf{a}}\left( {\bf{s}} \right) \odot {\bf{b}}\left( {{\Phi ^{{\rm{LoS}}}}} \right) + \\
 &\ \ \ \ \ \ \ \ \ \ \ \ \sum\limits_{q \in \Psi } {{\alpha _q}{\bf{a}}\left( {{\bf{e}}_q} \right) \odot {\bf{b}}\left( {{\Phi ^{{\rm{NLoS}}}}\left( {{\bf{e}}_q} \right)} \right)},
 \end{aligned}
 \end{equation}
 where ${\bf{b}}\left( {{\Phi ^{{\rm{LoS}}}}} \right) \in {\left\{ {0,1} \right\}^{M \times 1}}$ and ${{\bf{b}}\left( {{\Phi ^{{\rm{NLoS}}}}\left( {{\bf{e}}_q} \right)} \right)} \in {\left\{ {0,1} \right\}^{M \times 1}}$ are binary vectors that indicate the visibility or invisibility of UE and scatterer $q$ to the BS array, respectively. For example, ${{\bf{b}}_m}( {{\Phi ^{{\rm{LoS}}}}} ) = 1$ means that UE is visible to array element $m$, and ${{\bf{b}}_m}( {{\Phi ^{{\rm{LoS}}}}} ) =0$ otherwise. However, when a priori knowledge of the actual environment is unavailable, the UE's or scatterers' invisibility/visibility can be modelled as a random process in the spatial domain along the array axis. In general, the invisibility/visibility of the UE or scatterers to the BS array is modelled as a Markov process or a birth-death process \cite{lopez2016novel,wu2017general,lopez2018novel,wang2021general}.

\subsection{Spatial Correlation Based Near-Field Modelling}\label{subsectionSpatialCorrelationModelling}
 Apart from the multi-path channel model in \eqref{multiPathChannelVR} that describes each decomposable path, spatial correlation matrix-based model is another widely used channel model \cite{mcnamara2002spatial}. The spatial correlation is an important approach to characterize the second-order channel statistics, which helps develop the Kronecker channel model \cite{forenza2006benefit} and the transmission strategy with the statistical CSI \cite{jorswieck2004channel}.
 The spatial correlation matrix of wireless SIMO or MISO channels is defined as $\mathbf{R} =\frac{1}{\varsigma}\mathbb{E}\{\mathbf{h} \mathbf{h}^H \}\in \mathbb{C}^{M\times M}$, where $\varsigma$ denotes the large-scale channel factor at the reference antenna element.

 For the conventional far-field UPW assumption, the spatial correlation based channel vector between the single-antenna UE and the $M$-dimensional antenna array is modelled as \cite{marzetta2010noncooperative,ngo2013energy}
 \begin{equation}\label{eq:SCSMmimo}
 {\bf{h}} = {\sqrt \varsigma} {{\bf{R}}^{1/2}}{\tilde {\bf h}},
 \end{equation}
 where ${\tilde {\bf h}} \sim \mathcal{CN}(\mathbf{0},\mathbf{I}_M)$ denotes a circularly symmetric complex Gaussian random vector. In particular, when $\mathbf{R}$ is an identity matrix, $\bf{h}$ reduces to the well-known independent and identically distributed (i.i.d.) Rayleigh fading channel \cite{chiani2003capacity}. It is observed from \eqref{eq:SCSMmimo} that all the channel entries share the common large-scale channel factor, and each scatterer is visible to the whole array. However, in XL-MIMO systems, due to the NUSW property and the existence of VR, \eqref{eq:SCSMmimo} should be modified to fit the new channel characteristics, as elaborated in the following.

\subsubsection{Differences Brought by NUSW}
 When the UE is located in the near-field region, different array elements experience different large-scale fading conditions, as reflected by the following modified model \cite{amiri2019message,amiri2020deep,may2020stochastic,feng2022mutual}
 \begin{equation}\label{eq:SCSMxlmimo}
 {\bf{h}} = \sqrt {\bm{\varsigma}}  \odot {{\bf{R}}^{1/2}}{\tilde {\bf h}},
 \end{equation}
 where ${\bm {\varsigma}} \in \mathbb{R}^{M\times 1}$. The $m$-th element of ${\bm {\varsigma}}$, denoted as ${{\bm{\varsigma }}_m}$, depends on the link distance ${r_m}$, which can be modelled as \cite{amiri2019message}
 \begin{equation}\label{eq:wkm}
 {{\bm{\varsigma }}_m} = \varepsilon r_m^\nu,
 \end{equation}
 where $\varepsilon$ is the attenuation coefficient, and $\nu$ is the path loss exponent. In this case, when $\mathbf{R}$ is an identity matrix, $\mathbf{h}$ becomes an independent and non-identically distributed Rayleigh fading channel.

 On the other hand, the NUSW characteristic can be reflected by the spatial correlation matrix $\mathbf{R}$. For example, for the conventional far-field UPW assumption, the $(m,n)$th element of $\mathbf{R}$ is expressed as \cite{abdi2002space}
 \begin{equation}\label{eq:oneRingPW}
 {{\bf{R}}_{m,n}} = \int_{\bar\theta-\Delta}^{\bar\theta+\Delta} e^{j\frac{2\pi}{\lambda}(m-n)d\sin\theta}f(\theta){\rm d}\theta,
 \end{equation}
 where $\bar\theta$ and $\Delta$ denote the mean and spread of scatterers' angles, respectively, and $f(\theta)$ is the power angular spectrum (PAS). It is observed that the conventional far-field UPW based spatial correlation in (\ref{eq:oneRingPW}) only depends on the PAS and the relative antenna locations $m-n$, which exhibits spatial wide-sense stationarity (SWSS). Furthermore, the authors in \cite{dong2022near} derive an integral expression for the near-field spatial correlation based on the NUSW model, to accurately characterize the XL-MIMO communication. It was revealed in \cite{dong2022near} that the near-field spatial correlation is actually determined by the {\it power location spectrum} (PLS), whose scatterer distribution is characterized by both the scatterer's angles and the distances from the antenna array.
 The element of the near-field spatial correlation based on NUSW model can be expressed as \cite{dong2022near}
 \begin{equation}\label{eq:oneRingSW}
 {\mathbf{R}}_{m,n} = \int_{{\bf q}\in{\mathcal Q}} \frac{r^2(\bf q)}{r_{m}(\mathbf{q})r_{n}(\mathbf{q})} e^{j\frac{2\pi}{\lambda}(r_{m}(\mathbf{q})-r_{n}(\mathbf{q}))}f({\bf q}){\rm d}{\bf q},
 \end{equation}
 where $\mathcal Q$ denotes the set of scatterers, $r(\bf q)$ denotes the distance between scatterer ${\bf q}$ and the array reference element, $r_{m}(\mathbf{q})$ denotes the distance between scatterer ${\bf q}$ and the $m$-th array element, and $f\left( {\bf{q}} \right)$ represents the PLS of the scattering environment. Note that such a NUSW characteristic renders SWSS no longer valid for the near-field spatial correlation. Moreover, for the XL-SIMO communications, a closed-form expression of the near-field spatial correlation was derived by considering the generalized one-ring model in \cite{dong2022near}.

 \subsubsection{Differences Brought by VR}
 By taking into account the VR, the channel model in \eqref{eq:SCSMmimo} is modified as \cite{rodrigues2020low,amiri2018extremely,marinello2020antenna}
 \begin{equation}\label{eq:hVR}
 {\bf{h}} = \sqrt {\bm{\varsigma }}  \odot {\left( {{{\bf{R}}^{{\rm{VR}}}}} \right)^{1/2}}{\tilde {\bf h}},
 \end{equation}
 where ${{{\bf{R}}^{{\rm{VR}}}}}$ denotes the spatial correlation matrix considering the VR. When the BS-side VR w.r.t. scatterers cannot cover the whole array, some channel entries have zero value. Then, the spatial correlation matrix ${{\bf{R}}^{{\rm{VR}}}}$ can be modelled as \cite{ali2019linear,croisfelt2021accelerated,yang2020uplink}
 \begin{equation} \label{eq:Rk}
 {{\bf{R}}^{{\rm{VR}}}} = {{\bf{D}}^{1/2}}{\bf{R}}{{\bf{D}}^{1/2}},
 \end{equation}
 where $\mathbf{D}\in \{0,1\}^{M\times M}$ is a determined diagonal matrix, which indicates whether each antenna element is seen by the UE.
 %for the NLoS links.
 Specifically, ${\mathbf{D}}_{m,m}=1$ represents that the UE is visible to the $m$-th array element, and ${\mathbf{D}}_{m,m}=0$ indicates otherwise. Besides, the spatial correlation matrix $\mathbf{R}$ in \eqref{eq:Rk} is defined based on the conventional far-field UPW-based model in \eqref{eq:oneRingPW} \cite{ali2019linear,yang2020uplink}. Furthermore, a non-stationary channel model based on the double scattering MIMO channel \cite{shin2003capacity} was introduced in \cite{amiri2022distributed,Han2023Towards}. In this model, the VR is modelled based on the visibility between the UE and UE-side scatterers, as well as the visibility between the BS array and BS-side scatterers.

 \begin{table*}[!t]	
    \centering
	\caption{Spatial Correlation Matrices for Different Models}\label{table:SpatialCorrelationMatrices}
		\begin{tabular}{|c|c|}
			\hline
			{\bf Model }                            & {\bf Spatial Correlation Matrix }             \\ \hline\hline
			Far-field UPW model                     &  ${{\bf{R}}_{m,n}} = \int_{\bar \theta  - \Delta }^{\bar \theta  + \Delta } {{e^{j\frac{{2\pi }}{\lambda }(m - n)d\sin \theta }}} f(\theta ){\rm d}\theta $
            \\ \hline
			Near-field NUSW model                   &  ${{\bf{R}}_{m,n}} = \int_{{\bf{q}} \in {\cal Q}} {\frac{{{r^2}({\bf{q}})}}{{{r_m}({\bf{q}}){r_n}({\bf{q}})}}} {e^{j\frac{{2\pi }}{\lambda }({r_m}({\bf{q}}) - {r_n}({\bf{q}}))}}f({\bf{q}}){\rm d}{\bf{q}}$
            \\ \hline
			{\makecell[c]{Near-field NUSW model \\ considering BS-side VR w.r.t. scatterers}}   &  ${\bf{R}}_{m,n}^{{\rm{VR}}} = \int_{{\bf{q}} \in {\cal Q}}\mathbb{E}  \left[ {{{\bf{b}}_m}\left( {{\Phi ^{{\rm{NLoS}}}}({\bf{q}})} \right){{\bf{b}}_n}\left( {{\Phi ^{{\rm{NLoS}}}}({\bf{q}})} \right)} \right]\times\frac{r^2(\bf q)}{r_{m}(\mathbf{q})r_{n}(\mathbf{q})} e^{j\frac{2\pi}{\lambda}(r_{m}(\mathbf{q})-r_{n}(\mathbf{q}))}f({\bf q}){\rm d}{\bf q}$
            \\ \hline
	 \end{tabular}
\end{table*}

 In addition, the authors in \cite{dong2023nearfield} developed a new integral form for near-field spatial correlation in \eqref{eq:hVR}, which takes into account both the NUSW model and the BS-side VR w.r.t. scatterers. The spatial correlation matrix is expressed by
\begin{equation}\label{eq:oneRingSWVR}
\begin{aligned}
{\bf{R}}_{m,n}^{{\rm{VR}}} = &\int_{{\bf{q}} \in {\cal Q}}\mathbb{E}  \left[ {{{\bf{b}}_m}\left( {{\Phi ^{{\rm{NLoS}}}}({\bf{q}})} \right){{\bf{b}}_n}\left( {{\Phi ^{{\rm{NLoS}}}}({\bf{q}})} \right)} \right]\\
&\times\frac{r^2(\bf q)}{r_{m}(\mathbf{q})r_{n}(\mathbf{q})} e^{j\frac{2\pi}{\lambda}(r_{m}(\mathbf{q})-r_{n}(\mathbf{q}))}f({\bf q}){\rm d}{\bf q},
\end{aligned}
\end{equation}
where ${\bf{b}}_m\left( {\Phi ^{{\rm{NLoS}}}(\mathbf{q})} \right)$ describes the invisibility/visibility of scatterer ${\bf q}$ to the $m$-th array element.
To unveil the evolution of VR, the authors in \cite{dong2023nearfield} proposed a two-stage homogeneous Markov process to model the BS-side VR w.r.t. scatterers. It was shown that the SWSS is no longer valid for the near-field spatial correlation considering the partial visibility in \eqref{eq:oneRingSWVR}. A comparison of the spatial correlation matrices for different models is summarized in Table~\ref{table:SpatialCorrelationMatrices}.

\subsection{Extensions of Near-Field Modelling}
 In this subsection, we discuss some extensions of near-field modelling for XL-MIMO.

 \textbf{Uniform planar array (UPA) and modular XL-MIMO:} UPA-based XL-MIMO can be deployed to enable three-dimensional (3D) spatial resolution, and the above near-field modelling for the ULA can be extended to UPA, by considering the two-dimensional (2D) signal directions. Specifically, denote by $M = {M_H}{M_V}$ the number of UPA elements, with $M_H$ and $M_V$ denoting the number of array elements per row and per column, respectively. Let $d_H$ and $d_V$ denote the antenna spacing along the horizontal and vertical directions, respectively. Further denote by ${{\bf{\hat u}}_H}$ and ${{\bf{\hat u}}_V}$ the direction vectors of the UPA along the horizontal and vertical dimensions, respectively, with $\left\| {{{{\bf{\hat u}}}_H}} \right\| = \left\| {{{{\bf{\hat u}}}_V}} \right\| = 1$. By indexing the array element row-by-row, the corresponding row and column of array element $m$ are ${m_V} = \lceil {\frac{m}{{{M_H}}}} \rceil$, ${m_H} = \bmod \left( {m,{M_H}} \right)$, respectively. Let the array center be the reference point $\bf p$. The location of the $m$-th array element is ${{\bf{p}}_m} = {\bf{p}} + {\delta _{{m_V}}}{d_V}{{\bf{\hat u}}_V} + {\delta _{{m_H}}}{d_H}{{\bf{\hat u}}_H}$, with ${\delta _{{m_V}}} = (2{m_V} - {M_V} - 1)/2$ and ${\delta _{{m_H}}} = (2{m_H} - {M_H} - 1)/2$, respectively. The distance between the signal source $\bf s$ and array element $m$ is ${r_m} = \left\| {{{\bf{p}}_m} - {\bf{s}}} \right\| = \left\| {{\bf{p}} - {\bf{s}} + {\delta _{{m_V}}}{d_V}{{{\bf{\hat u}}}_V} + {\delta _{{m_H}}}{d_H}{{{\bf{\hat u}}}_H}} \right\|$. By substituting $r_m$ into \eqref{generalChannelSIMO}, the general near-field array response vector of UPA can be obtained. In this case, the near-field channel between the source and the antenna array depends on the link distance and the 2D signal direction pair.

\begin{figure}[!t]
 \centering
 \centerline{\includegraphics[width=3.2in,height=2.5in]{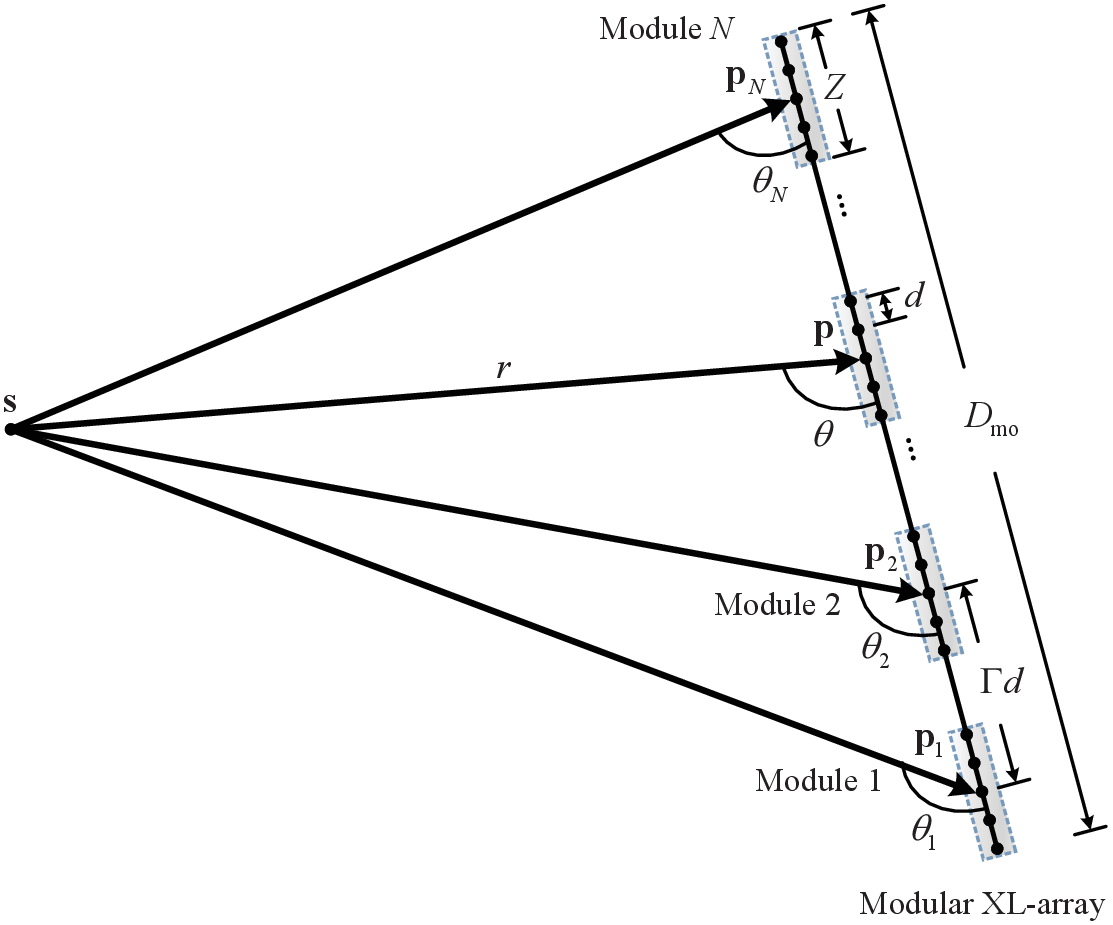}}
 \caption{An illustration of modular XL-array.}
 \label{fig:modularArrayULA}
 \end{figure}

 On the other hand, besides the conventional collocated XL-array architecture, near-field modelling for the new modular XL-array architecture has been pursued in \cite{li2022near,li2022modular}. Compared to the collocated XL-array where adjacent elements are separated by half-wavelength, the inter-module spacing is typically much larger than the signal wavelength, thus achieving a larger array aperture, as illustrated in Fig.~\ref{fig:modularArrayULA}. This renders the modular XL-array exhibit stronger near-field effect than the collocated counterpart. Let $Z$ and $D_{\rm mo}$ denote the physical dimensions of each module and the whole modular XL-array, respectively. Further denote by $\Gamma d$ the inter-module separation between the reference points of adjacent modules, with $\Gamma$ being a module separation parameter and $d$ denoting the antenna spacing within each module. In particular, when the source is located in the near-field region of the whole modular XL-array but the far-field region of each module, i.e., $2{Z^2}/\lambda  \le r < 2D_{{\rm{mo}}}^2/\lambda $, a simplified subarray based USW model with distinct angles was developed in \cite{li2023near,li2023multi}. Moreover, when the source is located in the near-field region of the whole array but the extended far-field region of each module, specified by the region $\max \left\{ {5{D_{{\rm{mo}}}},4Z{D_{{\rm{mo}}}}/\lambda } \right\} \le r < 2D_{{\rm{mo}}}^2/\lambda$, the AoAs of all modules are approximately equal, i.e., ${\theta}_n \approx \theta$, $\forall n$, where $\theta_n$ denotes the AoA at the reference point ${\bf p}_n$ of module $n$. In this case, the subarray based USW model with common angle can be used. It was also found that under the near-field subarray based USW model with common angle, the array response vector of the modular XL-array can be expressed as the Kronecker product of the array response vectors of a sparse array and a collocated array.

 \textbf{Polarization mismatch:} The polarization effect can be applied to increase the capacity of the system by creating independent channels, and the research on the polarization effect mainly focused on the dual-polarization \cite{li2014frequency,pereda2018experimental} and triple polarization \cite{yuan2021electromagnetic,zhang2023pattern,sanguinetti2023wavenumber}. As discussed in Section \ref{subSectionNewCharacteristics}, the array elements will experience different mismatch due to the distinct AoAs in the near-field region, whose effect cannot be reflected by simply multiplying the common loss coefficient as in the far-field region \cite{bjornson2020power}. By accurately modelling the variations of wave propagation distances, projected aperture, and losses of the polarization mismatch across array elements, the free-space LoS channel is established in \cite{bjornson2020power} based on the electric field between the transmitter and receiver. With such a channel model, the normalized antenna array gain, defined as the total received power of $M$ antennas divided by the received power of $M$ reference antennas (as if they were all located at the origin), is introduced in \cite{bjornson2021primer}, so as to characterize the array gain difference in the near- and far-field regions. In the far-field region, all the array elements capture the same power as the reference antenna, and the normalized antenna array gain is equal to one. Besides, it was shown that the normalized antenna array gain approaches one when the link distance is beyond the {\it Bj{\"o}rnson distance} \cite{bjornson2021primer}, given by ${r_{\rm{B}}} = 2{A_d}\sqrt M$, where ${A_d}$ is the diagonal dimension of each antenna. This implies that when the link distance is smaller than the Bj{\"o}rnson distance, it is necessary for near-field modelling to consider the losses of the polarization mismatch across array elements \cite{bjornson2020power,bjornson2021primer}. The impact of polarization has also been considered for the modelling of the LIS \cite{dardari2020communicating}.

 \textbf{Spatial-wideband effect:} In practice, XL-MIMO could be accompanied by high-frequency communication systems, such as millimeter wave (mmWave) and THz bands \cite{cui2023nearRainbow,yuan2023deep}. For wideband mmWave or THz XL-MIMO communications with the significantly increased system bandwidth $B$, the maximum propagation delay between different array elements is comparable to or even exceeds the symbol duration (inversely proportional to $B$). In this case, different array elements may even receive distinct symbols at the same sampling time, thus giving rise to the unsynchronized reception, which is known as {\it spatial-wideband} effect \cite{wang2018spatial,wang2018spatialMag,xu2022near,wu20233D}, and such an effect causes {\it beam squint/split} issue in the frequency domain. To include the impact of the variation of the propagation delay across array elements, the channel impulse response from the source to antenna $m$ can be expressed as
 \begin{equation}\label{generalChannelSIMOTimeDelay}
 {h_m}\left( t \right) = {\alpha} \sqrt {\frac{{{U_m}}}{U}} \frac{r}{{{r_m}}}{e^{ - j\frac{{2\pi }}{\lambda }\left( {{r_m} - r} \right)}}\delta \left( {t - {\tau _m}} \right),
 \end{equation}
 where ${\tau _m} = {r_m}/c$ denotes the propagation delay from the source to antenna $m$, and $U$ follows the same definition in \eqref{generalComplexChannelCoefficient}. On the other hand, when $D \ll c/B$, the propagation delays of different array elements are approximately equal \cite{wang2018spatial,wang2023cram}. In particular, the beam squint/split effect can be mitigated by utilizing the true-time-delay (TTD) line, which can introduce a programmable true time delay to compensate for the propagation delay among array elements, thus enabling frequency-dependent phase shift \cite{wang2018spatialMag,zhang2023nearbeamfocusing,myers2021infocus,wang2023beamfocusing,zhang2023deep}.

 \textbf{Green's Function:} Note that the above near-field channel modelling based on array response vector is applicable only for XL-MIMO with discrete array architecture. One generalized modelling method is based on the Green's function, which can be applied for both the discrete- and continuous-aperture XL-MIMO. The Maxwell's equations can depict the relationship between the current distribution ${\bf{J}}\left( {\bf{r}} \right)$ and the electric field ${\bf{E}}\left( {\bf{r}} \right)$ as \cite{dardari2020communicating}
 \begin{equation}\label{Maxwell}
 {\nabla _{\bf{r}}}  \times {\nabla _{\bf{r}}}  \times {\bf{E}}\left( {\bf{r}} \right) - {\kappa ^2}{\bf{E}}\left( {\bf{r}} \right) = j\kappa {Z_0}{\bf{J}}\left( {\bf{r}} \right),
 \end{equation}
 where $\bf{r}$ is an arbitrary point, $\kappa =2\pi /\lambda $ is the wavenumber, $Z_0=376.73$ $\Omega $ is the intrinsic impedance of spatial medium, and $\nabla _{\bf{r}}$ is the first-order partial derivative operator with respect to $\bf{r}$. Note that \eqref{Maxwell} can be solved numerically by the Green's function $\bf{G}\left( \bf{r},\bf{s} \right)$ \cite{sanguinetti2023wavenumber,yuan2021electromagnetic,zhang2023pattern} as
 \begin{equation}\label{Relation}
 {\bf{E}}\left( {\bf{r}} \right) = \int_{{V_S}} {{\bf{G}}\left( {{\bf{r}},{\bf{s}}} \right){\bf{J}}\left( {\bf{s}} \right)\mathrm{d}{\bf{s}}},
 \end{equation}
 where $\bf{J}\left( \bf{s} \right) $ is the current distribution at the transmitter point $\bf{s}$, and $V_S$ is the transmitter volume. It is observed that the Green's function can link the current distribution at the transmitter and the electric field at the receiver. One generalized approach is the dyadic Green's function, which can depict the polarization effect between the transmitter and receiver \cite{sanguinetti2023wavenumber,yuan2021electromagnetic,zhang2023pattern} as
 \begin{equation}\label{Dyadic}
 {\bf{G}}\left( {{\bf{r}},{\bf{s}}} \right) = \frac{{j\kappa {Z_0}}}{{4\pi }}\frac{{{e^{-j\kappa \left\| {{\bf{r}} - {\bf{s}}} \right\|}}}}{{\left\| {{\bf{r}} - {\bf{s}}} \right\|}}\left( {{{\bf{I}}_3} + \frac{{{\nabla _{\bf{r}}}\nabla _{\bf{r}}^H}}{{{\kappa ^2}}}} \right),
 \end{equation}
 where ${{\bf{I}}_3}$ denotes a $3\times 3$ identity matrix. Specifically, $\mathbf{G}\left( \mathbf{r},\mathbf{s} \right) \in \mathbb{C} ^{3\times 3}$ in \eqref{Dyadic} can be expressed as \cite{yuan2021electromagnetic}
 \begin{equation}\label{GPolar}
 \mathbf{G}=\left[ \begin{matrix}
	G_{xx}&		G_{xy}&		G_{xz}\\
	G_{yx}&		G_{yy}&		G_{yz}\\
	G_{zx}&		G_{zy}&		G_{zz}\\
 \end{matrix} \right],
 \end{equation}
 where $G_{ab}$ denotes the scalar Green's function between polarization direction $a$ of ${\bf r}$ and polarization direction $b$ of $\mathbf{s}$, with $a,b\in \left\{ x,y,z \right\}$. Thus, the effect of all polarization directions $x,y,z$ can be showcased. In \cite{yuan2021electromagnetic} and \cite{zhang2023pattern}, the authors studied the dyadic Green's function based channel for the discrete plane array and continuous surface, respectively. For far-field approximation $\left\| \bf{r}-\bf{s} \right\| \gg \lambda $, \eqref{Dyadic} can be approximated as
 \begin{equation}\label{FarField}
 \begin{aligned}
 {\bf{G}}\left( {{\bf{r}},{\bf{s}}} \right) \approx \frac{{j\kappa {Z_0}}}{{4\pi }}\frac{{{e^{-j\kappa \left\| {{\bf{r}} - {\bf{s}}} \right\|}}}}{{\left\| {{\bf{r}} - {\bf{s}}} \right\|}}\left( {{{\bf{I}}_3} - {\bf{\hat p}}{{{\bf{\hat p}}}^H}} \right),
 \end{aligned}
 \end{equation}
 with ${\bf{\hat p}}=\left( \bf{r}-\bf{s} \right) /\left\| \bf{r}-\bf{s} \right\|$. Moreover, one simplified approach without the polarization effect is the scalar Green's function as
 \begin{equation}\label{Scalar}
 {\bf{G}}\left( {{\bf{r}},{\bf{s}}} \right) = \frac{{j\kappa {Z_0}}}{{4\pi }}\frac{{{e^{-j\kappa \left\| {{\bf{r}} - {\bf{s}}} \right\|}}}}{{\left\| {{\bf{r}} - {\bf{s}}} \right\|}}.
 \end{equation}
 The authors in \cite{pizzo2020spatially} and \cite{pizzo2022spatial} studied the scalar Green's function based channel for the continuous surface.

 \textbf{Electromagnetic Information Theory (EIT):} As an emerging research topic, EIT has attracted significant research interest recently \cite{zhu2022electromagnetic,wan2023mutual,zhang2023pattern}. Specifically, EIT is a research field that combines the EM theory and the information theory to exploit the potentials through EM waves. Note that the classical information theory relies on the spatially discrete modelling and mismatches the continuous EM fields. Thus, EIT is expected to uncover the EM theoretical capacity bounds. The authors in \cite{wan2023mutual} considered two continuous regions over random EM fields. Then, the mutual information between the parallel linear transmitter/receiver was derived based on the Mercer expansion. More specifically, the capacity bounds based on parallel infinite-length linear transmitter/receiver, infinite-length linear transmitter/receiver, parallel linear infinite-length transmitter and finite-length receiver, and finite-length transmitter/receiver are derived.

 On the basis of the methods in \cite{wan2023mutual}, the authors in \cite{zhang2023pattern} studied the mutual information for the scenario where a single BS equipped with the continuous surface served multiple users equipped with the continuous surface. Then, the electric current density distribution at the transmitter was optimized to maximize the sum capacity. As observed, EIT can be applied to depict the theoretical capacity for the EM based XL-MIMO systems and to optimally design the XL-MIMO systems. To further promote the analysis and practical implementation for XL-MIMO systems, the EIT for more practical scenarios, such as the scenario with multiple UEs, should be investigated in the future. Besides, the polarization effect also should be introduced to construct the EIT analysis framework.

\subsection{Near-Field Channel Measurements}
\begin{table*}[ht]
\centering
\caption{A summary of near-field channel measurement campaigns}
\label{table:XL-MIMOChannelMeasurement}
\begin{tabular}{|c|c|c|c|c|c|}
\hline
 {\bf Reference} & \makecell[c]{{\bf Frequency} \\{\bf (GHz)} }& {\bf Bandwidth} & {\bf Scenario} & {\bf Antenna Configuration} & {\bf Measured Channel Parameters} \\
 \hline \hline
 \cite{li2015channel}& 1.4725 & 91 MHz &Outdoor  & \makecell[l]{ TX: 128-element virtual ULA\\RX: a bi-conical antenna} & NUSW: PAS \\ \hline

% \cite{sangodoyin2015cluster}& 2.53 & 20 MHz & Outdoor & \makecell[l]{ TX: (16$\times $60)-element virtual UCA \\RX: 32-element practical UCA} &  RMS AS, correlation\\ \hline

\cite{payami2012channel} & 2.6 & 50 MHz & Outdoor & \makecell[l]{ TX: single antenna\\RX: 128-element virtual ULA}  & \makecell[c]{ NUSW: PAS\\Non-stationarity: channel gain, K-factor, PAS} \\ \hline

\cite{willhammar2020channel} & 2.6 & 40 MHz & \makecell[c]{Indoor\\Outdoor} & \makecell[l]{ TX: single antenna\\RX: 128-element practical UCA}  & \makecell[c]{Channel hardening: channel gain \\standard deviation}\\ \hline

\cite{wang2017variation} & 3.5 & 200 MHz& Outdoor  &\makecell[l]{ TX:  256-element virtual UPA  \\RX:  16-element practical array}  &   Non-stationarity: PAS, RMS AS\\ \hline

\makecell[c]{\cite{zheng2022ultra,zheng2023measurements}} & 5.3 & 160 MHz& Outdoor & \makecell[l]{ TX:  8 omnidirectional antennas \\RX: (128$\times $8)-element practical ULA}   &\makecell[c]{ NUSW: PAS\\Non-stationarity: PDP, RMS DS, RMS AS\\ Channel hardening: channel gain \\standard deviation} \\ \hline

 \cite{martinez2016massive,martinez2018experimental}& 5.8 & 100 MHz& \makecell[c]{ Outdoor\\Indoor} & \makecell[l]{ TX:  128-element practical ULA\\RX: 2 antennas}   &\makecell[c]{ Non-stationarity: channel gain\\Channel hardening: channel gain \\ standard deviation} \\ \hline

  \cite{huang2017multi}& 11/16/28/38 & 2/2/4/4 GHz& Indoor & \makecell[l]{ TX: (51$\times$51)-/(76$\times$76)-/(91$\times$91)-/\\ \quad\quad(121$\times$121)-element vertical UPA \\RX: an omnidirectional biconical antenna}   &\makecell[c]{ NUSW: PAS\\Non-stationarity: PDP, PAS, RMS DS, RMS AS}\\ \hline

\cite{chen2017measurement}& 15 & 4 GHz& Outdoor & \makecell[l]{ TX: an omnidirectional
antenna \\RX: $40\times 40$-element virtual UPA}   &\makecell[c]{Non-stationarity:  K-factor, RMS DS, RMS AS}\\ \hline

\cite{ai2017indoor}& 26 & 0.2 GHz& Indoor & \makecell[l]{ TX: $64$-/$64$-/$128$-element \\ \quad\quad vertical ULA/UPA/UPA \\RX: 4-/4-/1-element virtual \\
\quad\quad ULA/ULA/antenna}   &\makecell[c]{ Non-stationarity: path loss, PDP, RMS DS} \\ \hline

  \cite{cai2020trajectory}& 28 & 2 GHz& Indoor & \makecell[l]{ TX: 360-element virtual UCA \\RX: an omnidirectional antenna}   & \makecell[c]{ Non-stationarity: PDP} \\ \hline

  \cite{fan2016measured}& 29 & 2 GHz& Indoor & \makecell[l]{ TX:  a biconical antenna\\RX: 720-element virtual UCA}   & \makecell[c]{ NUSW: PAS\\Non-stationarity: path loss, PAS, PDP} \\ \hline

  \cite{yuan2022spatial}& 29.5 & 6 GHz& Indoor & \makecell[l]{ TX: 720-element virtual UCA \\RX: an omnidirectional biconical antenna}   & \makecell[c]{Non-stationarity: channel gain, PDP}\\ \hline

  \cite{zhao2017channel}& 32 & 1 GHz& Outdoor & \makecell[l]{ TX: an omnidirectional antenna \\RX: 250-element virtual UCA }   &\makecell[c]{Non-stationarity: path loss, K-factor,\\ RMS DS, RMS AS} \\ \hline

 \cite{wu201760}& 60 & 2 GHz& Indoor & \makecell[l]{ TX: (72$\times$25)-/(15$\times$15$\times$6)-element\\ \quad \quad virtual uniform array \\RX: 250-element virtual UCA }   & \makecell[c]{ NUSW: PAS\\Non-stationarity: PDP, PAS, RMS DS, RMS AS} \\ \hline

\end{tabular}
\end{table*}
The new characteristics of near-field NUSW, spatial non-stationarity and channel hardening have been validated via XL-MIMO channel measurements for sub-6 GHz \cite{willhammar2020channel,wang2017variation,feng2022mutual,wang2022characteristics,zheng2022ultra,zheng2023measurements,martinez2016massive,martinez2018experimental} and mmWave bands~\cite{huang2017multi,ai2017indoor,cai2020trajectory,fan2016measured,yuan2022spatial,zhao2017channel,wu201760}. The near-field channel measurement campaigns are summarized in Table \ref{table:XL-MIMOChannelMeasurement}.
 The typical carrier frequencies for channel measurements at sub-6 GHz include 1.4725, 2.6, 3.5, 5.3, and 5.8 GHz. The mmWave XL-MIMO channel measurements have been well studied at some typical frequency bands, e.g., 26, 28, 29, 29.5, 32, 38, and 60 GHz bands. Besides the carrier frequency, the bandwidth, type of antenna, array configuration, and measurement scenario are considered for XL-MIMO channel measurements. Most of XL-MIMO channel measurement campaigns adopt the virtual array architecture, i.e., the array virtually formed by sequentially re-positioning one single antenna in space, and a few measurement campaigns for sub-6 GHz adopt the practical array architecture~\cite{feng2022mutual,wang2022characteristics,zheng2022ultra,zheng2023measurements}. The ULA, UPA, and uniform cylindrical array (UCA) are the three most commonly used XL-array architectures in channel measurement campaigns. After setting up the configuration of the measurement system, the parameter estimation algorithms are utilized to extract the channel parameters from the calibrated measurement data. The investigated channel statistical properties consist of first- and second-order statistics, where the former include channel gain, path loss, shadow fading (SF), K-factor, power delay profile (PDP), PAS, power angle delay profile (PADP) and so on, and the latter include root mean square (RMS) delay spread (DS), RMS angle spread (AS), standard deviation of channel gain and so on.

 In \cite{li2015channel,payami2012channel,fan2016measured,zheng2022ultra,zheng2023measurements,huang2017multi,wu201760}, the AoA of the LoS path exhibits an angle offset across the antenna array, which indicates that the far-field UPW model is invalid for XL-MIMO communications. This feature is regarded as the near-field NUSW characteristic and can be observed through the PAS of the LoS path. Besides, the spatial non-stationarity characteristic can be observed through the significant variations of channel gain\cite{payami2012channel,martinez2016massive,martinez2018experimental,yuan2022spatial}, K-factor\cite{payami2012channel,chen2017measurement,zhao2017channel}, PDP\cite{zheng2022ultra,zheng2023measurements,huang2017multi,ai2017indoor,cai2020trajectory,fan2016measured,yuan2022spatial,wu201760}, PAS  \cite{payami2012channel,wang2017variation,fan2016measured,zheng2022ultra,zheng2023measurements,huang2017multi,wu201760}, RMS DS\cite{zheng2022ultra,zheng2023measurements,huang2017multi,chen2017measurement,ai2017indoor,zhao2017channel,wu201760}, and RMS AS\cite{wang2017variation,zhao2017channel,wu201760,zheng2022ultra,zheng2023measurements,huang2017multi,chen2017measurement} over the XL-array. Furthermore, the measured PDP and PAS over the array show the cluster birth-death property, where some clusters are visible to the whole array and others are only seen by the partial array.
% Besides, the RMS AS is used as the metric of the spatial non-stationarity characteristic \cite{feng2022mutual,wang2022characteristics,zheng2022ultra,zheng2023measurements,chen2017measurement,zhao2017channel,wu201760}.
 The channel hardening characteristic can be studied from the frequency and time domains, where the small channel gain standard deviation in frequency and time domains can be observed~\cite{willhammar2020channel,zheng2022ultra,zheng2023measurements,martinez2016massive,martinez2018experimental}. In particular, in contrast to the near-field measurement campaigns, the angle offset and variations of measured channel parameters utilized to validate spatial non-stationarity are not observable for the far-field measurement campaigns, and the far-field channels exhibit a larger channel gain standard deviation in frequency and time domains. It is also worth mentioning that many efforts have been devoted to considering the above characteristics for near-field channels. For example, according to the approved release-19 study item ``Study on Channel Modelling Enhancements for 7-24 GHz for NR'', 3GPP will validate the existing channel models with measurement data, at least for 7-24 GHz spectrum, and if necessary, the channel models will be adapted by taking into account the near-field propagation and spatial non-stationarity \cite{RP-234018,lin2023bridge}.

 \subsection{Lessons Learned}
 \subsubsection{Far-Field Versus Near-Field} The deployment of XL-MIMO and the continuously shrinking cell size lead to a paradigm shift from far-field communications to near-field communications. In the conventional far-field communications, the assumptions of UPW and spatial stationarity are typically used, under which the channel phases are modelled linearly and amplitudes are modelled uniformly across array elements. However, several new channel characteristics appear in near-field communications, such as NUSW and spatial non-stationarity, rendering the assumptions of linear phase and uniform amplitude no longer valid. Moreover, instead of only dependent on the signal direction as in the far-field channel, the near-field channel is dependent on both the signal direction and distance. As will become clearer later, these differences have profound impacts on XL-MIMO performance analysis and practical designs.

 \subsubsection{Near-Field Modelling}
 Accurate near-field modelling is the prerequisite of XL-MIMO communications. The basic array response vector modelling consists of near-field phase and amplitude modelling, which yield four array response vector models, i.e., UPW, NUPW, USW, and NUSW models, as summarized in Fig.~\ref{fig:arrayResponseVectorModelsFlowChart}. For near-field free-space XL-MIMO modelling, the direct method is to model the channel coefficient between each transmit-receive antenna pair individually, and the whole channel matrix can be obtained by stacking all the channel coefficients. Moreover, the near-field multi-path XL-MIMO channel can be obtained by separately modelling the LoS and NLoS channel components, where an effective modelling of the NLoS channel component is based on the bistatic radar equation. On the other hand, VR is typically used to characterize the spatial non-stationarity, and VR of XL-MIMO communications is classified into UE-side VR and BS-side VR. By considering the two VRs, the near-field channel can be obtained correspondingly.

 Spatial correlation matrix-based model is another widely used approach, where the differences brought by NUSW and VR should be considered to cater to the near-field channel characteristics. One important finding is that the near-field spatial correlation is determined by PLS rather than PAS as in the conventional far-field UPW model, and the SWSS is no longer valid. Moreover, the impacts of polarization mismatch and spatial-wideband effect are other factors to consider for accurate near-field modelling.

 %On the other hand, the near-field LoS XL-MIMO channel can be expressed as the outer product of the near-field transmit and receive array response vectors. The major difference between the two near-field LoS XL-MIMO modelling is the rank, where the former rank can be greater than one, while the latter rank is always equal to one.

% >>>>>>>>>>>>>SECTIONS III -  here >>>>>>>>>>>>
\section{Performance Analysis of XL-MIMO}\label{sectionPerformanceAnalysis}
 In this section, we focus on the performance analysis of near-field communication with XL-MIMO, including SNR scaling laws, beam focusing pattern, achievable rate, DoF, and near-field sensing.

\subsection{SNR Scaling Laws}
\begin{figure}[!t]
 \centering
 \centerline{\includegraphics[width=3.0in,height=2.45in]{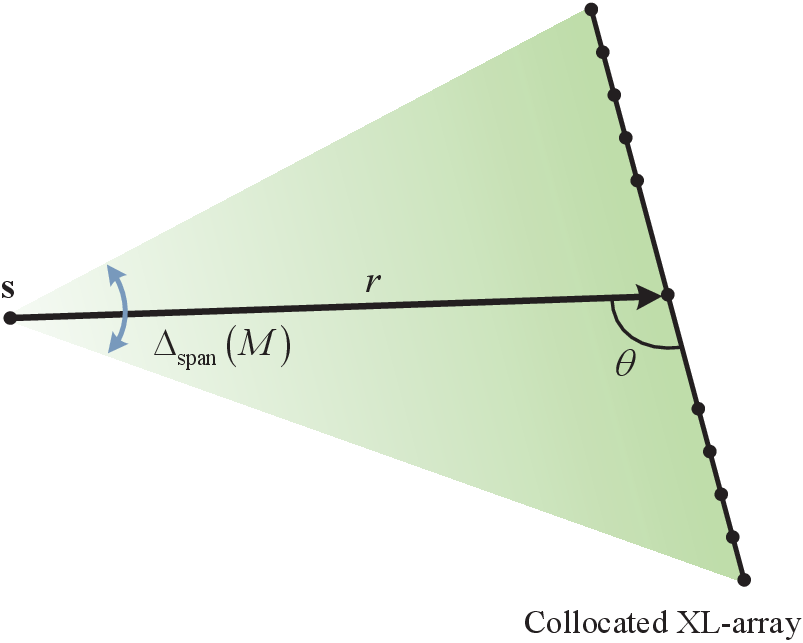}}
 \caption{An illustration of angular span in near-field communication with XL-array.}
 \label{fig:illustrationAngularSpan}
 \end{figure}

 We first discuss the SNR scaling law for single-user communication with collocated XL-ULA with isotropic elements. Specifically, we consider a free-space SIMO communication system shown in Fig.~\ref{fig:transmitReceiverArrayULA}, where the UE is located at ${\bf s}$. By substituting ${U_m} = U =  {\left( {\lambda /4\pi } \right)^2}$ into \eqref{arrayResponseVectorNUSWModel}, and applying the optimal maximal-ratio combining (MRC) beamforming, the resulting SNR can be expressed in closed-form as \cite{lu2021how}
 \begin{equation}\label{NUSWSNRSIMOCollocatedClosedForm}
 \begin{aligned}
 {\gamma ^{{\rm{NUSW}}}} = \bar P{\left\|\alpha {{{\bf{a}}^{{\rm{NUSW}}}}\left( {r,\theta } \right)} \right\|^2}
 = \frac{\bar P{{\lambda ^2}}}{{{{\left( {4\pi } \right)}^2}dr\sin \theta }}{\Delta _{{\rm{span}}}}\left( M \right),
 \end{aligned}
 \end{equation}
 where $\alpha$ follows the same definition below \eqref{generalComplexChannelCoefficient}, ${\bar P} \triangleq {P/{\sigma ^2}}$, with $P$ and ${\sigma^2}$ denoting the transmit and the noise power, respectively, and ${\Delta _{{\rm{span}}}}\left( M \right) \triangleq \arctan \left( {\frac{{Md}}{{2r\sin \theta }} + \cot \theta } \right) + \arctan \left( {\frac{{Md}}{{2r\sin \theta }} - \cot \theta } \right)$. The above result shows that with the near-field NUSW model, the resulting SNR scales with antenna number $M$ nonlinearly according to the parameter $\Delta _{\rm{span}} \left(M\right)$, termed {\it angular span} \cite{lu2021how}, rather than growing linearly with $M$ as in the far-field UPW model. Besides, a closer look at Fig.~\ref{fig:illustrationAngularSpan} shows that the angular span is the angle formed by the two line segments connecting the source with both ends of the antenna array. Therefore, for infinitely large-scale array such that $M \to \infty$, we have ${\Delta _{{\rm{span}}}}\left( M \right) \to \pi $. Then the resulting SNR in \eqref{NUSWSNRSIMOCollocatedClosedForm} reduces to
 \begin{equation}\label{NUSWAsymptoticSNRSIMO}
 \mathop {\lim }\limits_{M \to \infty } {\gamma ^{{\rm{NUSW}}}} = \frac{\bar P{{\lambda ^2}\pi }}{{{{\left( {4\pi } \right)}^2}dr\sin \theta }},
 \end{equation}
 which is a constant depending on the user's projected distance to the collocated XL-array ${r\sin \theta }$.

 As a comparison, when the far-field UPW array response vector  \eqref{arrayResponseVectorUPWModel} is used, the resulting SNR is given by
 \begin{equation}\label{UPWSNRSIMO}
 {\gamma ^{{\rm{UPW}}}} = \bar P{\left\|\alpha {{{\bf{a}}^{{\rm{UPW}}}}\left( \theta  \right)} \right\|^2} = \frac{\bar P{M{\lambda ^2}}}{{{{\left( {4\pi r} \right)}^2}}},
 \end{equation}
 which increases linearly with the antenna number $M$. As $M \to \infty$, the resulting SNR with the UPW model will go to infinity, which fails to comply with the law of power conservation. Note that when $r \gg Md/2$, it is verified that ${\gamma ^{{\rm{NUSW}}}} \approx {\gamma ^{{\rm{UPW}}}} = \bar PM{\lambda ^2}/{\left( {4\pi r} \right)^2}$ \cite{lu2021how}, which implies that the NUSW model generalizes the far-field UPW model.

\begin{figure}[!t]
 \centering
 \centerline{\includegraphics[width=3.5in,height=2.625in]{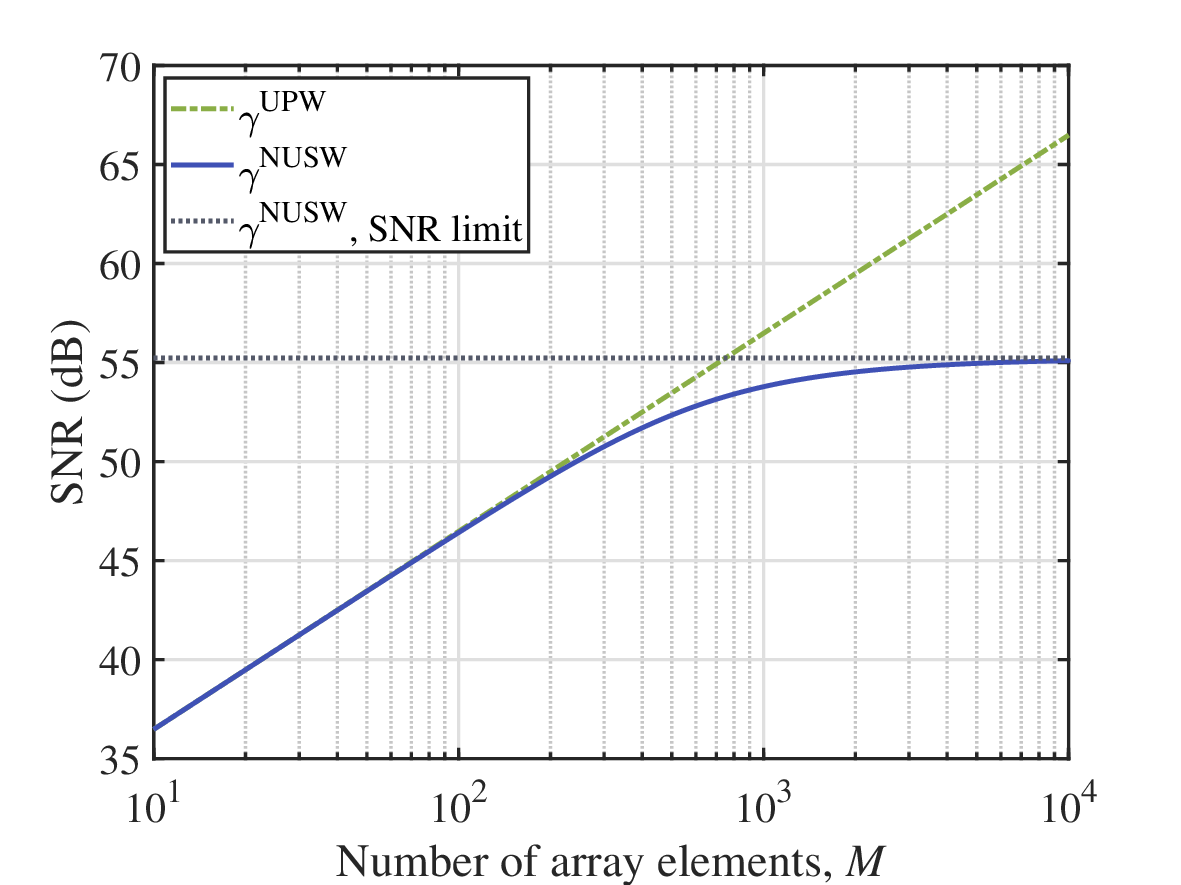}}
 \caption{SNR versus the number of antenna elements $M$ with NUSW and UPW models \cite{lu2021how}.}
 \label{fig:IntegralSNRApproximationLog}
 \end{figure}

 In Fig.~\ref{fig:IntegralSNRApproximationLog}, we compare the resulting SNRs for the near-field NUSW and the far-field UPW models. The locations of the source and the reference point of the array are ${\bf{s}} = {\left[ {0,0} \right]^T}$ m and ${\bf{p}} = {\left[ {15,0} \right]^T}$ m, respectively. The direction vector of the array is ${\bf{\hat u}} = {\left[ {0,1} \right]^T}$. The carrier frequency is 2.4 GHz, and the antenna separation is $d = {\lambda}/2 = 0.0628$ m. Besides, the transmit SNR is ${\bar P} = 90$ dB. It is observed that for relatively small antenna number $M$, ${\gamma ^{\rm{NUSW}}}$ matches well with ${\gamma ^{\rm{UPW}}}$. This is expected since the far-field UPW model gives a valid approximation. However, as $M$ further increases, quite different scaling laws are observed for the two SNR expressions, i.e., approaching a constant value versus increasing linearly and unbounded. The above result demonstrates that the proper spherical wavefront modelling is essential for the XL-array.

 Furthermore, by considering the variation of projected aperture across array elements, a closed-form SNR expression was derived for the general UPA \cite{lu2022communicating}, which is applicable for both the conventional discrete antenna array and the emerging continuous surfaces. As the antenna number $M \to \infty $, the received power of the continuous surface approaches to $P/2$. Such a result makes an intuitive sense since for the isotropic source, only half of the transmitted power will be captured by the infinitely large continuous surface, while the other half of the power will never reach the surface \cite{lu2022communicating,hu2018beyond}. By further taking into account the loss of the polarization mismatch for the continuous-aperture and discrete-aperture XL-array, the SNR expressions were derived in \cite{bjornson2020power} and \cite{zhi2023performance}, respectively. It was shown that in the presence of polarization mismatch, as $M \to \infty $, only $P/3$ can be captured by the continuous surface. Moreover, the power scaling law and the asymptotic analysis for the continuous-aperture LIS can be found in \cite{dardari2020communicating}.

 While the abovementioned works mainly focus on the standard collocated XL-array, preliminary efforts have been devoted to the new modular XL-array architecture in \cite{li2022near,li2022modular}. In \cite{li2022near}, the closed-form SNR with the NUSW model was derived for the modular XL-ULA, which depends on its geometric characteristics, such as the physical dimension  and the inter-module separation. When the inter-module separation is equal to the antenna spacing $d$, it was mathematically shown that the SNR of the modular XL-ULA degenerates to that of the collocated XL-ULA. Moreover, by properly modelling the variation of projected aperture across all modular elements, the SNR scaling law for the more general modular XL-UPA was analyzed in \cite{li2022modular}, for which the similar observation as the modular XL-ULA is obtained.

\subsection{Near-Field Beam Focusing Pattern}\label{nearFieldBeamFocusingPattern}
\begin{figure}[!t]
 \centering
 \centerline{\includegraphics[width=3.1in,height=2.3in]{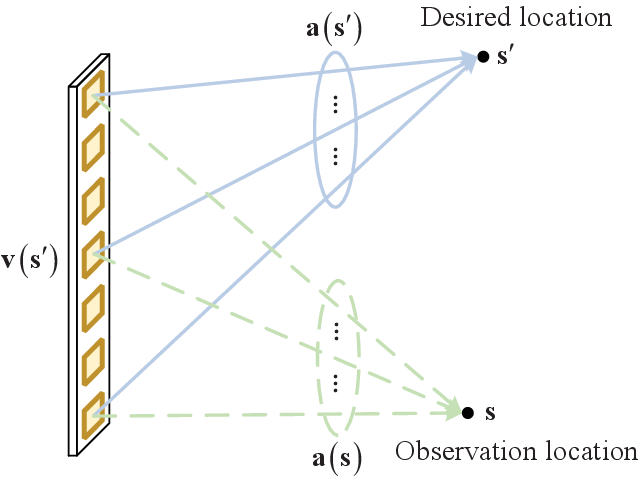}}
 \caption{Illustration of near-field beam focusing pattern.}
 \label{fig:beamFocusingPattern}
 \end{figure}

 In near-field XL-MIMO communications, the order of magnitude increase in antenna number brings enhanced spatial resolution beyond current massive MIMO systems \cite{bjornson2019massive,lu2021how}. For multi-user communications, one important aspect is the evolution from the far-field beam pattern to the {\it near-field beam focusing pattern}. Specifically, the beam pattern under the far-field UPW model describes the intensity distribution of a designed beam intended for a certain direction as a function of the observation direction. By contrast, the near-field beam focusing pattern is capable of characterizing the intensity distribution as a function of the observation location \cite{li2023multi,zhang2022beam,zhang20236g}. As illustrated in Fig.~\ref{fig:beamFocusingPattern}, let ${\bf{v}}\left( {\bf{s'}} \right)$ denote the beamforming vector designed for the desired location ${\bf {s'}}$, and ${\bf s}$ denote the actual observation location. The beam focusing pattern can be defined as \cite{li2023multi}
 \begin{equation}\label{beamFocusingPatternDefinition}
 G\left( {{\bf{s}};{\bf{s'}}} \right) \triangleq \frac{{\left| {{{\bf{v}}^H}\left( {{\bf{s'}}} \right){\bf{a}}\left( {\bf{s}} \right)} \right|}}{{\left\| {{\bf{v}}\left( {{\bf{s'}}} \right)} \right\|\left\| {{\bf{a}}\left( {\bf{s}} \right)} \right\|}},
 \end{equation}
 where ${\bf{a}}\left( {\bf{s}} \right)$ denotes the array response vector of the observation location ${\bf s}$. The observation location ${\bf s}$ can be located in either the near-field or the far-field regions. Besides, the choice of the beamforming vector ${\bf{v}}\left( {\bf{s'}} \right)$ is closely dependent on the available CSI and/or the pre-determined beam codebook, which includes the far- and near-field beamforming designs. In the following, similar to \cite{li2023multi}, depending on the observation location and the used beamforming vector, three beam focusing patterns are discussed.{\footnote[1]{In fact, we have another case for ``far-field observation with near-field beamforming'', which is not discussed here for brevity.}}

 \subsubsection{Far-Field Observation With Far-Field Beamforming}\label{FFObservationFFBeamforming}
 When the observation location $\bf{s}$ is in the far-field region, and the far-field UPW-based beamforming ${{\bf{v}}_{{\rm{FF}}}} = {{\bf{a}}^{{\rm{UPW}}}}\left( {\bf{s'}} \right)$ in \eqref{generalChannelSIMOUPW} is used, the beam focusing pattern in \eqref{beamFocusingPatternDefinition} reduces to the conventional far-field beam pattern. For ease of comparison for different array architectures, we consider an antenna array with a total of $NM$ elements, and the far-field beam pattern is
 \begin{equation}\label{beamPatternFarField}
 {G_{{\rm{FF}},{\rm{FF}}}}\left( {{\bf{s}};{\bf{s'}}} \right) \triangleq \frac{1}{NM}\left| {{{\left( {{{\bf{a}}^{{\rm{UPW}}}}\left( {{\bf{s}}'} \right)} \right)}^H}{{\bf{a}}^{{\rm{UPW}}}}\left( {\bf{s}} \right)} \right|,
 \end{equation}
 where the first subscript FF represents the far-field observation, i.e., the far-field array response vector ${{{\bf{a}}^{{\rm{UPW}}}}\left( {\bf{s}} \right)}$ is used for the far-field observation point $\bf{s}$, and the second subscript FF represents the far-field beamforming for ${\bf v}$. For collocated ULA where adjacent elements are separated by $d = {\lambda}/2$, by substituting \eqref{arrayResponseVectorUPWModel} into \eqref{beamPatternFarField}, we have
 \begin{equation}\label{beamPatternFarFieldCollocatedULA}
 G_{{\rm{FF}},{\rm{FF}}}^{{\rm{co}}}\left( {\theta ;\theta '} \right) = \underbrace {\left| {\frac{{\sin \left( {\frac{\pi }{2}NM{\Delta _\theta }} \right)}}{{NM\sin \left( {\frac{\pi }{2}{\Delta _\theta }} \right)}}} \right|}_{\left| {{\Xi _{NM, \frac{1}{2}}}\left( {{\Delta _\theta }} \right)} \right|},
 \end{equation}
 where the superscript ``co'' refers to collocated ULA, ${\bar d} \triangleq d/{\lambda} = 1/2$, $\Delta_{\theta}  \triangleq \cos \theta'  - \cos \theta$, and ${\Xi _{\tilde M,\tilde d}}\left( {{\Delta _\theta }} \right) \triangleq \sin ( {\pi \tilde M\tilde d{\Delta _\theta }})/(\tilde M \sin ( {\pi \tilde d{\Delta _\theta }}))$ is the Dirichlet kernel function \cite{li2023multi}. On the other hand, for modular ULA with $N$ modules and each module consisting of $M$ elements, the far-field beam pattern is \cite{li2023multi},
 \begin{equation}\label{beamPatternFarFieldModularULA}
 G_{{\rm{FF}},{\rm{FF}}}^{{\rm{mo}}}\left( {\theta ;\theta '} \right) = \underbrace {\left| {\frac{{\sin \left( {\frac{\pi }{2}N\Gamma {\Delta _\theta }} \right)}}{{N\sin \left( {\frac{\pi }{2}\Gamma {\Delta _\theta }} \right)}}} \right|}_{\left| {{\Xi _{N,\frac{\Gamma }{2}}}\left( {{\Delta _\theta }} \right)} \right|}\underbrace {\left| {\frac{{\sin \left( {\frac{\pi }{2}M{\Delta _\theta }} \right)}}{{M\sin \left( {\frac{\pi }{2}{\Delta _\theta }} \right)}}} \right|}_{\left| {{\Xi _{M,\frac{1}{2}}}\left( {{\Delta _\theta }} \right)} \right|},
 \end{equation}
 where the superscript ``mo'' refers to the modular ULA, and $\Gamma \ge M$ is the module separation parameter, with $\Gamma d$ being the inter-module separation between the reference points of adjacent modules, as illustrated in Fig.~\ref{fig:modularArrayULA}. Moreover, for sparse ULA where adjacent elements are separated by $I\lambda /2$, with $I >1$, the far-field beam pattern is
 \begin{equation}\label{beamPatternFarFieldSparseULA}
 G_{{\rm{FF}},{\rm{FF}}}^{{\rm{sp}}}\left( {\theta ;\theta '} \right) = \underbrace {\left| {\frac{{\sin \left( {\frac{\pi }{2}NMI{\Delta _\theta }} \right)}}{{NM\sin \left( {\frac{\pi }{2}I{\Delta _\theta }} \right)}}} \right|}_{{\Xi _{NM,\frac{I}{2}}}\left( {{\Delta _\theta }} \right)},
 \end{equation}
 where the superscript ``sp" refers to sparse ULA. It is observed that the far-field beam patterns of the three array architectures are only determined by the difference of two spatial frequencies, i.e., ${{\Delta _\theta }}$, while irrespective of the link distance. Besides, the null-to-null beam width of ${\Xi _{\tilde M,\tilde d}}\left( {{\Delta _\theta }} \right)$ can be obtained by letting $\pi \tilde M\tilde d{\Delta _\theta } =  \pm \pi$, given by $2/\tilde M\tilde d$. By defining half of the null-to-null beam width as the {\it angular resolution}, for the three architectures, we have
 \begin{equation}\label{angularResolutionFarField}
 \left\{ \begin{split}
 &\omega _\theta ^{{\rm{co}}} = \frac{2}{NM},\\
 &\omega _\theta ^{{\rm{mo}}} = \frac{2}{{N\Gamma}},\\
 &\omega _\theta ^{{\rm{sp}}} = \frac{2}{{NMI}},
 \end{split} \right.
 \end{equation}
 i.e., the increase in antenna number $NM$ helps improve the angular resolution. In particular, by letting $\Gamma = M$, modular array reduces to collocated array. Therefore, with the same number of array elements, modular array provides a higher angular resolution than the conventional collocated array since $\Gamma > M$. However, it is worth mentioning that the undesired grating lobes will appear in the beam pattern when ${\tilde d} > 1/2$ \cite{balanis2016antenna}. Since ${\Gamma}/2 > 1/2$ in \eqref{beamPatternFarFieldModularULA}, the improvement of angular resolution of the modular array is in fact at the cost of grating lobes, with the adjacent grating lobes separated by $2/\Gamma $. Fortunately, the grating lobes are suppressed to certain extent by the envelope of the other term ${{\Xi _{M,\frac{1}{2}}}\left( {{\Delta _\theta }} \right)}$. On the other hand, since $I > 1$, grating lobes also exist for sparse array, with the adjacent grating lobes separated by $2/I$ \cite{wang2023can}. Besides, the sparse array also includes the collocated array as a special case when $I = 1$.

\begin{figure}[t]
 \centering
 \centerline{\includegraphics[width=3.5in,height=2.625in]{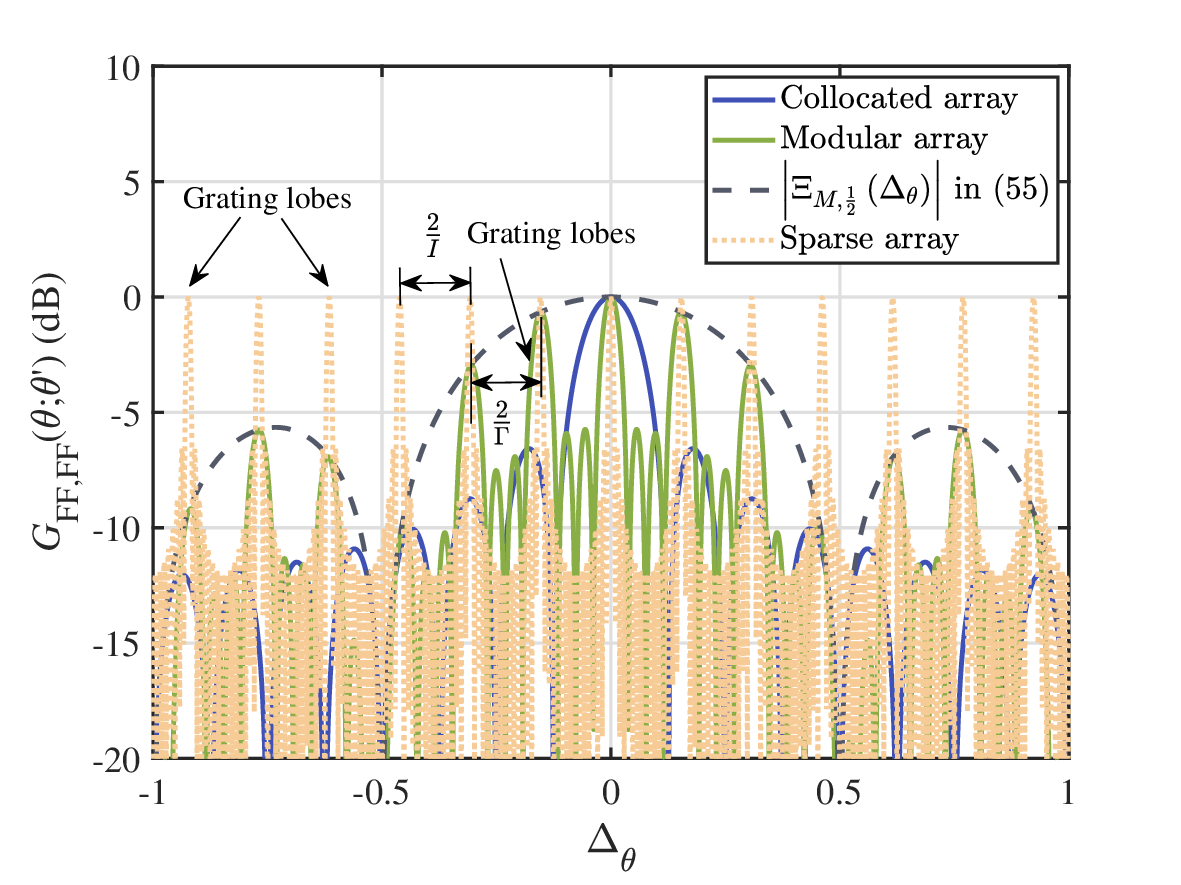}}
 \caption{Comparisons of far-field beam pattern for collocated, modular and sparse array architectures.}
 \label{fig:farFieldBeamPattern}
 \end{figure}

 Fig.~\ref{fig:farFieldBeamPattern} shows the comparisons of far-field beam pattern for collocated, modular and sparse array architectures. The total number of array elements is $NM = 16$, with $N=4$ and $M=4$, respectively. The module separation parameter for modular array is $\Gamma = 13$, and the antenna separation parameter for sparse array is $I = 13$. For the considered setup, it is observed that the angular resolution of modular array is superior to collocated array, but inferior to sparse array, as can be inferred from \eqref{angularResolutionFarField}. It is also observed that the undesired grating lobes exist in both modular and sparse arrays. Fortunately, the grating lobes of modular array is suppressed to certain extent by the envelope of the other term ${{\Xi _{M,\frac{1}{2}}}\left( {{\Delta _\theta }} \right)}$ in \eqref{beamPatternFarFieldModularULA}, while for sparse array, the amplitude and bandwidth of grating lobes are equal to those of the main lobe.

 \subsubsection{Near-Field Observation With Far-Field Beamforming}
 When the observation location $\bf s$ lies in the near-field region, while the far-field beamforming is used for the desired locations $\bf s'$, the near-field beam focusing pattern is
 \begin{equation}\label{beamFocusingPatternFarField}
 {G_{{\rm{NF,FF}}}}\left( {{\bf{s}}; {\bf{s'}}} \right) \buildrel \Delta \over = \frac{{\left| {{{\left( {{{\bf{a}}^{{\rm{UPW}}}}\left( {{\bf{s}}'} \right)} \right)}^H}{\bf{a}}\left( {\bf{s}} \right)} \right|}}{{\left\| {{{\bf{a}}^{{\rm{UPW}}}}\left( {{\bf{s}}'} \right)} \right\|\left\| {{\bf{a}}\left( {\bf{s}} \right)} \right\|}}.
 \end{equation}
 For collocated ULA, by using the USW-based array response vector in \eqref{arrayResponseVectorUSWModel} for the observation location, we have
 \begin{equation}\label{beamFocusingPatternFarFieldCollocatedULA}
 G_{{\rm{NF}},{\rm{FF}}}^{{\rm{co}}}\left( {r,\theta ;\theta '} \right) = \frac{1}{{NM}}\left| {\sum\limits_{m = 1}^{NM} {{e^{ - j\frac{{2\pi }}{\lambda }\left( {{r_m} - md\cos \theta '} \right)}}} } \right|.
 \end{equation}
 On the other hand, by using the subarray based USW model with different angles for the observation point, the near-field beam focusing pattern of the modular ULA is \cite{li2023multi}
 \begin{equation}\label{beamFocusingPatternFarFieldModularULADiff}
 \begin{aligned}
 G_{{\rm{NF,FF}}}^{{\rm{mo}}}\left( {r,\theta ;\theta '} \right) = \frac{1}{N}\left| {\sum\limits_{n = 1}^N {{e^{ - j\frac{{2\pi }}{\lambda }\left( {{r_n} - \left( {n - 1} \right)\Gamma d\cos \theta '} \right)}}}  \times } \right.\\
 \Bigg. {{\Xi _{M, \frac{1}{2}}}\left( {\cos \theta ' - \cos {\theta _n}} \right)} \Bigg|,
 \end{aligned}
 \end{equation}
 where $r_n$ denotes the distance between ${\bf s}$ and ${\bf p}_n$.

 It is observed that different from the far-field beam pattern discussed in Section \ref{FFObservationFFBeamforming}, the near-field observation pattern under the far-field beamforming design depends on the specific observation location $\left({r, \theta}\right)$ and the intended beamforming direction $\theta'$. Besides, when the observation location $\bf s$ coincides with the desired location $\bf s'$, \eqref{beamFocusingPatternFarField} characterizes the beamforming gain loss due to the mismatch between the near-field channel and the far-field beamforming. In \cite{cui2021near}, the distance metric termed {\it effective Rayleigh distance} was introduced, which is defined as the minimum link distance so that the normalized beamforming gain is no smaller than a certain threshold. By setting the threshold as $0.95$, the effective Rayleigh distance is given by \cite{cui2021near}
 \begin{equation}\label{effectiveRayleighDistance}
 {r_{{\rm{effRayl}}}}\left( \theta  \right) = \left( {0.367{{\sin }^2}\theta } \right)\frac{{2{D^2}}}{\lambda }.
 \end{equation}

 \begin{figure}[t]
 \centering
 \centerline{\includegraphics[width=3.5in,height=2.625in]{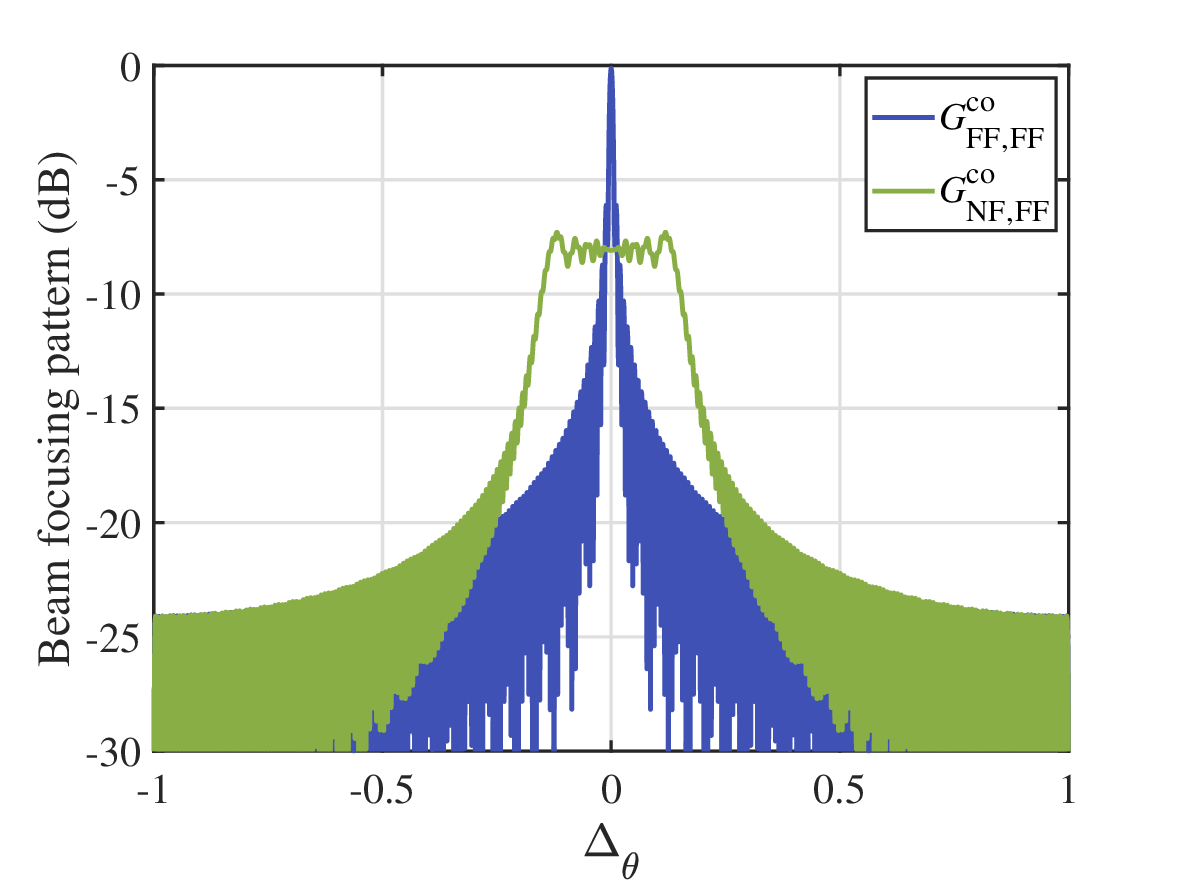}}
 \caption{{Illustration of energy spread effect.}}
 \label{fig:energySpreadEffect}
 \end{figure}
 In the following, the interference created by a far-field beamforming to the near-field observation location is studied, where the {\it energy spread effect} will appear in the beam focusing pattern \cite{cui2022channel,zhang2023mixed,li2023multi}. As an example, Fig. \ref{fig:energySpreadEffect} shows the beam pattern of collocated ULA versus the spatial frequency difference ${\Delta _\theta }$, by fixing the desired location $\left( {r',\theta '} \right) = \left( {2000\ {\mathrm m},\pi /2} \right)$. The near-field and far-field observation distances are $r = 50$ m and $r = 2000$ m, respectively. The number of antennas is 256. It is observed that when the observation location lies in the far-field region, corresponding to $r = 2000$ m, we have the far-field beam pattern, and the energy focuses on the desired beamforming direction. However, for the near-field observation location, an expanded beam width is observed for the near-field beam focusing pattern, i.e., the energy will spread towards the neighboring directions.

 The beam focusing pattern reflects the correlation between the near- and far-field channels. As a result, the energy spread effect will cause another key fact from an interference perspective, i.e., inter-user interference (IUI) between near-field user and far-field user \cite{zhang2023mixed}. Specifically, when the near-field user locates in the neighbor of the far-field user, a more complicated IUI issue arises as compared to the conventional far-field communications. On the other hand, the interesting inter-user channel correlation in the mixed-field communications also brings new design opportunities. For example, from the WPT perspective, the power leakage caused by discrete Fourier transform (DFT)-based codebook for serving far-field users can be exploited to charge the near-field energy-harvesting (EH) user. In \cite{zhang2023joint,zhang2023swipt}, the joint beam scheduling and power allocation was investigated for simultaneous wireless information and power transfer (SWIPT) in mixed-field channels to maximize the harvested sum-power at EH users under a minimum sum-rate constraint for far-field information-decoding users. An interesting result was revealed that for mixed-field SWIPT, the energy-harvesting user located in the near-field should always be scheduled to maximize the harvested sum-power, which is in sharp contrast with the convention far-field SWIPT, for which only information-decoding users are scheduled \cite{xu2014multiuser}.

 \subsubsection{Near-Field Observation With Near-Field Beamforming}
 When the observation location ${\bf s}$ lies in the near-field region, and the near-field beamforming is designed for the desired location $\bf s'$, i.e., ${{\bf{v}}_{{\rm{NF}}}} = {\bf{a}}\left( {{\bf{s}}'} \right)$, the beam focusing pattern in \eqref{beamFocusingPatternDefinition} is given by
 \begin{equation}\label{beamFocusingPatternNearField}
 {G_{{\rm{NF}},{\rm{NF}}}}\left( {{\bf{s}};{{\bf{s'}}  } } \right) \triangleq \frac{{\left| {{{\bf{a}}^H}\left( {{\bf{s}}'} \right){\bf{a}}\left( {\bf{s}} \right)} \right|}}{{\left\| {{\bf{a}}\left( {{\bf{s}}'} \right)} \right\|\left\| {{\bf{a}}\left( {\bf{s}} \right)} \right\|}}.
 \end{equation}
 For collocated XL-array and USW-based array response vector in \eqref{arrayResponseVectorUSWModel}, we have
 \begin{equation}\label{beamFocusingPatternNearField}
 {G_{{\rm{NF}},{\rm{NF}}}^{\rm co}}\left( {r,\theta ;r',\theta '} \right) = \frac{1}{NM}\left| {\sum\limits_{m = 1}^{NM} {{e^{j\frac{{2\pi }}{\lambda }\left( {{r'_m} - {r_m}} \right)}}} } \right|.
 \end{equation}
 For moderately large array, $r_m^{{\rm{second}}}$ is an effective approximation for phase modelling, and a closed-form near-field beam focusing pattern can be derived based on  \cite{li2023multi}. On the other hand, under the subarray based USW model with common angle, the closed-form near-field beam focusing pattern of modular XL-array was derived in \cite{li2023multi}. In particular, one important difference is that the near-field beam focusing pattern provides the spatial resolution over both the angular and distance domains. In order to quantify the angular and distance resolution, the {\it effective angular resolution} is defined as half of the approximated null-to-null beam width in the angular domain, and the {\it effective distance resolution} is half of the 3 dB beam width in the distance domain.
 \begin{itemize}[\IEEEsetlabelwidth{12)}]
 \item \textbf{Angular resolution:} The near-field beam focusing pattern gives comparable angular resolution as the far-field beam pattern. The effective angular resolution is approximated given in  \eqref{angularResolutionFarField}.
 \item \textbf{Distance resolution:}
     Let ${{r_{{\rm{hp}}}}\left( {\theta '} \right)}= 0.1{\sin ^2}\theta '\frac{{2{D^2}}}{\lambda }$ denote the {\it half power effective distance} \cite{li2023multi}. The effective distance resolution of the near-field beam focusing pattern is
     \begin{equation}\label{effectiveDistanceResolution}
     {\omega _{1/r}}\left( {\theta '} \right) = \frac{1}{{{r_{{\rm{hp}}}}\left( {\theta '} \right)}},
     \end{equation}
     and the two locations along the same direction can be separated when $\left| {1/r' - 1/r} \right| \ge 1/{r_{{\rm{hp}}}}\left( {\theta '} \right)$.
     Such a result is applicable to the three architectures, by using the corresponding physical dimension $D$. Compared to the effective angular resolution that is inversely proportional to array physical dimension, the effective distance resolution is inversely proportional to the square of array physical dimension.
 \end{itemize}

 The spatial resolution and grating lobes for the three array architectures are summarized in Table \ref{table:SpatialResolution}.

\begin{table*}[!t]
	\centering
	\caption{Spatial Resolution and Grating lobes for Different Array Architectures}\label{table:SpatialResolution}
	\resizebox{\textwidth}{!}{
		\begin{tabular}{|m{3.8cm}<{\centering}|m{2.4cm}<{\centering}|m{2.4cm}<{\centering}|l|m{3cm}<{\centering}|l }
			\hline
			{\bf Array Architectures} &
			{\bf Far-Field Angular Resolution } &
			{\bf Near-Field Angular Resolution} &
            \qquad\qquad{\bf Near-Field Distance Resolution} &
            {\bf Separation of Adjacent Grating Lobes}
%			{\begin{tabular}[c]{@{}c@{}} {\bf{Separation of}}\\  {\bf{Adjacent Grating Lobes}}\end{tabular}}
 \\ \hline\hline
			{{\begin{tabular}[l]{@{}l@{}}{\bf Collocated ULA}\\ $\bullet$ $NM$: number of elements \qquad\qquad \end{tabular}}}                          & $\omega _\theta ^{{\rm{co}}} = \frac{2}{NM}$                             &  $\omega _\theta ^{{\rm{co}}} = \frac{2}{NM}$    & {\begin{tabular}[c]{@{}l@{}} $\omega _{1/r}^{{\rm{co}}}\left( {\theta '} \right) = \frac{1}{{r_{{\rm{hp}}}^{{\rm{co}}}\left( {\theta '} \right)}}$,\\ $r_{{\rm{hp}}}^{{\rm{co}}}\left( {\theta '} \right) = 0.1{\sin ^2}\theta '\frac{{2D_{{\rm{co}}}^2}}{\lambda }$, ${D_{{\rm{co}}}} = \frac{{\left( {NM - 1} \right)\lambda }}{2}$\end{tabular}} & No grating lobes
\\ \hline
{{\begin{tabular}[l]{@{}l@{}}{\bf Modular ULA}\\ $\bullet$ $N$: number of modules\\ $\bullet$ $M$: number of elements\\ within each module \\  $\bullet$ $\Gamma$: module separation parameter  \end{tabular}}}                          & $\omega _\theta ^{{\rm{mo}}} = \frac{2}{{N\Gamma}}$                             & $\omega _\theta ^{{\rm{mo}}} = \frac{2}{{N\Gamma}}$ & {\begin{tabular}[c]{@{}l@{}}$\omega _{1/r}^{{\rm{co}}}\left( {\theta '} \right) = \frac{1}{{r_{{\rm{hp}}}^{{\rm{mo}}}\left( {\theta '} \right)}}$,\\ $r_{{\rm{hp}}}^{{\rm{mo}}}\left( {\theta '} \right) = 0.1{\sin ^2}\theta '\frac{{2D_{{\rm{mo}}}^2}}{\lambda }$, ${D_{{\rm{mo}}}} = \frac{{\left[ {\left( {N - 1} \right)\Gamma  + M - 1} \right]\lambda }}{2}$\end{tabular}} &$\frac{2}{\Gamma}$
\\ \hline
{{\begin{tabular}[l]{@{}l@{}}{\bf Sparse ULA}\\ $\bullet$ $NM$: number of elements \\  $\bullet$ $I$: antenna separation parameter  \end{tabular}}}                         & $\omega _\theta ^{{\rm{sp}}} = \frac{2}{{NMI}}$                             & $\omega _\theta ^{{\rm{sp}}} = \frac{2}{{NMI}}$    &{\begin{tabular}[l]{@{}l@{}}$\omega _{1/r}^{{\rm{co}}}\left( {\theta '} \right) = \frac{1}{{r_{{\rm{hp}}}^{{\rm{sp}}}\left( {\theta '} \right)}}$,\\ $r_{{\rm{hp}}}^{{\rm{sp}}}\left( {\theta '} \right) = 0.1{\sin ^2}\theta '\frac{{2D_{{\rm{sp}}}^2}}{\lambda }$, ${D_{{\rm{sp}}}} = \frac{{\left( {NM - 1} \right)I\lambda }}{2}$ \end{tabular}} &$\frac{2}{I}$
\\ \hline
		\end{tabular}
	}
\end{table*}

 \begin{figure}[!t]
 \centering
 \centerline{\includegraphics[width=3.5in,height=2.625in]{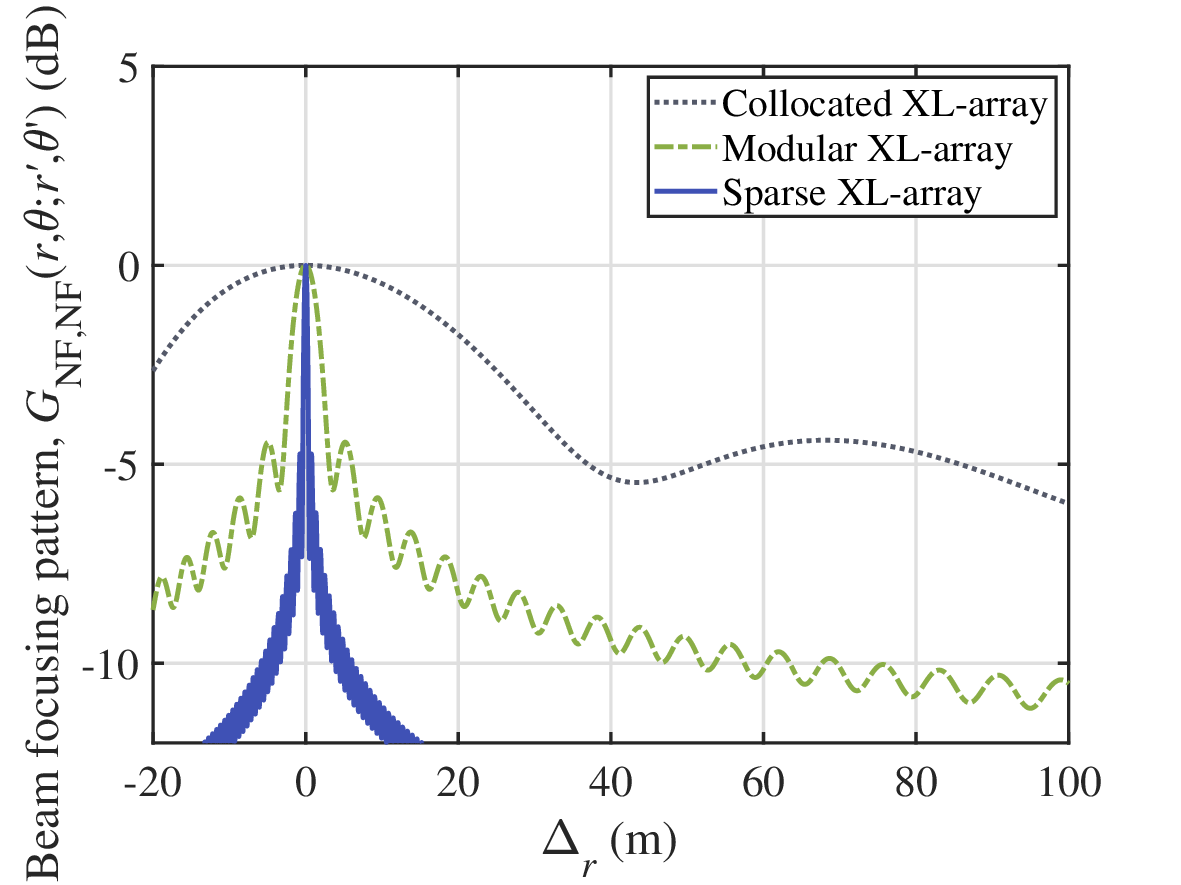}}
 \caption{Comparisons of beam focusing patterns for collocated, modular and sparse XL-array architectures.}
 \label{fig:sparseModularCollocatedArraySpatialResolution}
 \end{figure}

 Fig.~\ref{fig:sparseModularCollocatedArraySpatialResolution} compares the beam focusing pattern versus the distance separation ${\Delta _r} = r-r'$ for collocated, modular and sparse XL-array architectures, by fixing $r'= 200$ m and $\theta  = \theta' = \pi /2$. The total number of array elements is $NM = 512$, with $N=128$ and $M=4$, respectively, and the USW-based array response vector is used. The module separation parameter for modular XL-array is $\Gamma = 13$, and the antenna separation parameter for sparse XL-array is $I = 13$. It is observed that the near-field beam focusing patterns of all the three array architectures exhibit a general trend of decreasing, and the sparse XL-array possesses the highest distance resolution due to the narrowest beam width in the distance domain. This is expected since under the setup of $\Gamma = I$, sparse XL-array has the largest physical dimension. Besides, the distance resolution of modular XL-array is superior to that of collocated XL-array. Thus, XL-array provides not only the angular resolution as in the conventional MIMO and massive MIMO systems, but also the new distance resolution. The similar observations can be found in \cite{lu2021how,lu2021near,cui2022near}. This thus enables the possibility of XL-array to extract the propagation distance of EM waves in space, i.e., {\it spatial depth} \cite{bjornson2019massive}.

\subsection{Achievable Rate of Near-Field Communication}
 The enhanced spatial resolution of XL-MIMO brings a new DoF for IUI suppression, i.e., the IUI can be suppressed not only by the angular separation as in the conventional far-field UPW model, but also by the distance separation for users along the same direction \cite{lu2021how,lu2021near,zhang2022beam}. For example, in \cite{lu2021near}, the signal-to-interference-plus-noise ratio (SINR) performance of three typical beamforming schemes, i.e., MRC, zero-forcing (ZF), and minimum mean-square error (MMSE) beamforming, were evaluated for multi-user near-field communications. Specifically, consider a multi-user uplink communication system, where the XL-array equipped with $M$ elements serves $K$ single-antenna users. Let ${{\bf{h}}_k} \in {{\mathbb C}^{M \times 1}}$ denote the multi-path channel of user $k$, which can be obtained based on \eqref{generalMultiPathChannelMISOSIMO}, and ${P_k}$ denote the transmit power of user $k$. By applying the receive beamforming ${{\bf{v}}_k} \in {{\mathbb C}^{M \times 1}}$ to user $k$, with $\left\| {{{\bf{v}}_k}} \right\| = 1$, the resulting SINR of user $k$ is
 \begin{equation}\label{SINRMultiUserCommunication}
 {\gamma _k} = \frac{{{{\bar P}_k}{{\left| {{\bf{v}}_k^H{{\bf{h}}_k}} \right|}^2}}}{{\sum\limits_{i = 1,i \ne k}^K {{{\bar P}_i}{{\left| {{\bf{v}}_k^H{{\bf{h}}_i}} \right|}^2} + 1} }},\ \forall k,
 \end{equation}
 where ${{\bar P}_k} \triangleq {P_k}/{\sigma ^2}$ denotes the transmit SNR of user $k$. Then the achievable sum rate in bits/second/Hz (bps/Hz) is
 \begin{equation}\label{sumRateMultiUserCommunication}
 {R_{{\rm{sum}}}} = \sum\limits_{k = 1}^K {{{\log }_2}\left( {1 + {\gamma _k}} \right)}.
 \end{equation}

 In particular, for the special case of two users, the closed-form SINR  expressions were derived in \cite{lu2021near}, with the expressions of user $k$, $k=1,2$, given by
 \begin{equation}\label{SINRMRCZZFMMSE}
 {\gamma _k} = \left\{ \begin{split}
 &{\bar P_k}{\left\| {{{\bf{h}}_k}} \right\|^2}\left( {1 - \frac{{{{\bar P}_{k'}}{{\left\| {{{\bf{h}}_{k'}}} \right\|}^2}{\rho _{kk'}}}}{{{{\bar P}_{k'}}{{\left\| {{{\bf{h}}_{k'}}} \right\|}^2}{\rho _{kk'}} + 1}}} \right),\ {\rm MRC},\\
 &{{\bar P}_k}{\left\| {{{\bf{h}}_k}} \right\|^2}\left( {1 - {\rho _{kk'}}} \right),\ \ \ \ \ \ \ \ \ \ \ \ \ \ \ \ \ \ \ \ \ {\rm ZF},\\
 &{{\bar P}_k}{\left\| {{{\bf{h}}_k}} \right\|^2}\left( {1 - \frac{{{{\bar P}_{k'}}{{\left\| {{{\bf{h}}_{k'}}} \right\|}^2}{\rho _{kk'}}}}{{{{\bar P}_{k'}}{{\left\| {{{\bf{h}}_{k'}}} \right\|}^2} + 1}}} \right),\ \ \ \ \ \ {\rm MMSE},
 \end{split} \right.
 \end{equation}
 where $k' \ne k$, ${\rho _{kk'}} \triangleq \frac{{{{\left| {{\bf{h}}_k^H{{\bf{h}}_{k'}}} \right|}^2}}}{{{{\left\| {{{\bf{h}}_k}} \right\|}^2}{{\left\| {{{\bf{h}}_{k'}}} \right\|}^2}}}$, with $0 \le {{\rho}_{kk'}} \le 1$, accounts for the channel's squared-correlation coefficient between users $k$ and $k'$. For free-space LoS propagation, ${\rho _{kk'}}$ is related to the beam focusing pattern presented in Section \ref{nearFieldBeamFocusingPattern}, via ${\rho _{kk'}} = G_{{\rm{NF}},{\rm{NF}}}^2\left( {{r_k},{\theta _k};{r_{k'}},{\theta _{k'}}} \right)$. It is observed from \eqref{SINRMRCZZFMMSE} that the SINR of user $k$ can be expressed as the SNR of the single-user system minus the penalty term due to the existence of IUI, where the penalty term varies for the three beamforming schemes. Besides, the increase of ${\rho _{kk'}}$ deteriorates the SINR of all the three beamforming schemes, and it can be shown that MMSE beamforming yields the best SINR performance.

 \begin{figure}[!t]
 \centering
 \centerline{\includegraphics[width=3.5in,height=2.625in]{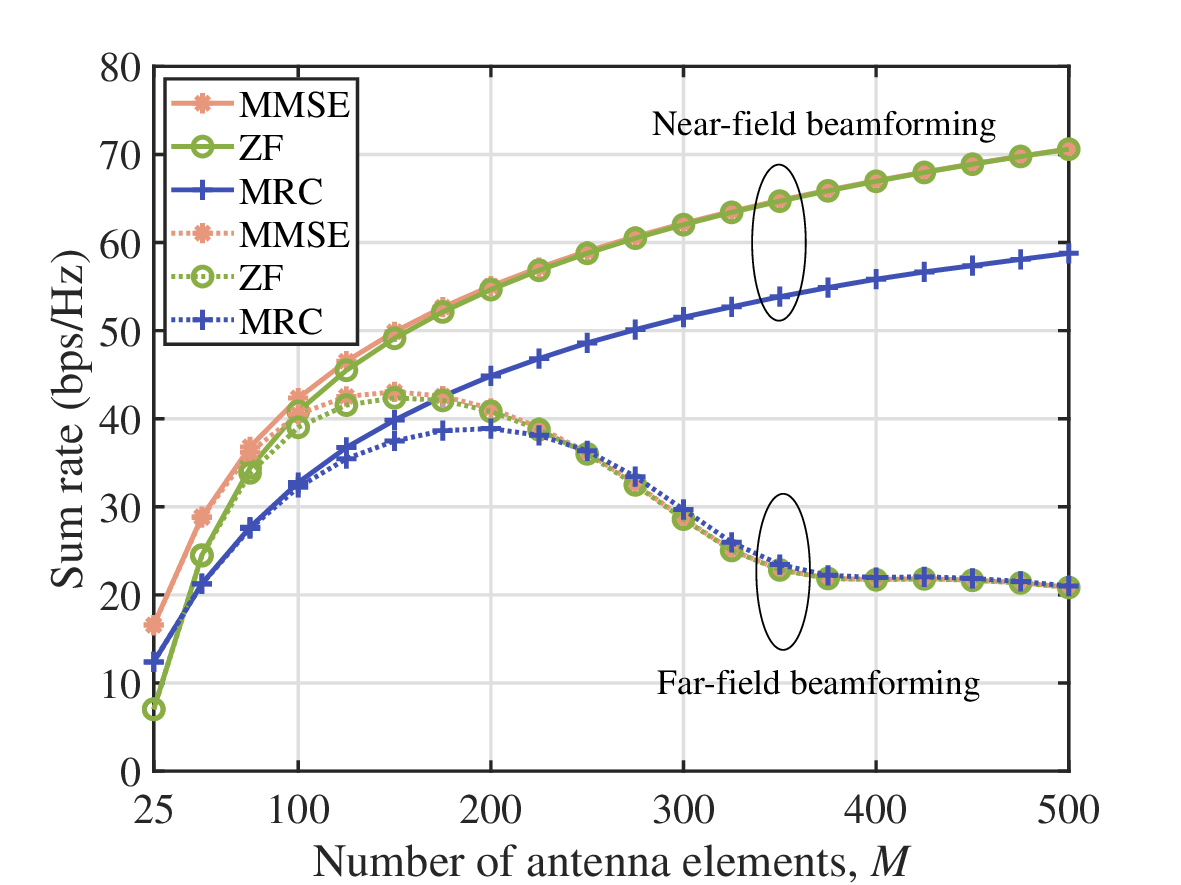}}
 \caption{Sum rate versus the antenna number $M$ for near- and far-field MMSE, ZF, and MRC beamforming.}
 \label{fig:sumRateVersusNumberofAntennaElements}
 \end{figure}

 Fig.~\ref{fig:sumRateVersusNumberofAntennaElements} shows the sum rate of the SIMO system versus antenna number $M$, where the multi-path channels of users are based on the near-field NUSW model in \eqref{generalMultiPathChannelMISOSIMO}. For comparison, the NUSW-based near-field and UPW-based far-field receive beamforming designs are respectively considered, including MRC, ZF, and MMSE. The carrier frequency and transmit SNR of each user are the same as Fig.~\ref{fig:IntegralSNRApproximationLog}. $K = 10$ users are uniformly distributed in the circular area with center $r_c = 600$ m and radius $r_{\max} = 200$ m, i.e., ${r_k} \in \left[ {{r_c} - {r_{\max }},{r_c} - {r_{\max }}} \right]$ and ${\theta _k} \in \left[ { - {\theta _{\max }},{\theta _{\max }}} \right]$, with ${\theta _{\max }} = \arcsin \left( {{r_{\max }}/{r_c}} \right)$. Besides, the multi-path channel \eqref{generalMultiPathChannelMISOSIMO} is considered, where each user has one LoS channel path and $Q = 9$ NLoS channel paths. The NLoS channel component follows the bistatic radar equation based modelling, and the locations of scatterers are randomly distributed in the area given by  ${r_q} \in \left[ {200,500} \right]$ m and ${\theta _q} \in \left[ { - {{60}^ \circ },{{60}^ \circ }} \right]$. The RCS of each scatterer is uniformly distributed in $\left[1,10\right]$ m$^2$. It is observed that for relatively small antenna number $M$, all the three far-field beamforming schemes give the similar performance as the near-field beamforming counterparts. This is expected since when $M$ is small, the users and scatterers are located in the far-field region, for which the far-field UPW model is a valid approximation for NUSW model. However, as $M$ increases, the far-field beamforming schemes give much worse performance than the near-field beamforming schemes, since the far-field beamforming schemes are designed based on the far-field UPW channels, which fail to match with the actual near-field NUSW channels. It is also observed that for large $M$, the performance of the near-field ZF beamforming is comparable to that of the near-field MMSE beamforming. This is due to the fact that the IUI is dominated over the noise in this case.

 Furthermore, by taking into account VR, the authors in \cite{ali2019linear} studied the SINR performance by deriving the approximate SINR expressions under conjugate beamforming and ZF precoders, which are functions of the VR size of each UE and the size of the overlapping VR regions. In \cite{yang2020uplink}, the authors proposed a subarray-based system architecture for XL-MIMO system, by considering the spatial non-stationary channel. The tight closed-form uplink spectral efficiency approximations with linear receivers were derived, including MRC and linear MMSE receivers. Then the subarray phase coefficient design was proposed to maximize the achievable spectral efficiency. To resolve the undesired grating lobe issue suffered by modular XL-array, an efficient user grouping method was developed for multi-user transmission scheduling in \cite{li2023multi}, which avoids allocating the same time-frequency resource block to users located within the grating lobes, so as to maximize the achievable sum rate. Moreover, from a network perspective, it is worthy of analyzing the network-wide XL-MIMO performance over the conventional network-wide massive MIMO \cite{chen2022theoretical} in the future, such as network area spectral efficiency.

 Regarding the near-field power control to achieve higher spectral efficiency, there exist both advantages and challenges compared to the far-field counterpart. On one hand, the near-field beam focusing pattern brought by spherical wave renders it possible to separate users along the same direction but different distances. This enables more flexible multi-user scheduling and power control for achieving higher spectral efficiency. On the other hand, due to the near-field spatial non-stationarity, different users may have distinct visibility region \cite{amiri2018extremely,xu2023resource}, which can be exploited for user grouping, e.g., assigning the same time-frequency resource block to users with the non-overlapping visibility region, so as to simplify the near-field power control. It is also worth mentioning that near-field power control faces several challenges \cite{Liu2023cell,liu2023double}. For example, for the mixed near- and far-field user communications, the energy-spread effect results in a more complicated IUI, and hence a more complicated power control strategy is required. Besides, the grating lobes appearing in modular and sparse arrays should be carefully considered for near-field power control. Last but not least, the difference between near- and far-field power control schemes becomes non-negligible for the cases of near-field user communications and mixed near- and far-field user communications. In particular, when all users are located in the far-field region, the near-field power control scheme reduces to the far-field counterpart, i.e., no mismatch occurs in this case.

 It is also worth mentioning that spectral efficiency can be improved by exploiting the polarization effect. In particular, compared to the single-polarization multi-user system, the dual- and triple-polarization counterparts suffer from both the IUI and cross-polarization interference, rendering near-field precoding more complicated. In \cite{wei2023tri}, to reap the benefit of polarized multi-user communication, a triple-polarization near-field channel was modelled with the dyadic Green's function. Based on the constructed channel model, a user-cluster-based precoding scheme and a two-layer precoding scheme were proposed to mitigate the cross-polarization interference and IUI. The results showed that the triple-polarization system yields a higher spectral efficiency than the dual- and single-polarization systems.

\subsection{DoF}
 In wireless communications, DoF, also known as the spatial multiplexing gain, is defined as the asymptotic slope of the capacity versus SNR \cite{heath2018foundations}. In general, DoF is applied to denote the number of independent data streams that can be simultaneously supported by channels, and a higher system capacity performance can be obtained with a larger DoF. For the conventional massive MIMO communications with UPW-based channel modelling, DoF is limited by the number of channel paths, which is only one for the LoS scenario. However, in the near-field XL-MIMO communications, a superior DoF performance can be achieved, thanks to the NUSW characteristics with both angle and distance resolution capabilities.

 Considerable research efforts have been devoted to exploiting the DoF performance for XL-MIMO systems \cite{dardari2020communicating,yuan2021electromagnetic,jiang2022electromagnetic,xie2023performance,pizzo2020degrees,ouyang2023near}. For example, the authors in \cite{dardari2020communicating} studied the DoF performance for continuous-aperture surface based XL-MIMO. Firstly, the authors modelled the channel based on the dyadic Green's function under the far-field approximation as \eqref{FarField}. Then, achievable DoF expression was derived based on the 2D sampling theory arguments as shown in \cite[Sec. IV]{dardari2020communicating}. Besides, the authors in \cite{yuan2021electromagnetic} studied the effective degrees-of-freedom (EDoF) performance for the discrete antenna array, where the transmitter and receiver are equipped with $M_t$ and $M_r$ antennas, respectively. The concept of EDoF was originally introduced in \cite{muharemovic2008antenna} to approximate the MIMO channel capacity in the low-SNR regime as ${\rm{EDoF}} \times [{\log _2} {\frac{{{E_b}}}{N_0}}  - {\log _2} {\frac{{{E_b}}}{N_0}}_{\min }]$, where ${\frac{{{E_b}}}{{{N_0}}}}$ denotes the bit energy over the noise power spectral density, and ${{{\frac{{{E_b}}}{{{N_0}}}}_{\min }}}$ denotes the minimum value for reliable communications. In \cite{yuan2021electromagnetic}, by modelling the channel $\mathbf{H}\in \mathbb{C} ^{{M_r}\times {M_t}}$ based on the dyadic Green's function, the EDoF is approximately given by
 \begin{equation}\label{EDoF}
 {\rm{EDoF}} \approx {\left( {\frac{{{\rm{tr}}\left( {{\bf{\bar H}}} \right)}}{{{{\left\| {{\bf{\bar H}}} \right\|}_F}}}} \right)^2},
 \end{equation}
 where ${\bf{\bar H}} = {\bf{H}}{{\bf{H}}^H} \in \mathbb{C} ^{{M_r}\times {M_r}}$.

 Furthermore, EDoF for both the continuous surface and discrete array was analytically studied in \cite{jiang2022electromagnetic,xie2023performance}. In \cite{jiang2022electromagnetic}, for the discrete ULA based transmitter and receiver, the channel ${\bf{H}} \in {\mathbb C}^{{M_r} \times {M_t}}$ was modelled based on the scalar Green's function, and the EDoF was computed as \eqref{EDoF}. Moreover, the EDoF result for the discrete ULA is extended to the continuous-aperture ULA, by exploiting the auto-correlation kernel function. Specifically, the transmitter and receiver are equipped with the continuous-aperture ULA, with the lengths being $D_T$ and $D_R$, respectively. Besides, let ${\cal T}$ and ${\cal R}$ denote the region of the transmitter and receiver, respectively. The scalar Green's function between any two arbitrary locations ${\bf{r}}_T \in {\cal T}$ and ${\bf{r}}_R\in {\cal R}$ was denoted as $G\left( {\bf{r}}_R,{\bf{r}}_T \right)$. Then, the auto-correlation kernel $K\left( {\bf{r}}_T, {{\bf{r}}'_T} \right) $, which correlates two arbitrary locations ${\bf{r}}_T,{{\bf{r}}'_T} \in {\cal T}$ at the transmitter region, was defined as
 \begin{equation}\label{Kernel}
 K\left( {{{\bf{r}}_T},{\bf{r}}'_T} \right) = \int_{\cal R} {{G^H}\left( {{{\bf{r}}_R},{{\bf{r}}_T}} \right)} G\left( {{{\bf{r}}_R},{\bf{r}}'_T} \right){\rm d}{{\bf{r}}_R}.
 \end{equation}
 Note that the continuous array scenario can be viewed as the asymptotic scenario for the discrete array scenario with $M_r,M_t \to \infty$. As such, the $(n_1,n_2)$-th element of ${\bf{\bar H}} ={\bf{H}}{{\bf{H}}^H}$ would have the asymptotic form as
 \begin{equation}\label{RAs}
 \left[ {\bf{\bar H}} \right] _{n_1,n_2}\rightarrow \frac{{M_r^2}}{D_{R}^{2}}\left| K\left( {{{\bf{r}}_T},{\bf{r}}'_T} \right) \right|^2.
 \end{equation}
 Then the EDoF for the continuous-aperture ULA is derived as \cite{jiang2022electromagnetic}
 \begin{equation}\label{EDoFCon}
 \begin{aligned}
 {\rm{EDoF}}_{{\rm{con}}} &= \mathop {\lim }\limits_{{M_t},{M_r} \to \infty } \left( {\frac{{{\rm{tr}}\left( {{\bf{\bar H}}} \right)}}{{{{\left\| {{\bf{\bar H}}} \right\|}_F}}}} \right)^2\\
 &= \frac{{{{\left( {\int_{\cal T} {\int_{\cal R} {{{\left| {G\left( {{{\bf{r}}_R},{{\bf{r}}_T}} \right)} \right|}^2}{\rm{d}}{{\bf{r}}_R}{\rm{d}}{{\bf{r}}_T}} } } \right)}^2}}}{{\int_{\cal T} {\int_{\cal T} {{{\left| {K\left( {{{\bf{r}}_T},{{\bf{r}}'_T}} \right)} \right|}^2}{\rm{d}}{{\bf{r}}_T}{\rm{d}}{{\bf{r}}'_T}} } }},
 \end{aligned}
 \end{equation}
 and such a result holds for both the near-field and far-field regions. It is also worth mentioning that some elegant approximate DoF expressions were obtained in \cite{miller2019waves,miller2000communicating}. For example, the DoF for the ULA-based transmitter and receiver can be approximated as
 \begin{equation}\label{ApproximationDoFULA}
 {\rm{DoF}}_{\rm ULA} \approx \frac{{{D_R}{D_T}}}{{\lambda r}},
 \end{equation}
 where $D_T$ and $D_R$ are the physical dimensions of the transmitter and receiver, respectively, and $r$ denotes the distance between the transmitter and receiver. On the other hand, the DoF for the UPA-based transmitter and receiver can be approximately computed as
 \begin{equation}\label{ApproximationDoFUPA}
 {\rm{DoF}}_{{\rm{UPA}}} \approx \frac{{{A_r}{A_t}}}{{{\lambda ^2}{r^2}}},
 \end{equation}
 where $A_T$ and $A_R$ denote the areas of the transmitter and receiver, respectively. Besides, it would be interesting to explore the accurate DoF performance to provide guidance for the practical design of XL-MIMO systems in the future.

\subsection{Near-Field XL-MIMO Sensing}
 Besides the basic wireless communications, XL-MIMO can be leveraged to support the various applications, such as sensing \cite{wang2022snr,wang2023cram,wang2023near,wang2023terahertz}, localization  \cite{friedlander2019localization,guidi2021radio,wymeersch2022radio,tian2023low,wu2023source,chen2023cramer,hua2023near}, and tracking \cite{guerra2021near}. In \cite{wang2022snr}, the sensing SNR expressions for the NUSW model were derived in closed-form for both XL-MIMO radar and XL-phased-array radar modes, and more practical sensing SNR scaling laws were observed as compared to those for the existing UPW model. Subsequently, another important metric of radar sensing, i.e., Cram{\'e}r-Rao bound (CRB), was studied for the near-field sensing in \cite{wang2023cram}, whose closed-form expressions were respectively derived for the above two radar modes based on the USW model. It was mathematically revealed that as the transmit array size goes to infinity, the CRB for angle estimation tends to a certain limit for the near-field XL-MIMO radar, rather than decreasing indefinitely as in the conventional UPW model. In particular, such a saturation limit for the near-field XL-MIMO radar is observed since for the near-field spherical wave, the phase curvature of the signal along the far-end elements becomes diminishing as the array goes very large. In this case, further increasing the array size does not contribute additional gain for CRB. Besides, thanks to the capability to resolve the propagation distance with spherical wavefront, the position rather than only the direction can be inferred for near-field sensing with one single antenna array. For example, the possibility of directly positioning the signal source based on the wavefront curvature was investigated in \cite{guidi2021radio}. By taking into account the VR of the signal source, the localization with the XL-array can be found in \cite{tian2023low,wu2023source}. Moreover, based on the accurate EM propagation model, the authors in \cite{chen2023cramer} proposed a generic near-field positioning model that considered three different observed electric field types and the universality of the terminal position. The CRBs for the three electric field observation types were then derived, and an improved estimation accuracy of dimensions parallel to the receiving antenna surface can be observed for the case of multiple receiving antennas. On the other hand, the curvature information encapsulated in the spherical wavefront can be exploited to achieve the signal source tracking for inferring its position and moving velocity \cite{guerra2021near}, wherein the accuracy of different Bayesian tracking algorithms was evaluated.

 Besides the conventional collocated array, sparse array can also be used for near-field sensing \cite{yang2023enhancing,yan2023improved}. In particular, benefiting from the improved spatial resolution, a superior sensing capability can be achieved for sparse array, thus enabling a better discrimination and characterization of practical EM environments \cite{yang2023enhancing}. Besides, in \cite{yan2023improved}, two types of non-uniform sparse arrays were designed for mixed near- and far-field source localization, which outperform the conventional collocated array in terms of angle and distance estimation. It is also worth mentioning that the grating lobes appearing in modular and sparse arrays may lead to the angular ambiguity problem, i.e., notable estimation error occurs when the target locates within the grating lobes. Such an issue can be alleviated if prior information about the target location is obtained by performing spatial filtering. In summary, near-field sensing with modular and sparse arrays in the presence of grating lobes remains an open problem.

 \subsection{Lessons Learned}
 Near-field XL-MIMO communications lead to quite different performance from conventional far-field communications, as reflected by SNR scaling laws, near-field beam focusing pattern, achievable rate, and DoF. It is revealed that for free-space LoS SIMO communication, the resulting SNR scales with antenna number nonlinearly according to the angular span, rather than increasing linearly as in conventional far-field UPW model. Besides, the significantly enhanced spatial resolution leads to the evolution from far-field beam pattern to near-field beam focusing pattern, and XL-MIMO provides not only the angular resolution as in the conventional MIMO or massive MIMO systems, but also the distance resolution. Such an improvement of resolution enables new opportunities for various applications, such as IUI suppression, WPT, sensing, localization and tracking. Moreover, compared to far-field MIMO communications, a superior DoF performance can be achieved in near-field XL-MIMO communications, which can be leveraged for spatial multiplexing and capacity enhancement.

% >>>>>>>>>>>>>SECTIONS IV -  here >>>>>>>>>>>>
\section{XL-MIMO Design}\label{sectionXLMIMODesign}
 In this section, we present practical XL-MIMO design issues in near-field beam codebook, beam training, channel estimation, DAM, cost-efficient and low-complexity implementation. First, to fully reap the promising beamforming gain brought by XL-MIMO, efficient near-field beam training methods based on carefully designed codebook are indispensable to establish strong initial link between the BS and user before conducting channel estimation and data transmission. Moreover, channel estimation for obtaining the complete CSI facilitates the signal processing design and performance analysis. Second, the super spatial resolution of XL-MIMO motivates a novel DAM transmission technology that enables an inter-symbol interference (ISI)-free communication. Last, practical issues of cost and signal processing complexity with respect to XL-MIMO implementation are discussed in this section.

 Let $\mathcal{W}=\{\mathbf{w}_{1}, \cdots, \mathbf{w}_{N}\}$ denote a beam codebook to be designed that includes $N$ beam codewords. Then, the aim of XL-array beam training is to find the best codeword in the designed beam codebook $\mathcal{W}$ that achieves the maximum received SNR at the user, which can be mathematically formulated as
 \begin{equation}\label{Eq:beamtrainig}
 \begin{aligned}
 \mathbf{w}^{\text{opt}}=&\arg\max  |y(\mathbf{w}_m)|^2 \\
 &~\text{s.t.} ~~{\mathbf{w}_m \in \mathcal{W}},
 \end{aligned}
 \end{equation}
 where $y({{\bf{w}}_m}) = {{\bf{h}}^H}{{\bf{w}}_m}x + z$, with $x$ and $z$ denoting the pilot signal and receiver noise, respectively.
 Generally speaking, the beam training performance can be characterized by the following three performance metrics:
  \begin{itemize}
  \item[1)] \emph{Beam training overhead}, which specifies the number of required training symbols for determining the best beam codeword, while its scaling order is proportional to the number of required training symbols.
  \item[2)] \emph{Beam training success rate}, which denotes the probability that the best beam codeword can be found by a beam training method,
  \item[3)] \emph{Achievable rate}, which characterizes the achievable rate when the XL-array applies the obtained beam codeword for data transmission.
  \end{itemize}

  Note that both the success rate and achievable rate indicate the accuracy of the considered beam training methods, while the achievable rate is more representative since it is directly related to the ultimate objective of beam training, i.e., maximizing the received SNR. As shown in \eqref{Eq:beamtrainig}, the codebook design and beam-training method critically determine the near-field beam training performance, which will be elaborated in the following.

\subsection{Near-Field Beam Codebook Design}\label{beamCode}
 In this subsection, we elaborate the basic codebook design for near-field beam-training. Recall that in the conventional far-field beam training, the DFT-based codebook in the angular domain has been widely used for determining the user angle. Let $\vartheta \triangleq 2d\cos \theta /\lambda  = \cos \theta $ denote the spatial angle, with $d = \lambda/2$. The far-field DFT-based codebook can be mathematically denoted as \cite{you2020fast}
 \begin{equation}
 \mathcal{W}_{\rm{ang}}=\left[ \mathbf{a}(\vartheta_{1}), \mathbf{a}(\vartheta_{2}),\cdots,\mathbf{a}(\vartheta_{N})\right],
 \end{equation}
 with each codeword pointing towards a specific spatial angle ${\vartheta _n} = \frac{{2n - N + 1}}{N}$, $n = 0, \cdots ,N - 1$, given by
 \begin{equation}\label{Eq:far-field}
 \mathbf{a}(\vartheta_{n})=\frac{1}{\sqrt{N}}\left[1, e^{j \pi \vartheta_{n}},\cdots, e^{j \pi (N-1)\vartheta_{n}}\right]^{T}.
 \end{equation}

 However, this codebook is not best suited to the near-field beam training for XL-arrays. Specifically, when the DFT-based codebook is adopted for the exhaustive-search beam training, the user may receive high signal energy at multiple beam codewords due to the energy spread effect, as illustrated in Fig.~\ref{fig:energySpreadEffect}, thus rendering it incapable of finding the best beam. This can be intuitively explained, since the DFT-based codebook is designed for matching the user direction only, while the optimal near-field beamforming is jointly determined by both the user direction and distance, hence leading to the channel mismatch issue and performance loss.

 To address the above issue, several new codebooks dedicated to near-field beam training have been proposed, which spans the beam codewords in different domains.
 \begin{figure*}[t]
 \centering
 \centerline{\includegraphics[width=6.1in,height=2.0in]{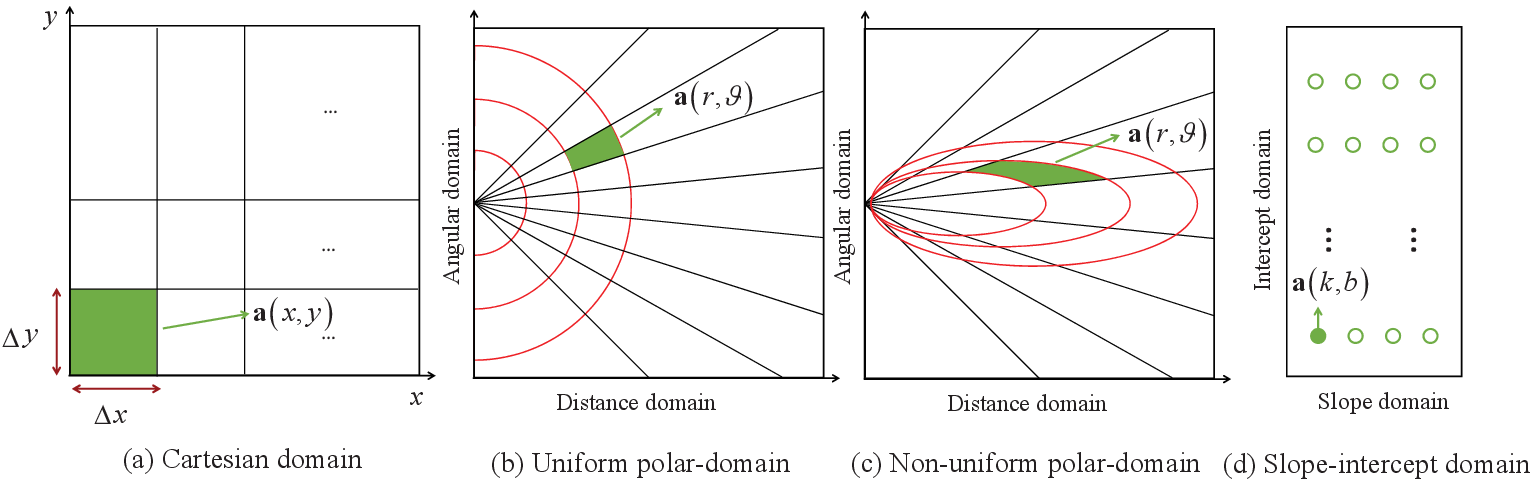}}
 \caption{{Illustration of different near-field beam codebook designs.}
 \label{fig:codebook}}
 \end{figure*}

\subsubsection{Cartesian Domain}
Cartesian-domain codebook is designed to cover the  entire 2D plane by uniformly sampling in the $x$-$y$ coordinate system with sampling steps being $\Delta x$ and $\Delta y$, respectively \cite{han2020channel,wei2022codebook}, as illustrated in Fig.~\ref{fig:codebook}(a). As such, the Cartesian-domain codebook is given by
\begin{equation}
	\mathcal{W}_{\text{car}}\triangleq[\mathbf{a}(x,y)|(x,y)\in \Theta],
\end{equation}
where each column of $\mathcal{W}_{\text{car}}$ denotes a near-field codeword that steers towards its targeted location $(x,y)$ in the $x$-$y$ plane, and $\Theta=\{(x,y)|x=x_{\rm min},x_{\rm min}+\Delta x,\cdots,x_{\rm max};y=y_{\rm min},y_{\rm min}+\Delta y,\cdots,y_{\rm max}\}$.

Note that the dimension of the Cartesian-domain codebook is prohibitively high, which is determined by the product of the number of sampled points on the $x$- and $y$-axes. This thus imposes great challenges to the subsequent beam training method design, resulting in long training overhead.

\subsubsection{Polar Domain}
The codewords in polar-domain codebook is obtained by sampling both the angular and distance domains, which has been recently proposed for near-field beam training \cite{cui2022channel}.
Specifically, let ${N}$ and ${S}$ denote the numbers of angle and distance samples in these two domains. Then the polar-domain codebook can be represented as
\begin{equation}
\begin{aligned} \mathcal{W}_{\rm{pol}}\triangleq[\mathbf{a}(r_{1,1},\vartheta_1)&,\cdots,\mathbf{a}(r_{1,S},\vartheta_1),\cdots,\\
	&\mathbf{a}(r_{N,1},\vartheta_N),\cdots,\mathbf{a}(r_{N,S},\vartheta_N)],
\end{aligned}
\end{equation}
where each column of $\mathcal{W}_{\text{pol}}$ represents a near-field codeword targeting the sampled spatial angle $\vartheta_{n}$ and distance $r_{n,s}$.

For the polar-domain sampling, one straightforward approach is taking a similar idea as in far-field codebooks, that is, \emph{uniformly} sampling the spatial angular and distance domains as illustrated in Fig.~\ref{fig:codebook}(b). In particular, the spatial angular domain can be divided into $N$ samples similarly as in the DFT-based codebook to cover the entire angular domain, while the number of distance domain uniform samples is determined by the channel representation accuracy. Generally speaking, the more samples in the distance domain, the more accurate the channel representation and the higher the training overhead. To exploit this fact, a more efficient \emph{non-uniform} polar-domain codebook was proposed in \cite{cui2022channel}, which leverages the compressed sensing (CS) framework for designing the sampling method of the polar-domain codebook. Specifically, to satisfy the channel recovery accuracy, the column-wise codeword correlation, defined as $\psi  = \mathop {\max }\limits_{p \ne q} |{{\bf{a}}^H}\left( {{r_p},{\vartheta _p}} \right){\bf{a}}\left( {{r_q},{\vartheta _q}} \right)|$, should be set as small as possible. Although $\psi$ is highly complicated, it can be obtained in a more tractable form as below by using the Fresnel approximation
\begin{equation}\label{Eq:correlation}
\begin{aligned}
\psi&= f\left(r_p,r_q,\vartheta_p,\vartheta_q\right)\\
& \approx \left|\frac{1}{M} \sum_{m=-(M-1) / 2}^{(M-1) / 2} e^{j m \pi\left(\vartheta_p-\vartheta_q\right)+j{\frac{2\pi}{\lambda}} m^2 d^2\left(\frac{1-\vartheta_p^2}{2 r_p}-\frac{1-\vartheta_q^2}{2 r_q}\right)}\right|.
\end{aligned}
\end{equation}

Define the curve $\frac{1-\vartheta^2}{r}=\frac{1}{\upsilon}$ as the distance ring $\upsilon$, which is marked as red curves in Fig. \ref{fig:codebook}(c). Then, for the locations sampled along the same distance ring $\upsilon$, i.e., $\frac{1-\vartheta_p^2}{r_p}=\frac{1-\vartheta_q^2}{r_q}=\frac{1}{\upsilon}$, the column correlation function reduces to $f\left(r_p, r_q, \vartheta_p, \vartheta_q\right) = f({\vartheta _p} - {\vartheta _q}) =\left|\frac{1}{M} \sum_{m=-(M-1) / 2}^{(M-1) / 2} e^{j m \pi\left(\vartheta_p-\vartheta_q\right)}\right| = \left| {{\Xi _{M,1/2}}({\vartheta _p} - {\vartheta _q})} \right|$, which only depends on the difference of spatial angles. As such, it can be easily shown that the spatial angular domain along the same distance ring should be uniformly sampled to minimize the column correlation, similar as that in the DFT-based codebook. On the other hand, given the same spatial angles, the correlation function $f(r_p,r_q,\vartheta,\vartheta)$ can be approximated as
\begin{equation}
	f\left(r_p,r_q,\vartheta,\vartheta\right) \approx|G(\Lambda )|=\left|\frac{C(\Lambda)+j S(\Lambda)}{\Lambda}\right|,
\end{equation}
where $\Lambda=\sqrt{\frac{M^2 d^2\left(1-\vartheta^2\right)}{2 \lambda}\left|\frac{1}{r_p}-\frac{1}{r_q}\right|}$, $C\left( \Lambda \right) = \int_0^{\Lambda} {\cos (\frac{\pi }{2}{t^2})} {\rm{d}}t$ and $S\left( \Lambda \right) = \int_0^{\Lambda} {\sin (\frac{\pi }{2}{t^2})} {\rm d}t$ denote the Fresnel integrals \cite{gradshteyn2007table}.  As such, the sampled distances are set as follows
\begin{equation}\label{Eq:distSamp}
	r_{n,s}=\frac{1}{s}Z_{\Delta}(1-\vartheta^2_{n}),  ~~s=0,1,2,\cdots,
\end{equation}
where $Z_{\Delta}=\frac{D^2}{2\lambda\Lambda^2_{\Delta}}$ is the threshold distance defined to limit the correlation between two near-field steering vectors. For example, if the correlation is set  lower than $\Delta=0.5$, then $\Lambda\ge 1.6$ which can be obtained by solving $|G(\Lambda_{0.5})|=0.5$. It can be easily observed from \eqref{Eq:distSamp} that, in contrast to the spatial angular-domain uniform sampling, the distance domain is more densely sampled in the short-range region and less sampled otherwise. This can be intuitively explained, since the distance has a more prominent effect on the phase variations when the distance is shorter, while the near-field spherical wavefront reduces to the planar wave when the distance is sufficiently large. Compared with the uniform polar-domain sampling method, the non-uniform counterpart is more efficient, since it requires less codewords for channel representation.

\subsubsection{Slope-Intercept Domain}
 The fundamental idea of slope-intercept-domain codebook origins from the linear frequency modulation signal in continuous wave radar, which shares the same structure as the near-field PBW-based array response vector in \eqref{arrayResponseVectorPBWModel} \cite{shi2023hierarchical}. As a result, the quadratic term and linear term can be represented as slope $k$ and intercept $b$, respectively. Then, the slope-intercept-domain codebook can be generated by uniformly quantizing the slope $k$ and intercept $b$, which is given by
\begin{equation}
	\mathcal{W}_{\text{s+i}}\triangleq[\mathbf{a}(k,b)|k\in[{k_{\rm min},k_{\rm max}}], b\in [-1,1]],
\end{equation}
where $(k,b)$ corresponds to one point in Fig.~\ref{fig:codebook}(d). When the slope and intercept quantization is properly chosen, this slope-intercept-domain codebook may include a smaller number of codewords than the polar-domain codebook, while its involved beam training design is more complicated.

\subsection{Near-Field Beam Training}
In this subsection, we elaborate efficient near-field beam training algorithms for the narrowband and wideband XL-MIMO communication systems, respectively. It is worth noting that most of beam training methods in this subsection are designed based on the (non-uniform) polar-domain codebook, which may not suit for other forms of codebooks such as the slope-intercept domain codebook.
 \begin{figure*}[t]
 \centering
 \centerline{\includegraphics[width=6.0in,height=5.05in]{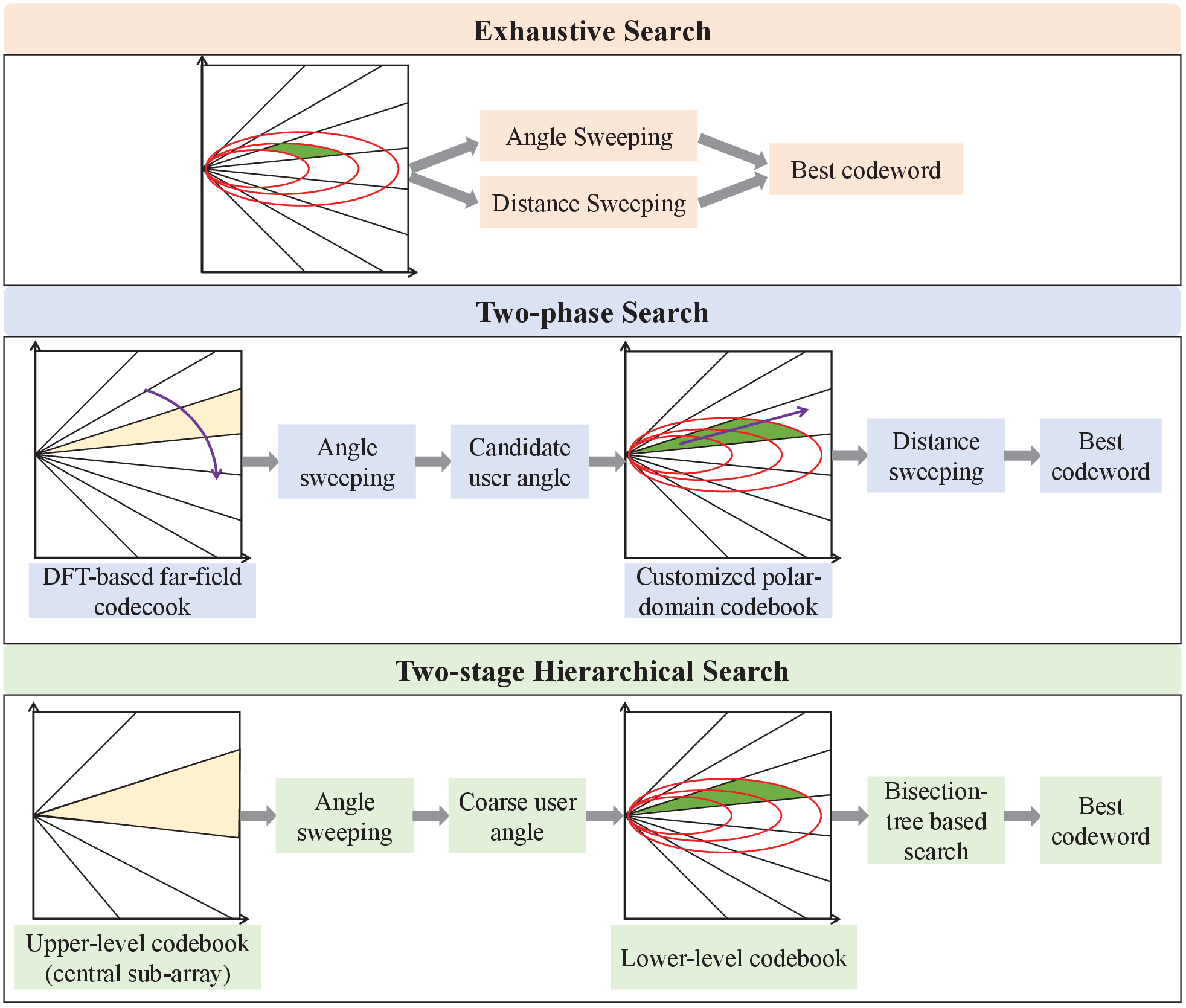}}
 \caption{Illustration of different near-field beam training methods.}
 \label{fig:beamTraining}
 \end{figure*}

\subsubsection{Narrowband Near-Field Beam Training}\label{Narrowband}
For narrowband XL-array systems in high-frequency bands, it is usually assumed that there exists a single dominant LoS path between the XL-array and user. Various beam training approaches have been proposed in the literature for finding the best beam codeword with low training overhead.

\textbf{Exhaustive search:}
Given the non-uniform polar-domain codebook \cite{cui2022channel}, the most straightforward near-field beam training method is performing a 2D exhaustive search over all possible beam codewords in the angular-distance domains, as illustrated in Fig.~\ref{fig:beamTraining}. This method, however, will incur a prohibitively high beam training overhead, which is the product of the numbers of sampled angles and distances, i.e., $NS$. For example, when $N=512$, $S=6$, the total training overhead is $3072$ training symbols.

\begin{figure}[t]
\centering
\centerline{\includegraphics[width=3.4in,height=2.55in]{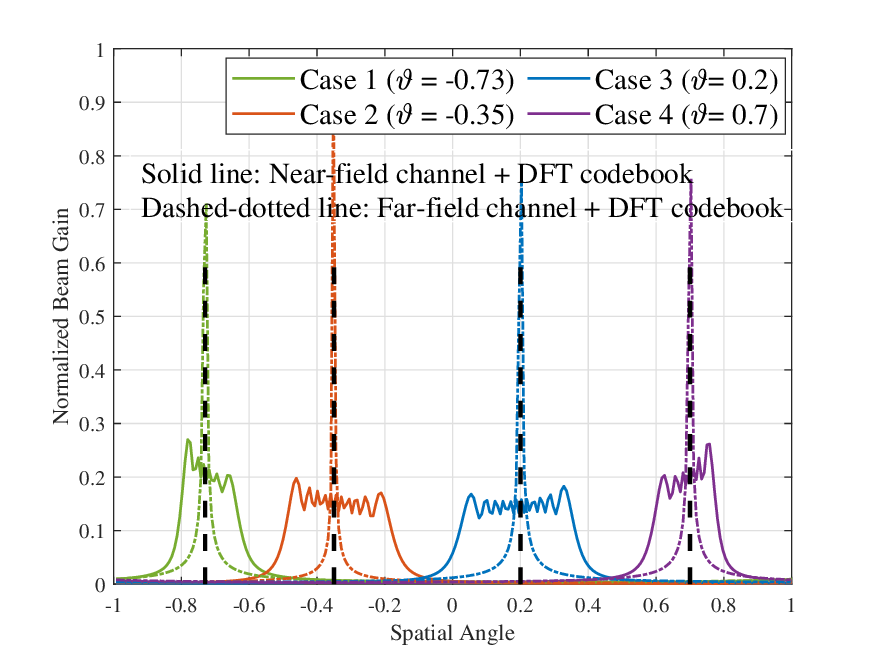}}	
\caption{Illustration of angle-in-the-middle effect \cite{zhang2022fast}, where the solid lines denote the normalized beam gains when the far-field DFT codebook is applied to the near-field channel under four spatial angles (marked by black dashed lines), and the dotted lines denote the normalized beam gains when the far-field DFT codebook is applied to the far-field channel under four spatial angles.}
\label{fig:obs1}
\end{figure}

\textbf{Two-phase search:} To reduce the training overhead of the 2D exhaustive search, one efficient approach is sequentially estimating the user angle and distance in two phases. To this end, the authors in \cite{zhang2022fast} revealed a key ``\emph{angle-in-the-middle}" observation as illustrated in Fig. \ref{fig:obs1}, where the true spatial angle approximately locates in the middle of the dominant angular region with sufficiently high beam powers, when the conventional DFT-based far-field beams are applied for beam training. As such, the best polar-domain codeword can be identified by firstly estimating the user spatial angle using the DFT-based codebook, and then resolving the user distance along the estimated user angle based on the polar-domain codebook, as illustrated in Fig.~\ref{fig:beamTraining}. To improve the angle estimation accuracy, $K$ spatial angles in the middle of the dominant-angle region can be selected as candidate angles, which is termed as the \emph{middle-$K$} angle selection scheme.
This two-phase near-field beam training method only entails a training overhead of $N+KS$, hence leaving more time for data transmission than the 2D exhaustive search method. For example, when $N=512$, $S=6$, and $K=3$, the overall training overhead is $530$ training symbols, which is much less than 3072 as required by the exhaustive search method.

The above two-phase search method was further extended in \cite{wu2023near}, where the authors proposed to jointly estimate the user angle and distance based on the DFT codebook only, which achieved reduced training overhead and enhanced distance estimation accuracy. To this end, the authors analyzed the received beam pattern at the user when far-field beamforming vectors are used for beam scanning. An interesting result was shown that this beam pattern actually contains useful user angle and distance information. Specifically, the user angle can be estimated based on the dominant angular region, while the user distance can be inferred from the width of dominant angular region, which is monotonically decreasing with the user distance given fixed user angle. Then, two efficient beam training schemes were proposed. The first scheme estimated the user angle based on the dominant angular region and decided the user distance by leveraging its width, while the second scheme estimated the user distance by minimizing a defined power ratio mean square error to improve the distance estimation accuracy. Similar to \cite{zhang2022fast}, the middle-$K$ angle selection scheme was applied to improve the angle estimation accuracy. Hence, the total beam training overhead of the joint estimation scheme is only $N+K$. For example, when $N=512$ and $K=3$, the total training overhead is $515$ training symbols.

\textbf{Hierarchical search:} It is worth mentioning that the training overhead of the two-phase beam training scheme is still \emph{proportional} to the number of array antennas, which is still very large for the XL-array. This thus motivates another line of research that targets at designing hierarchical beam training schemes for near-field communications. Thus, the new hierarchical codebook tailored to near-field communications needs to be developed. Unlike the far-field hierarchical codebook design that concerns the spatial angular domain only, the near-field hierarchical codebook design is more challenging, since it needs to take into account both the spatial angular- and distance-domains. Specifically, similar to the far-field hierarchical codebook design, the ideal near-field hierarchical codebook design should satisfy the following two basic criteria:

 \begin{itemize}
 \item[1)] {\emph{Intra-layer beam coverage:}} For each hierarchical layer, the union of beam patterns of its all codewords should cover the whole angular- and distance-domains.
 \item[2)] {\emph{Inter-layer beam coverage:}} For an arbitrary codeword
 within a layer, its beam coverage should be covered by the union of those of several codewords in the next layer, in both the angular- and distance-domains.
\end{itemize}

If given the ideal near-field hierarchical codebook in Section \ref{beamCode}, near-field hierarchical beam training can be easily performed: the XL-array can first generate polar-domain wide beams to find the coarse user angle and distance, and then gradually refine it using narrower beams. However, the ideal near-field hierarchical codebook design is much more challenging than that in the far-field case, since each codeword spans its beam pattern in both the angular and distance domain, which renders the above two criteria difficult to be satisfied.

Several efficient approaches have been proposed in the existing literature to address the above issues \cite{shi2023hierarchical,chen2023hierarchical,lu2022hierarchical}.
For example, the authors in \cite{chen2023hierarchical} proposed an adaptive near-field hierarchical codebook, in which the lower-layer codewords are designed to cover the Fresnel region by using a steering beam gain calculation method, while the upper-layer codewords are devised to generate an arbitrary beam pattern at target locations by using the beam rotation and beam relocation techniques. On the other hand, the authors in \cite{lu2022hierarchical} showed that the ideal hierarchical near-field beam pattern can only be realized by the fully digital beamforming, which, however, incurs unaffordable energy and hardware cost for XL-arrays. To tackle this difficulty, a hybrid digital-analog structure was proposed to design the near-field hierarchical codebook. This is achieved by formulating and solving a multi-resolution near-field codewords design problem to minimize the error between the ideal beam pattern in the near-field hierarchical codebook and that achieved by practical codewords based on hybrid beamforming structure. Besides the near-field hierarchical codebook design in the polar domain, the authors in \cite{shi2023hierarchical} studied a new joint spatial-angular and slope-intercept representation for the near-field channel model, based on which a novel spatial-chirp beam-aided codebook was proposed for designing the near-field hierarchical codebook under similar criteria in the polar domain. Moreover, a \emph{hybrid} hierarchical codebook design was developed in \cite{wu2023two} for facilitating fast near-field beam training. Specifically, it first employs a central subarray of the XL-array to estimate a coarse spatial angle by using the conventional far-field hierarchical codebook. Then, the fine-grained user angle-and-distance pair is progressively resolved by devising a polar-domain hierarchical codebook, where the antenna deactivation scheme is used for guaranteeing the angle coverage and the distance domain is progressively divided in a binary-tree manner, as illustrated in Fig.~\ref{fig:beamTraining}. By this way, the original linear order $\mathcal{O}(N)$ of the two-phase near-field beam training  can be reduced to an order of $\mathcal{O}(\log(N))$.

\textbf{Learning-based search:}
Moreover, in \cite{liu2023deep}, a deep-learning aided beam training scheme was investigated, in which the best near-field codeword is predicted by utilizing powerful neural networks with measurements of the beam gains given conventional far-field wide beams. Specifically, the DFT-based far-field wide beams are first generated and applied to acquire the measurements of various received signals for neural networks input, based on which two neural networks, i.e., the direction and distance networks, are leveraged to obtain the optimal polar-domain near-field beam.

\subsubsection{Wideband Near-Field Beam Training}
For wideband XL-array systems, the near-field beam training design becomes more difficult due to the following two reasons. First, the multi-path channel model needs to be considered for relatively low-frequency band systems, which calls for new and efficient near-field beam training for multi-path channels. Second, the large bandwidth and XL-array aperture introduce the so-called \emph{near-field beam split effect} \cite{cui2023nearRainbow}. Specifically, the beams generated at different frequencies with spherical wavefronts are generally dispersed at different locations due to use of the frequency-independent phase shifters at the XL-array. Note that the near-field beam split effect is more severe than that in the far-field case, since the beam disperses in both the angular and distance domains. This effect renders the near-field beam training more challenging, since the beams cannot be well focused at the targeted location for each codeword as in narrow-band beam training (see Section~\ref{Narrowband}). Fortunately, it was shown that using TTD lines can introduce specific time delays to signals and thereby create frequency-dependent phase shifts \cite{zhang2023nearbeamfocusing,myers2021infocus,wang2023beamfocusing,zhang2023deep}, so that both the angular and distance coverage region of the beams under the beam split effect can be flexibly controlled \cite{cui2021near}. In other words, the XL-array beamformer based on TTD devices can flexibly tune multiple beams towards multiple locations by one RF chain. This thus motivates the design of an efficient near-field rainbow-based beam training for wideband XL-array systems \cite{cui2023nearRainbow}, which generates multiple beams  pointing to all  sampled angles with the same distance in each time, while  the XL-array controls the beam sweep  over different distances at different time for determining the best distance. Therefore, the wideband near-field beam training with controllable beam split achieves a low overhead of $S$, hence significantly reducing the training overhead of narrow-band near-field beam training. The comparison of beam training overhead of different methods is summarized in Table~\ref{table:beamTrainingOverhead}. %where the parameters for the example are $N = 512$, $S = 6$, $N_L = 128$, and $T= 4$, respectively.
% $T$ is the number of narrow beams each large beam contain
\begin{table*}[t]
	\caption{{Comparison of beam training overhead of different methods}}
	\label{table:beamTrainingOverhead}
	\centering
	\begin{tabular}{|c|c|c|c|}
		\hline
		&{\bf{Beam Training Method}} & \begin{tabular}[c]{@{}c@{}} \bf{Training Overhead}\\ (number of required training symbols)\end{tabular}  & \begin{tabular}[c]{@{}c@{}} \bf{Example}\\ $N = 512$, $S=6$, $K=3$,\\ $N_L = 128$, $T= 4$\end{tabular}\\
		\hline
		\multirow{7}{*}{\textbf{Narrowband}}&Exhaustive-search based near-field beam training \cite{cui2022channel} & $NS$&$3072$ \\
		\cline{2-4}
			&Two-phase near-field beam training \cite{zhang2022fast}  & $N+KS$&$530$ \\
        \cline{2-4}
            &Near-field beam training with DFT codebook \cite{wu2023near}&
$N+K$&$515$\\
			\cline{2-4}
			&Two-stage hierarchical near-field beam training \cite{wu2023two} & $2\log_2(N_{L})+4\log_2(\frac{N}{N_{L}})$ &$22$\\
		\cline{2-4}
			&Deep learning based near-field beam training  \cite{liu2023deep} & $N/T+S$ &$134$\\
			\cline{2-4}
			&Far-field exhaustive-search based beam training  & $N$ &$512$\\
		\cline{2-4}
			&Far-field hierarchical beam training & $2\log_2(N)$ &$18$\\
		\hline
		\textbf{Wideband}&Near-field rainbow beam training \cite{cui2023nearRainbow}& $S$ &$6$\\
		\hline
	\end{tabular}
\end{table*}

\subsection{Channel Estimation}
 The previous subsection discusses the codebook based near-field beam training methods, which aim to select the best beam for establishing a high-quality link between the BS and user, yet without requiring the explicit CSI. In this subsection, we consider the channel estimation for XL-MIMO with the purpose of obtaining the complete channel matrices, which is beneficial to implementing further signal processing design and performance analysis. However, as discussed above, XL-MIMO embraces quite different channel characteristics from the conventional massive MIMO. As a result, it is important to develop efficient channel estimation schemes, so as to match the channel characteristics of XL-MIMO with high accuracy and acceptable complexity \cite{zhu2023sub}.

\subsubsection{Channel Estimation Based on Statistical Characteristics}
 As discussed in Section~\ref{subsectionSpatialCorrelationModelling}, one widely used near-field modelling for XL-MIMO is to model the channel based on the statistical characteristics, such as the channel correlation matrix. Correspondingly, channel estimation schemes based on statistical characteristics can be implemented. Based on the channel correlation matrix, MMSE channel estimation scheme has been widely applied in massive MIMO \cite{zhang2020prospective,wang2022uplink}, which can realize lower normalized mean-square error (NMSE) compared to the least-squares (LS) estimator. However, for XL-MIMO communications, MMSE channel estimation scheme would lead to extremely high computational complexity. Besides, it is also difficult to obtain complete knowledge of the spatial correlation matrix, which is of extremely high dimension. In this case, the LS channel estimation scheme can typically be applied for channel estimation, which does not require prior knowledge of channel statistics. However, the LS channel estimation scheme leads to a higher NMSE performance compared to the MMSE channel estimation scheme, especially at low SNR. Thus, it is necessary to balance the accuracy and complexity to design efficient statistical characteristics based estimation schemes.

 To tackle this issue, the authors in \cite{demir2022channel} proposed a subspace based channel estimation scheme called reduced-subspace LS (RS-LS), where only the orthonormal eigenvectors matrix of the spatial correlation matrix instead of the whole spatial correlation matrix was applied. The proposed RS-LS channel estimation scheme outperformed the conventional LS channel estimation scheme and approached the MMSE channel estimation scheme. The study of low-complexity estimation schemes based on some key channel statistical characteristics deserves more investigation in the future.

\subsubsection{Channel Estimation Exploiting Sparsity}
 The near-field XL-MIMO channel exhibits some interesting sparsity characteristics, which can be exploited to implement efficient channel estimation. For the conventional far-field communications, the channel exhibits the angular-domain sparsity, and an angular-domain representation for the far-field channel can be implemented, where a Fourier transform matrix sampled from the angular space can be utilized to construct the angular-domain sparsity. Based on the angular-domain sparsity, several CS based channel estimation schemes can be applied. However, this angular-domain sparsity is no longer  valid in XL-MIMO communications. Instead, the channel embraces the sparsity in the polar-domain, and the polar-domain representation for the near-field channel was proposed in \cite{cui2022channel}. A new transform matrix was designed which was composed of several near-field steering vectors sampled in the angular-distance domain. Based on this new transform matrix, an on-grid polar-domain simultaneous orthogonal matching pursuit and an off-grid polar-domain simultaneous iterative gridless weighted channel estimation schemes were then proposed in \cite{cui2022channel} to estimate the near-field channel.

 For near-field wideband XL-MIMO systems, channel estimation faces a further challenge due to the existence of beam split effect, i.e., the beams generated at different frequencies would disperse in both the angular and distance domains. To tackle this issue, the authors in \cite{cui2023nearWideband} first revealed that the sparse set of near-field channels exhibits a linear relationship with the frequency in both the angular and distance domains, i.e., the bilinear pattern of the near-field beam split effect. Based on this characteristic, a bilinear pattern detection based algorithm was proposed to estimate the near-field wideband channels, by exploiting the polar-domain sparsity. Besides, by exploiting the channel sparsity, a federated learning based approach was proposed in \cite{elbir2023near} as a model-free solution for near-field wideband channel estimation.

 Moreover, the XL-MIMO channel also exhibits the spatial non-stationary characteristic, where the VR could be considered for the near-field channel estimation \cite{han2020channel,han2022localization,han2020deep,tian2023low,liu2023location,iimori2022joint,iimori2022grant}. In \cite{han2020channel}, the VRs of subarrays and scatterers were introduced to describe the channel non-stationarity, and the subarray-wise and scatterer-wise channel estimation schemes were proposed to estimate the spatial non-stationary channel from the perspectives of scatterer and subarray, respectively. For the subarray-based analog beamforming structure, the authors in \cite{tian2023low} proposed an element-wise VR identification scheme, and the results showed that its achievable spectral efficiency is close to that with perfect CSI. In \cite{liu2023location}, VR is assumed to be location-independent, and the VRs of some beacon users are known a priori. By training these beacon users, the VR can be obtained through the Voronoi cell partition or the proposed VR-net when the location of one user serves as the input. Besides, by exploiting the block-wise sparsity caused by spatial non-stationarity, joint activity and channel estimation was studied in \cite{iimori2022joint,iimori2022grant}.

\subsubsection{Channel Knowledge Map}
 The large-dimensional channel matrix of XL-MIMO renders it more challenging to acquire the instantaneous CSI as compared to the conventional massive MIMO. To this end, channel knowledge map (CKM) for environment-aware communications was proposed in \cite{zeng2021toward}, which facilitates or even avoids the real-time CSI acquisition in XL-MIMO communications. Specifically, CKM is a site-specific database, tagged with the locations of the transmitters and/or receivers \cite{zeng2021toward}, and the stored channel knowledge includes complex-valued channel coefficients, AoAs/AoDs, delays and Doppler frequencies, where the real channel-related data required for CKM construction can be collected from the offline and online measurements \cite{zeng2021toward,wu2021environment,wu2023environment,ding2021environment,zeng2023tutorial}. As such, CKM is able to enable environment-awareness communications and provide the location-specific channel knowledge, thus circumventing the prohibitive pilot overhead and complicated channel estimation. This provides an efficient and low-complexity CSI acquisition method for near-field XL-MIMO communications.

\subsubsection{Machine Learning Based Estimation Schemes}
 To further improve the efficiency of channel estimation schemes, machine learning can be utilized to design efficient and intelligent channel estimation schemes. Based on the existing multiple residual dense network (MRDN), the authors in \cite{Lei2023CE} proposed a polar-domain MRDN (P-MRDN) based channel estimation scheme relied on the polar-domain sparsity. Furthermore, with atrous spatial pyramid pooling-based residual dense network (ASPP-RDN), a polar-domain multi-scale residual dense network (P-MSRDN) based channel estimation scheme was considered to further improve the accuracy. Numerical results found that the P-MRDN and P-MSRDN based channel estimation schemes could efficiently capture the polar-domain sparsity and outperform the conventional MRDN based channel estimation scheme. In the future, more machine learning based channel estimation schemes can be designed to exploit the channel characteristics of XL-MIMO and enable intelligent processing solutions tailored for XL-MIMO.

\subsection{Delay Alignment Modulation}
 Orthogonal frequency-division multiplexing (OFDM) has been a dominant wideband transmission technology in 4G, 5G, and WiFi wireless networks. However, OFDM also faces many well-known challenges, e.g., high out-of-band (OOB) emission \cite{nissel2017filter}, sensitivity to carrier frequency offset (CFO) \cite{goldsmith2005wireless,heath2018foundations}, and high peak-to-average-power ratio (PAPR) \cite{cho2010mimo,han2005overview}. Though various techniques have been proposed to resolve these issues, such as windowing, filter bank multi-carrier (FBMC) \cite{nissel2017filter}, single-carrier discrete Fourier transform-spread OFDM (DFT-s-OFDM), and orthogonal time frequency space (OTFS) \cite{hadani2017orthogonal}, they either incur performance loss or exacerbate the signal processing complexity.

 Motivated by the super spatial resolution of XL-MIMO and the multi-path sparsity of mmWave/THz channels, a novel transmission technology termed delay alignment modulation was recently proposed in \cite{lu2022delay}. The key idea of DAM is {\it path delay pre-/post-compensation} and {\it path-based beamforming}. In particular, the unprecedented spatial resolution of XL-MIMO and ISAC endow the transmitter/receiver with the capability of extracting the features of each multi-path, e.g., AoA/AoD, propagation delay, and Doppler frequency \cite{lu2023Manipulating,zhang2021enabling}. In this context, by judiciously performing delay pre-/post-compensation matching the respective multi-paths, and in conjunction with path-based beamforming, DAM enables all multi-path signal components to reach the receiver concurrently and constructively. As a result, the original time-dispersive channel can be transformed into the simple additive white Gaussian noise (AWGN) channel, yet free from the sophisticated channel equalization or multi-carrier transmissions \cite{lu2022delay,lu2023Manipulating}. It is also worth mentioning that though DAM involves path delay compensation, it is quite different from TTD-based design, since the two techniques focus on different issues, where the former aims to address the ISI issue by exploiting the spatial-delay processing without requiring additional hardware, while the latter targets for mitigating the beam squint/split effect.

\begin{figure}[!t]
 \centering
 \centerline{\includegraphics[width=3.1in,height=2.5in]{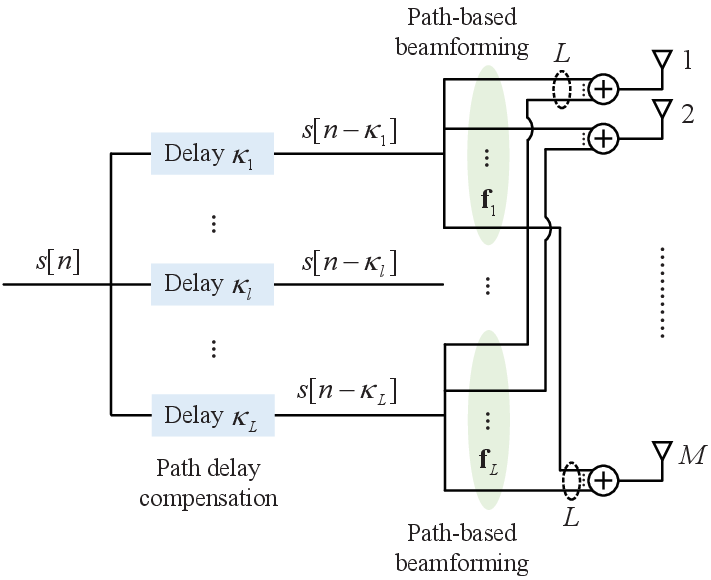}}
 \caption{Transmitter architecture of single-carrier delay alignment modulation that exploits the super spatial resolution of XL-MIMO and multi-path sparsity of high-frequency channels \cite{lu2022delay}.}
 \label{fig:DAMBlockDiagram}
 \end{figure}

 To illustrate the effect of DAM, we consider a basic XL-MISO communication system, where there exists $L$ temporal-resolvable multi-paths between the transmitter and the receiver. Under the quasi-static block fading model, the channel impulse response within each channel coherence block can be expressed as
 \begin{equation}\label{multiPathChannel}
 {\bf{h}}\left[ n \right] = \sum\limits_{l = 1}^L {{{\bf{h}}_l}\delta \left[ {n - {n_l}} \right]},
 \end{equation}
 where ${{\bf{h}}_l} \in {{\mathbb C}^{M \times 1}}$ denotes the channel vector of multi-path $l$ and it can be obtained based on \eqref{generalComplexChannelCoefficient}, and $n_l$ denotes its discretized delay. For transmitter-side single-carrier DAM, the transmitter architecture is illustrated in Fig.~\ref{fig:DAMBlockDiagram}, where path delay compensation simply means time shift of symbol sequence, without requiring the additional hardware like TTD lines. The transmitted signal is \cite{lu2022delay,lu2023Manipulating}
 \begin{equation}\label{transmitSignalDAM}
 {\bf{x}}\left[ n \right] = \sum\limits_{l = 1}^L {{{\bf{f}}_l}s\left[ {n - {\kappa _l}} \right]},
 \end{equation}
 where $s\left[ n \right]$ denotes the i.i.d. transmitted data symbol, with ${\mathbb E}[ {{{\left| {s[n]} \right|}^2}}] = 1$, ${{\bf{f}}_l} \in {{\mathbb C}^{M \times 1}}$ denotes the path-based beamforming associated with multi-path $l$, and ${\kappa _l} \ge 0$ denotes the deliberately introduced path delay pre-compensation for multi-path $l$, with ${\kappa _l} \ne {\kappa _{l'}}$, $\forall l \ne l'$. The power of \eqref{transmitSignalDAM} is $ {\mathbb E}[ {{{\left| {{\bf{x}}\left[ n \right]} \right|}^2}} ] = \sum\nolimits_{l = 1}^L {{{\left\| {{{\bf{f}}_l}} \right\|}^2}}  \le P$, where $P$ is the maximum available power at the transmitter.

 By setting ${\kappa _l} = {n_{\max }} - {n_l}$, $\forall l$, with ${n_{\max }} \triangleq \mathop {\max }\limits_{1 \le l \le L} {n_l}$ being the maximum delay among all the $L$ multi-paths, the received signal can be expressed as
 \begin{equation}\label{receivedSignalDAM}
 \begin{aligned}
 y\left[ n \right] &= {{\bf{h}}^H}\left[ n \right] \circledast  {\bf{x}}\left[ n \right] + z\left[ n \right]\\
 &= \underbrace {\left( {\sum\limits_{l = 1}^L {{\bf{h}}_l^H{{\bf{f}}_l}} } \right)s\left[ {n - {n_{\max }}} \right]}_{{\rm{desired\ signal}}} +  \\
 &\ \ \ \ \underbrace {\sum\limits_{l = 1}^L {\sum\limits_{l' = 1}^L {{\bf{h}}_l^H{{\bf{f}}_{l'}}s\left[ {n - {n_{\max }} + {n_{l'}} - {n_l}} \right]} } }_{{\rm{ISI}}} + z\left[ n \right],
 \end{aligned}
 \end{equation}
 where $z\left[ n \right] \sim {\cal CN}\left( {0,{\sigma ^2}} \right)$ denotes the AWGN at the receiver. When the receiver is synchronized to the delay $n_{\max}$, the first term in \eqref{receivedSignalDAM} is the desired signal, which is contributed by all the $L$ multi-path signal components, and the second term leads to the undesired ISI.

 For the asymptotic case when the number of transmit antennas $M$ is much lager than that of multi-paths $L$, i.e., $M \gg L$, it was shown in  \cite{lu2023Manipulating} that the multi-path channel vectors tend to be asymptotically orthogonal, which helps ease the path-based beamforming design.
 Specifically, with the low-complexity path-based maximal-ratio transmission (MRT) beamforming ${{\bf{f}}_l} = \sqrt P {{\bf{h}}_l}/\sqrt {\sum\nolimits_{i = 1}^L {{{\left\| {{{\bf{h}}_i}} \right\|}^2}} } $, the received signal in \eqref{receivedSignalDAM} reduces to
 \begin{equation}\label{receivedSignalDAMMRT}
 y\left[ n \right] \to \left( {\sqrt {P\sum\nolimits_{l = 1}^L {{{\left\| {{{\bf{h}}_l}} \right\|}^2}} } } \right)s\left[ {n - {n_{\max }}} \right] + z\left[ n \right].
 \end{equation}
 The result shows that there only exists the symbol sequence $s\left[n\right]$ delayed by $n_{\max}$ in the received signal, and the coherent combining of all the $L$ multi-path signal components is achieved. This implies that the original time-dispersive channel can be transformed into the simple ISI-free AWGN channel, without the aid of traditional ISI-mitigation techniques such as  channel equalization or multi-carrier transmissions.

 It is worth mentioning that when channel vectors are non-orthogonal, the ISI-free AWGN channel is still attainable, by properly designing the path-based ZF beamforming such that the ISI is completely eliminated, i.e.,
 \begin{equation}\label{ZFCondition}
 {\bf{h}}_l^H{{\bf{f}}_{l'}} = 0,\ \forall l \ne l',
 \end{equation}
 which are feasible almost surely when $M \ge L$. In this case, the received signal in \eqref{receivedSignalDAM} reduces to \cite{lu2023Manipulating}
 \begin{equation}\label{receivedSignalDAMZF}
 y\left[ n \right] = \left( {\sum\limits_{l = 1}^L {{\bf{h}}_l^H{\bf{H}}_l^ \bot {{\bf{b}}_l}} } \right)s\left[ {n - {n_{\max }}} \right] + z\left[ n \right],
 \end{equation}
 where ${\bf{H}}_l^ \bot  \in {{\mathbb C}^{M \times \left( {M - L + 1} \right)}}$ is an orthonormal basis for the orthogonal complement of ${{\bf{H}}_l}$, with ${{\bf{H}}_l} \in {{\mathbb C}^{M \times L}} \triangleq [{{\bf{h}}_1}, \cdots ,{{\bf{h}}_{l - 1}},{{\bf{h}}_{l + 1}}, \cdots ,{{\bf{h}}_L}]$, and ${{\bf{b}}_l} \in {{\mathbb C}^{\left( {M - L + 1} \right) \times 1}}$ denotes the new path-based beamforming vector to be designed. As such, an ISI-free AWGN channel can be similarly achieved, provided that $M \ge L$. On the other hand, when certain ISI is tolerable, path-based MRT and MMSE beamforming schemes were further developed for DAM in \cite{lu2022delay}. Moreover, the comparisons of single-carrier DAM, OFDM, and OTFS are summarized in Table.~\ref{table:DAMOFDMOTFSComparison}.

 \begin{table*}[htbp]\scriptsize
 \caption{Comparison of single-carrier DAM, OFDM, and OTFS}
 \begin{center}
 \centering
 \begin{tabular}{|l|c|c|c|c|}
 \hline
 \centering
  & \bf Single-Carrier DAM & \bf OFDM & \bf OTFS \\
 \hline
 \centering
 \bf Signal representation & Time domain & Time-frequency domain  & Delay-Doppler domain \\
 \hline
 \centering
 \bf Signal detection latency & Low   & Relatively high & High\\
 \hline
 \centering
 \bf Receiver complexity & Low  & Relatively high   & High\\
 \hline
 \centering
 \bf PAPR & Low & High & Relatively high \\
 \hline
 \centering
 \bf CP overhead & Low & High & Relatively high \\
 \hline
 \end{tabular}
 \end{center}
 \label{table:DAMOFDMOTFSComparison}
 \end{table*}

 In addition to the above perfect DAM targeting for zero delay spread, the {\it generic DAM} technique was further developed in \cite{lu2023Manipulating} for manipulating the channel delay spread to a certain value, which enables a flexible framework for efficient single- or multi-carrier transmissions. Furthermore, to address the Doppler shift and ISI issues in the more general time-variant frequency-selective channels, DAM can be extended to {\it delay-Doppler alignment modulation} (DDAM) to achieve the Doppler-ISI dual mitigation \cite{lu2023delayDoppler,xiao2023exploiting}, where the time-variant frequency-selective channels can also be transformed into time-invariant ISI-free channels. Besides the mobile scenarios, DDAM is also applicable to near-field and far-field quasi-static scenarios by reducing to DAM. In particular, for the asymptotic case when $M \gg L$, it was shown in \cite{lu2023delayDoppler} that DDAM is able to achieve the time-invariant ISI-free communications with the simple delay-Doppler compensation and path-based MRT beamforming. Moreover, the result showed that DDAM yields a better spectral efficiency than OTFS as $M$ increases, yet with much lower receiver complexity and detection latency. Despite the appealing advantages, DAM also faces several practical challenges. For example, accurate CSI is crucial to DAM for performing path delay pre-compensation and path-based beamforming, and a first attempt towards channel estimation for DAM can be found in \cite{ding2022channel}. However, it only focuses on the single-carrier DAM, more research endeavors are thus needed to develop efficient channel estimation methods for multi-carrier DAM and the more general DDAM. Besides, the fractional delay brings a new challenge to distinguish multi-path components, rendering perfect DAM difficult to achieve in practice. Moreover, practical low-cost hybrid analog/digital path-based beamforming architecture deserves future investigation.

 \subsection{Cost-Efficient and Low-Complexity Implementation}
 Practical implementation of XL-MIMO involves the cost and signal processing complexity issues. On one hand, the deployment of XL-MIMO renders the hardware cost issue more prominent, e.g., expensive RF chains. On the other hand, XL-MIMO suffers from the complicated signal processing due to the large-dimensional channel. As such, many efforts have been endeavored for cost-efficient and low-complexity XL-MIMO design.

 \subsubsection{Cost- and Energy-Efficient Implementation}
 The utilization of low-cost and low-resolution devices brings the opportunity to reduce the cost and energy expenditure of XL-MIMO. Besides, the spatial non-stationarity characteristic renders antenna selection suitable for addressing the cost and energy issues.

 \paragraph{Low-Resolution ADCs and Mixed-ADCs}
 To alleviate the high cost and power consumption of high-resolution ADCs, the simple low-resolution ADCs can be used, where fewer bits are applied to digitize the received signal \cite{zhang2016on,zhang2018on,jacobsson2017throughput}, thus decreasing the signal processing complexity of XL-MIMO and meeting the demands of green communications. However, the low-resolution ADCs are at the cost of performance degradation. Fortunately, the mixed-ADCs are proposed to balance the spectral efficiency and energy efficiency \cite{zhang2017performance,zhang2016mixed,zhang2019mixed}. For mixed-ADCs, a large number of low-resolution ADCs and a small number of high-resolution ADCs are used, and such a setting can balance the trade-off between the cost and performance for XL-MIMO. Note that existing mixed ADC designs for the conventional massive MIMO communications are based on the far-field UPW assumption. For XL-MIMO communications, it would be interesting and potential to implement the low-resolution or mixed ADC designs that consider the NUSW and spatial non-stationarity characteristics.

 \paragraph{Antenna Selection}\label{AntennaSelection}
 In practice, due to the spatial non-stationarity characteristic of XL-MIMO, not all antennas are equally important to the UEs. To reduce the circuit cost and computational complexity, antenna selection can be applied, whose idea is to select partial antennas to serve the UE. Extensive research efforts have been devoted to antenna selection for XL-MIMO communications \cite{marinello2020antenna,de2021quasi,ubiali2020xl,ubiali2021energy}.

 In \cite{marinello2020antenna}, the authors proposed four antenna selection schemes to select a suitable subset of antennas for serving UEs, based on the long-term channel fading parameters. A simple antenna selection scheme was first proposed based on the highest received normalized power (HRNP) criterion, and three heuristic schemes were considered, including local search, genetic algorithm (GA), and particle swarm optimization based on the HRNP active antennas set as the initial solution. The results showed that GA based antenna selection scheme usually achieves the best energy efficiency performance. Based on a subarray switching architecture, the authors in \cite{de2021quasi} proposed the GA-based near-optimal and low-complexity antenna selection schemes, where joint antenna selection and power allocation were optimized to maximize the spectral efficiency. Compared to the benchmarking schemes, the two GA-based optimization schemes achieve higher spectral efficiency performance, especially in crowded XL-MIMO scenarios.

 Moreover, the flexible antenna selection (FAS) and fixed subarray selection (FSS) algorithms were proposed in \cite{ubiali2021energy}. Different from FAS that directly selected the suitable antenna subset, FSS partitioned the whole antenna array into multiple subarrays of fixed size, and then the antenna selection was implemented in a subarray manner. The results showed that the FSS algorithm yields comparable performance as FAS, while reducing the computational complexity and hardware implementation. Note that antenna selection also helps reduce the computational complexity, since only the subset of BS antennas are activated for uplink/downlink transmissions \cite{ubiali2020xl}. For example, in \cite{goisferreira2021}, a low-complexity antenna selection algorithm based on matching pursuit was proposed, which achieved a compromise between bit error rate and computational complexity.

\subsubsection{Low Signal Processing Complexity Implementation}
 In addition to hardware cost and energy expenditure, XL-MIMO also faces the high signal processing complexity. One direct approach is to develop low-complexity algorithms to reduce the signal processing requirement at the central processing unit (CPU). Alternatively, the distributed processing architecture can be exploited to assign the task to multiple local processing units (LPUs).

\paragraph{Low-Complexity Algorithms}\label{Low-Complexity}
 In order to reduce the high signal processing complexity caused by large-dimensional channel matrix, the randomized Kaczmarz algorithm (rKA) was introduced in XL-MIMO to approximate the performance of regularized zero-forcing (RZF) in  \cite{croisfelt2021accelerated}, where rKA solves the linear equation in a cyclic and iterative manner, thus avoiding the high computational complexity of matrix inversion. Based on rKA, the authors in \cite{xu2023low} proposed a new mode of randomization termed sampling without replacement randomized Kaczmarz algorithm (SwoR-rKA), which improves the probability of choosing users with better channel conditions and achieves a better convergence performance than rKA with equal user probabilities. In \cite{ribeiro2021low}, the authors proposed a new low-complexity ZF precoding scheme termed mean-angle based zero-forcing. By partitioning the antenna array into multiple subarrays and grouping the users according to their elevation angles, the ZF precoding can be approximated by the Kronecker product of two low-dimensional UPW based ZF precoding vectors, thus reducing the high computational complexity suffered by the classic ZF precoding in XL-MIMO communications. Besides the ZF/RZF scheme, the low-complexity variational message passing (VMP) receiver was proposed for multi-user XL-MIMO communications in \cite{amiri2019message}, together with a MRC processing for  initialization. It was shown that since no matrix inversion is involved, the computational complexity of VMP only linearly scales with the number of array elements and users, which is appealing for XL-MIMO communications.

 On the other hand, a distance-based user scheduling scheme was proposed for multi-user XL-MIMO communication in \cite{gonzalez2021low}, where the {\it equivalent distance} was used to approximate the equivalent channel gains for determining the user priority. Such a distance-based user scheduling achieves a comparable performance as the benchmarking scheme of ZF beamforming with successive user selection, while significantly alleviating the computational cost. Last but not least, distributed optimization algorithms, such as the dual decomposition method and the alternating direction method of multipliers, can be implemented in a parallel and low-complexity manner \cite{liu2024survey}, and thus are suitable for near-field XL-MIMO systems. The applications of these low-complexity distributed optimization algorithms in near-field XL-MIMO communications deserve further studies.

\paragraph{Distributed Processing}
 Instead of centralized processing at the CPU, the distributed processing is an alternative architecture for XL-MIMO communications to alleviate the processing complexity \cite{amiri2018extremely,amiri2022distributed,amiri2022uncoordinated,wang2020expectation}. Specifically, the XL-array is partitioned into multiple subarrays, where each subarray is equipped with a LPU and connected to the CPU. The signal processing task is then distributed to parallel LPUs, thus avoiding the large-dimensional signal processing. In \cite{amiri2018extremely}, the authors proposed a distributed receiver architecture based on distributed linear data fusion. The users were firstly detected per subarray, and the detected signals from each individual subarray were then fused at the CPU, so as to perform the final hard decision. Besides, the distributed XL-MIMO detector based on expectation propagation was proposed in \cite{wang2020expectation}, which achieved a balance between the system performance and practical implementation of XL-MIMO. In \cite{rodrigues2020low}, the authors modelled the VR under two power normalization schemes for the spatially non-stationary channel and proposed a low-complexity signal detection algorithm. By exploiting the non-stationarity, the algorithm achieved signal detection through subarray-wise processing and data fusion at the central unit.

 The distributed processing also can be applied to resolve the random access (RA) and pilot allocation in high user-density scenarios \cite{marinello2022exploring,nishimura2022fairness,alves2022crowded,alves2023noma}. Different from the centralized strongest user collision resolution (SUCRe) protocol, a 2-step non-overlapping VR XL-MIMO (NOVR-XL) RA protocol was proposed in \cite{marinello2022exploring} based on the distributed processing. Such a distributed access protocol only needs 2 steps to seek UEs with non-overlapping VRs to be scheduled in the same payload data, which reduces the access latency and improves the sum rate. Apart from the 2-step RA protocol, the works \cite{nishimura2022fairness,alves2022crowded,alves2023noma} improved the RA performance for XL-MIMO systems based on the 4-step protocol. %In \cite{nishimura2022fairness}, the authors proposed the access class barring with power control (ACBPC) protocol, which aims to provide fair access alongside the entire cell area. The LPU decodes the signals with different transmit power for each UE, which can improve connectivity and excessive access delays for edge users.
 In \cite{alves2022crowded,alves2023noma}, more than one inactive UEs were selected which can be served at the same resource through a collision resolution based on non-orthogonal multiple access (NOMA), and LPUs decoded the UEs' signals sharing the same pilot sequence and VR via successive interference cancellation strategy. The output signals are then equally combined in an equal gain combiner at the CPU, where the UEs can be identified and allocated the payload pilot. Compared to the centralized RA protocol, the distributed RA protocol with parallel processing can achieve a lower access latency and higher connectivity performance.

\subsection{Lessons Learned}
 XL-MIMO communications face various practical design issues. To fully reap the benefits of XL-MIMO, new beam codebooks dedicated to near-field beam training are essential, so as to match the new channel characteristics. Compared to far-field beam training, near-field beam training involves the codebook search in the angular-distance domains, and it is important to devise efficient beam training methods. To acquire the complete CSI, channel estimation for XL-MIMO can exploit the characteristic of spatial non-stationarity to reduce the complexity, or use the new technology of CKM. To address the issues of hardware cost and energy expenditure, the low-resolution ADCs or mixed-ADCs can be applied, together with appropriate antenna selection schemes. Besides, some low-complexity algorithms and the distributed processing architecture can be utilized to alleviate the signal processing complexity of XL-MIMO. Moreover, by exploiting the super spatial resolution brought by XL-MIMO and the multi-path sparsity of mmWave/THz channels, the new transmission technology of DAM enables ISI-free communication, without resorting to the conventional channel equalization or multi-carrier transmissions, which may complement with existing transmission technologies.

% >>>>>>>>>>>>>SECTIONS V -  here >>>>>>>>>>>>
\section{Conclusion and Future Directions}\label{sectionFutureDirections}
 \subsection{Conclusion}
 The release of 6G visions by 3GPP identifies the requirements for 6G networks, and XL-MIMO is expected to enhance the network capacity, coverage, connection density, sensing-related capabilities, and localization, thanks to its unprecedentedly high spectral efficiency and spatial resolution. Specifically, XL-MIMO is able to alleviate network congestion and improve the quality of service for users in high-density urban areas, and extend network coverage and provide high-speed connectivity to previously underserved communities in remote areas. The enhanced spatial resolution and multiplexing of XL-MIMO empowers ultra-dense connectivity, which contributes to a variety of new use cases, such as smart home, wearables, agricultures, and factories. Besides, the super spatial resolution brought by XL-MIMO lays foundation for high-accuracy wireless localization and sensing, thus enabling the future network new capabilities beyond communications. However, instead of a simple increase in antenna number or size, the evolution from MIMO and massive MIMO to XL-MIMO fundamentally changes channel characteristics and leads to a paradigm shift from far-field communications to near-field communications. In this article, we provided a comprehensive tutorial overview on near-field XL-MIMO communications, by focusing on the near-field modelling, performance analysis, and practical design issues. We first presented the near-field modelling of XL-MIMO communications, spanning from near-field array response vector, free-space LoS XL-MIMO and multi-path XL-MIMO modelling to spatial correlation based near-field modelling, as well as some important extensions and channel measurements. The performance analysis of XL-MIMO was then presented, including SNR scaling laws, near-field beam focusing pattern, achievable rate, DoF, and near-field sensing. Furthermore, we reviewed the practical design issues in near-field beam codebook and beam training, channel estimation, and the new transmission technology of DAM, followed by the discussion of cost- and energy-efficient implementation for XL-MIMO. In summary, near-field XL-MIMO communications still face new design challenges, and more dedicated research efforts are needed to devise innovative solutions. It is hoped that this article will provide fresh motivation and useful resources to inspire future research on near-field XL-MIMO communications.

\subsection{Future Directions}
 Finally, some possible future directions for XL-MIMO communications that deserve further investigation are discussed.

\subsubsection{Multi-Cell Near-Field XL-MIMO Communications}
 Most existing studies on near-field XL-MIMO communications only focus on single-cell systems. Compared to single-cell XL-MIMO communications, multi-cell XL-MIMO communications face the additional inter-cell interference. Depending on the XL-MIMO physical dimension and cell size, the UE served by the current cell can be located in the far-field or near-field region of the adjacent cell. In particular, when the cell edge UE is located in the near-field region of the adjacent cell, it will suffer from more severe inter-cell interference caused by the near-field beam focusing. As such, the near-field multi-cell interference analysis is important to investigate for multi-cell XL-MIMO communications. Besides, to address the issue of inter-cell interference, several techniques have been proposed, including CoMP and coordinated scheduling/beamforming \cite{3GPPTR36.814,gesbert2010multi}. However, it is difficult to directly apply the CoMP transmission to multi-cell XL-MIMO communications, due to a higher computation complexity and larger amount of information exchange among coordinated BSs as compared to existing MIMO and massive MIMO systems. Moreover, the CoMP transmission relies on the accurate CSI to mitigate or exploit the interference, while channel estimation for multi-cell XL-MIMO communications is practically challenging. In particular, limited by the finite number of orthogonal pilots, pilot contamination remains a critical issue in multi-cell XL-MIMO communications \cite{taniguchi2021resource,ubiali2023improving}, as in the conventional massive MIMO systems \cite{chen2022theoretical}. As a result, more research efforts are needed to investigate efficient multi-cell XL-MIMO channel estimation with pilot contamination, as well as the low-complexity interference mitigation methods.

\subsubsection{Multi-Path Near-Field Beam Training and Beam Tracking}
 Most of the existing works on near-field beam training have focused on the single LoS path setup in low mobility scenarios, while multiple NLoS paths and/or high mobility scenarios need to be studied in future work.

 Multi-path beam training is particularly important when there are multiple scatterers between the XL-array and UE. Among others, one straightforward approach for this scenario is searching over all possible locations over the non-uniform polar domain, similar to the LoS case. However, to reduce the training overhead, the near-field beam training methods designed for the LoS case cannot be directly applied. For example, the dominant angular regions of different paths may be heavily overlapped, thus making it incapable of resolving the candidate angle for different paths. Moreover, how to design the hierarchical near-field beam training for multi-path channels remains an open problem. Hence, new near-field beam training methods tailored to the multi-path case need to be developed in future work.

 Furthermore, for near-field high-mobility UEs, it is necessary to devise efficient near-field beam tracking methods to maintain high-quality links over time. Otherwise, a slight beam misalignment may result in considerable performance loss. Besides, compared to far-field beam tracking over the angular domain, the near-field beam tracking is more challenging, since it needs to keep track of both the UE angle and distance. These thus call for developing efficient near-field beam tracking methods in future work. For example, the (extended) Kalman filter based methods can be leveraged to predict the best near-field beam based on the estimated and predicted UE position and velocity \cite{zhao2023sensing}. Moreover, for mobile UEs, the UE beams in previous time slots can be used to effectively estimate the UE angle and distance in the present time slot \cite{zeng2023ckm}. The key challenge lies in how to choose the optimal beam based on the estimated parameters to meet the beam tracking accuracy requirement.

\subsubsection{Near-Field Hybrid Active and Passive Communications With XL-MIMO}
 IRS is a promising technology to configure the wireless propagation environment in favor of signal transmission and sensing, by judiciously tuning the phase shifts and/or amplitude of reflecting elements \cite{wu2021intelligentTutorial}. Besides, fully passive metal reflectors can be deployed for coverage enhancement, blind-zone compensation and rank improvement, by properly adjusting their orientations \cite{yu2023wireless}. Compared to the semi-passive IRS, the fully passive metal reflectors have the appealing features of ultra low cost, maintenance-free and full compatibility with existing wireless networks, though without the capability of dynamic adjustment as IRS. In general, to fully reap the benefits brought by IRS or metal reflector, their physical dimensions should be sufficiently large, rendering the UEs/scatterers also likely located in their near-field regions \cite{feng2023near}. This thus leads to the near-field hybrid active and passive communications, and more research efforts are needed to devise efficient near-field channel estimation methods, as well as practical active and passive beamforming designs. On the other hand, symbiotic radio with passive ambient backscattering devices achieves spectrum- and energy-efficient communications \cite{xu2023mimo}, which is appealing for XL-MIMO communications. When symbiotic radio meets XL-MIMO, accurate near-field modelling and effective near-field beamforming designs to compensate for the severe double-fading attenuation of passive backscattering links are worthy of investigation in future work.

\subsubsection{Near-Field ISAC With XL-MIMO}
 The unprecedented spatial resolution brought by XL-MIMO provides new opportunities for high-precision ISAC, which in turn facilitates XL-MIMO communications. For example, XL-MIMO may help to improve the sensing performance like estimation accuracy and sensing SNR. On the other hand, the combination of XL-MIMO and ISAC with extremely high resolution is expected to bring a paradigm shift for channel estimation, i.e., from estimating the composite channels superimposed by multi-path channel components to extracting the path information of each individual channel component, such as AoA/AoD, delay, Doppler frequency \cite{zhang2021enabling}. By leveraging the sensing-aided near-field beam training, e.g., angle and distance information, the search space of candidate beams can be significantly narrowed, thus reducing the training overhead. Moreover, with the enhanced near-field sensing, ISAC endows XL-MIMO the capability of high-accuracy localization and tracking for supporting various applications, such as autonomous driving and smart manufacturing. However, near-field ISAC with XL-MIMO faces the high signal processing complexity and hardware cost. Thus, low-complexity yet efficient sensing algorithms deserve future studies.

% >>>>>>>>>>>>> appendices -  here >>>>>>>>>>>>
%\begin{appendices}
%\end{appendices}
%\ifCLASSOPTIONcaptionsoff
%  \newpage
%\fi

\bibliographystyle{IEEEtran}
\bibliography{refXLMIMOTutorial}

\end{document}